# A Sociotechnical View of Algorithmic Fairness

*Mateusz Dolata* ✉, Department of Informatics, University of Zurich, dolata@ifi.uzh.ch

*Stefan Feuerriegel,* LMU Munich School of Management, LMU Munich, feuerriegel@lmu.de

*Gerhard Schwabe,* Department of Informatics, University of Zurich, schwabe@ifi.uzh.ch

**Abstract**

Algorithmic fairness has been framed as a newly emerging technology that mitigates systemic discrimination in automated decision-making, providing opportunities to improve fairness in information systems (IS). However, based on a state-of-the-art literature review, we argue that fairness is an inherently social concept and that technologies for algorithmic fairness should therefore be approached through a sociotechnical lens. We advance the discourse on algorithmic fairness as a sociotechnical phenomenon. Our research objective is to *embed AF in the sociotechnical view of IS*. Specifically, we elaborate on why outcomes of a system that uses algorithmic means to assure fairness depends on mutual influences between technical and social structures. This perspective can generate new insights that integrate knowledge from both technical fields and social studies. Further, it spurs new directions for IS debates. We contribute as follows: First, we problematize fundamental assumptions in the current discourse on algorithmic fairness based on a systematic analysis of 310 articles. Second, we respond to these assumptions by theorizing algorithmic fairness as a sociotechnical construct. Third, we propose directions for IS researchers to enhance their impacts by pursuing a unique understanding of sociotechnical algorithmic fairness. We call for and undertake a holistic approach to AF. A sociotechnical perspective on algorithmic fairness can yield holistic solutions to systemic biases and discrimination.

**Keywords**: algorithmic fairness, artificial intelligence, AI, decision support, information systems, IS, literature review, machine learning, ML, problematization, research agenda, sociotechnical perspective, state of the art, systematic review

# 1. Introduction

Biases in automated decision-making have negative implications for different stakeholders: (1) individuals and groups, who are at risk of discrimination and systematically inferior outcomes, (2) organizations, which may receive bad publicity and/or may suffer from legal consequences given that systematic discrimination is often penalized by law (White & Case, 2017), and (3) societies, which risk inflexible social stratification and political riots by those who have been discriminated against. Thus, it is critical to ensure fair decisions – especially those taken or supported by algorithms.

An unfair decision is often hard to detect and repair. This is particularly challenging when the decision is based on machine learning (ML). ML cannot provide meaningful explanations of how a decision was made. Thus, in case of errors, easy corrections are not possible (Mitchell, 2019). Static and rules-based algorithms can also generate biases, which remain obscured unless outcomes are systematically reviewed. Since algorithms are chosen based on their performance on the task at hand (e.g., predicting a risk) rather than on fairness, undesired effects often go unnoticed. Recent advances in algorithmic fairness (AF) claim to provide a remedy.

*Algorithmic fairness* is used to refer to technological solutions that prevent systematic harm (or benefits) to different subgroups in automated decision-making (Barocas & Selbst, 2016). From a technical perspective, AF seeks to mathematically quantify bias and, based on this metric, to mitigate discrimination in ML against subgroups. In recent years, several operationalizations and applications of AF have emerged in the information systems (IS) research (Ebrahimi & Hassanein, 2019; Feuerriegel et al., 2020; Haas, 2019; Martin, 2019; Rhue, 2019; van den Broek et al., 2019; Wang et al., 2019). These publications have often viewed unfairness in decision-making systems as a technical issue and have sought technical remedies. Thus, they have followed most other studies, which view AF as a technical discipline. However, unfairness in algorithmic decision-making is not solely a technical phenomenon: it has societal, organizational, and technical sources, is reinforced by both social and technical structures, and – as we argue – should be approached from a sociotechnical perspective.[1]

The sociotechnical perspective acknowledges that a system's outcomes depend on mutual influences between technical and social structures, as well as between instrumental and humanist values (Sarker et al., 2019). Decision-making does not happen in a purely technical context: persons are subjects *and* objects of decisions, or experience higher-level consequences thereof. However, a purely social framing may also be inappropriate if it does not consider how the proposed social solutions fit the multiple algorithms in decision-making processes. Algorithms not only support persons in taking a decision, but may also delude them, may trick them into a decision, or may simply be used as an excuse when a decision becomes unpopular (O'Neil, 2016). Thus, the social and technical components of decision-making become intertwined in many ways, demanding a broader, ecosystem-based perspective (Stahl, 2021). The sociotechnical perspective bears the potential to yield holistic solutions to the unfairness that emerges in the state-of-the-art decision-making configurations.

An effective approach to AF requires coordination and balance among technical innovation, political/legal actions, and social awareness. However, we do not yet have a unifying perspective that integrates technological and social efforts in the context of AF. Researchers across disciplines have proposed various solutions, focusing on specific aspects of AF, but have lacked a comprehensive, overarching framework to ensure coherence among the approaches. For instance, the political decision that forbids the collection of sensitive attributes (e.g., gender or ethnicity) does not align well with the algorithmic solutions that use exactly these attributes to assure that no group is systematically discriminated against. Thus, organizations have rarely considered AF to be relevant (Mulligan et al., 2019), because the proposed technological solutions are considered incomplete and thus not practically applicable. Motivated by this void, we explore the potential of adopting a sociotechnical perspective to understand the origins of algorithmic unfairness and to study its impacts in IS practice. Specifically, we contribute by (1) interrogating the premises that underlie purely technical and social perspectives, (2) discussing how a sociotechnical perspective yields new insights about AF's complex nature, and (3) showing how IS research can take a leading role in creating and implementing holistic AF solutions.

---

[1] We develop AF's meaning from a purely technological notion, as has been dominant to date, to a sociotechnical notion, positioning AF as a phenomenon. According to the former, AF comprises technological means (e.g., additional criteria implemented in the algorithm) that prevent systematic bias in an algorithm's output. In our proposed sociotechnical notion, AF describes the use of algorithmic as well as organizational or processual ways to assure that the application and output of whole decision processes that involve algorithms does not produce systematic discrimination and injustice. We use *fairness* to refer to the notion or construct being used to assess the quality of a decision or a state.

Our research objective is to *embed AF in the sociotechnical view of IS*. We address it according to the steps proposed by Alvesson and Sandberg (2011). First, we review the technical literature on AF, identifying underlying assumptions of current AF practice. We then investigate the premises behind positions that criticize current AF approaches as groundless or inadequate. Nonetheless, we acknowledge that fairness has emerged as an important social construct that may be compromised by a purely technical perspective. Simultaneously, we accept that technology will be involved in high-stakes decisions and fairness considerations. We regard AF as an inherently sociotechnical construct and elaborate on its roles in sociotechnical systems (Sarker et al., 2019). Further, we map the existing body of AF literature onto a sociotechnical perspective, identifying the points at which algorithmic (un)fairness can arise in sociotechnical systems, and then formulate research directions for IS: they show how a sociotechnical perspective can help address challenges of AF. Overall, we interrogate the fundamental assumptions that underlie current AF research and propose future research directions.

The remainder of this paper is structured as follows. In Section 2, we provide a background on the technical notion of AF (which dominates in the literature) and contrast it to the notions of fairness in other disciplines. In Section 3, we describe our overall methodology. In Section 4, we explore the premises of the literature on AF as a technical or a social construct. In Section 5, we argue that AF should be understood as a sociotechnical phenomenon. In Section 6, we employ this view to classify the origins of biases and propose directions for the IS discipline to mitigate bias.

## 2. Background on Fairness and Algorithmic Fairness

AF seeks to detect, quantify, and subsequently mitigate disparate harm (or benefits) across subgroups affected by automated decision-making (Barocas & Selbst, 2016). In this section, we review how *fairness* has been defined in AF (Section 3.1) and summarize the current discussion of how fairness has been conceptualized in other disciplines (Section 3.2). This allows us to discuss the relationships between AF and other notions of fairness.

### 2.1. Fairness as a Mathematical Construct

A technical approach to AF uses mathematical formalizations (or *notions*) of fairness. The question *What is fair?* is reduced to a single mathematical expression. The mathematical notion of fairness is integrated into algorithms as a mathematical constraint or directly into the objective function. For illustration, we use a recidivism risk assessment example in which an ML system assesses the risk that a prematurely released prisoner will commit another crime. Such systems are widely used in the U.S., with COMPAS as a typical example. However, these systems were found to show systemic discrimination against people of color, making clear that it is necessary to study AF (Angwin et al., 2016; O'Neil, 2016). Key to AF is that there are different mathematical notions of fairness (for overviews, see Barocas & Selbst, 2016; Friedler et al., 2019; Verma & Rubin, 2018). These are loosely grouped into various concepts of fairness across (i) groups or (ii) individuals.

Notions of fairness at the group level build on a predefined sensitive attribute (Barocas & Selbst, 2016) that describes membership in a protected group, against which discrimination must be prevented. In practice, the sensitive attribute is declared a priori (e.g., by policymakers or IS practitioners). Systems for recidivism risk assessment use race as a sensitive attribute to ensure fairness across people with different skin colors (Angwin et al., 2016; O'Neil, 2016). Based on this sensitive attribute, group-level notions of fairness interpret discrimination by how the prediction model's outcomes are distributed across groups: inside and outside the protected group. An exemplary notion of fairness, *statistical parity*, requires the likelihood of events to be equal across all groups. In our example, it requires the recidivism system to predict white individuals and people of color as posing a risk in the same ratio.

|  | True outcome | | |
|---|---|---|---|
|  | True $y = 1$ | False $y = 0$ |  |
| Predicted label: positive $\hat{y} = 1$ | True positive ($tp$) | False Positive ($fp$) $\rightarrow$ type-I error | Positive predictive value $ppv = P(y = 1 \mid \hat{y} = 1) = \frac{tp}{tp+fp}$ |
| Predicted label negative $\hat{y} = 0$ | False negative ($fn$) $\rightarrow$ type-II error | True negative ($tn$) | Negative predictive value $npv = P(y = 0 \mid \hat{y} = 0) = \frac{tn}{fn+tn}$ |
|  | True positive rate $tpr = P(\hat{y} = 1 \mid y = 1) = \frac{tp}{tp+fn}$ | False positive rate $fpr = P(\hat{y} = 1 \mid y = 0) = \frac{fp}{fp+tn}$ | Accuracy $= P(\hat{y} = y) = \frac{tp+tn}{tp+fp+fn+tn}$ |
|  | False negative rate $fnr = P(\hat{y} = 0 \mid y = 1) = \frac{fn}{tp+fn}$ | True negative rate $tnr = P(\hat{y} = 0 \mid y = 0) = \frac{tn}{fp+tn}$ |  |

*Figure 1. Algorithmic Fairness at the Group Level Defines Mathematical Notions Using the above Confusion Matrix*

Other group-level notions of fairness rely on prediction errors (Hardt & Price, 2016; Kleinberg et al., 2017; Zafar et al., 2017), whereby the confusion matrix from Figure 1 is considered. For instance, one notion, *equality of accuracy*, requires ML to attain equal prediction accuracies across both groups. Specifically, it requires the system to keep both the ratio of individuals who do not pose a risk (and are recommended for release from prison) and those who pose a risk (and are recommended to remain in prison) the same across white prisoners and those of color. As can be seen in the different notions, there is no universal operationalization of fairness; further, it is mathematically impossible to fulfill all the different notions of fairness at once (Chouldechova, 2017; Kleinberg et al., 2017). Thus, the IS designer must choose an appropriate mathematical notion for assuring an unbiased outcome, yet one for which little guidance is available. Explainable or interpretable ML (Caruana et al., 2020; Mitchell, 2019; Molnar, 2020; Senoner et al., 2021) can offer support by allowing practitioners to observe both the outputs of their models and the (potential) reasoning that support them. However, explainable ML often delivers insights that are not perceived as useful by persons (Molnar, 2020). Further, all notions of fairness share the same key requirement: within the data, the sensitive attribute must be available to algorithms, so that potential discrimination can be mitigated, even though collecting such data may be illegal owing to the sensitive nature of the attributes.

Notions of fairness at the individual level are based on the assumption that similarly situated individuals should be treated similarly (Dwork et al., 2012). This approach strives to ensure fairness independent of group membership. This requires a definition of similarity that is suitable in the given use case and provides the basis on which to perform pair-wise comparisons (rather than group-wise ones). In recidivism risk assessment, this requires two individuals whose relevant attributes (criminal history, sentence length, etc.) are equal to be subject to the same decision. As in group-level fairness, individual-level fairness requires access to sensitive attributes. It also leaves open how to specify which attributes are relevant or how to formalize relevant yet nonquantitative attributes (e.g., psychological instability or addiction).

## 2.2. Fairness as a Social Construct

*Philosophical* debates about fairness were originally driven by the question of the distribution of goods and rights and the utopian ideal of a fair ruler. Today's understanding of fairness suggests that all persons with equal gifts should have equal opportunities for advancement, regardless of their initial position in society (Rawls & Kelly, 2003). In short, equal distribution of chances for self-advancement to achieve equity in goods distribution (an individual's benefits are proportional to their input) dominates the current philosophical debate. Thus, it is unfair to prevent individuals from improving their situation by limiting chances – intentionally or otherwise – based on race, origin, or gender.

*Law, criminology*, *sociology*, and the *political sciences* often take a modern philosophical approach to addressing fairness. This leads to the emergence of restorative, transitional, or retributive justice concepts, in which society (re)establishes a fair distribution by repair, re-balancing of power, or punishment (Clark, 2008), in which fairness is neither given nor predefined. Instead, it is continually constructed (Lind et al., 1998). This line of reasoning stresses actionable notions of social fairness: constantly trying to balance several forces and preserve fairness. AF has emerged as a defining and affirming force. However, its mathematical limitations are new, compared to the social construct.

*Anthropology* focuses on the historical origins of fairness as an innate human value also observed in other primates (Brosnan, 2013). For instance, persons and primates who contribute the same amount of work as others but receive a lesser payoff are likely to stop working (Brosnan & de Waal, 2014). They tend to punish those engaged in the unequal distribution of goods (Brosnan, 2013). This suggests that the sense of fairness emerges from evolution toward collaborative societies. Since collaborative behaviors are relevant for survival, fairness propagates over generations (Hamann et al., 2011). Thus, primates may have a built-in fairness calculation mechanism. AF may look like an attempt to discover this mechanism, which is likely to include as yet unknown aspects that are hard to quantify.

*Neuroscience* looks for the origins of fairness in the human brain rather than in history. Brain processes concerning fairness occur in an evolutionarily old brain area (Vavra et al., 2017). This supports claims from anthropology and suggests that fairness is emotional and can be experienced as an intrinsic need (Decety & Yoder, 2016). A neurological rewards system is activated when goods distribution reaches a fair state (Vavra et al., 2017). However, this process is moderated by external factors, including similarity among affected individuals, situational identity, or salient personal goals (Bargh, 2017; Cohn et al., 2014). This may explain why people often disagree about what is fair. Thus, achieving fairness remains an ongoing process rather than a one-time challenge.

Finally, *psychology* sees fairness as a perception of an individual comparing themselves to other individuals whom they consider to be relational partners (i.e., as somehow similar to them) (Adams, 1963). This view was adopted (and extended) by organization science, with three aspects differentiated: distributive (who receives what?), reciprocal (is the treatment or benefit adequate to a person's input?), and procedural (how was the decision taken?). These aspects led to the formulation of *organizational justice* theory (Greenberg, 1986). Reciprocal fairness demonstrates that persons expect truthful and respectful treatment if they act accordingly (Bies, 2001). Procedural fairness requires that individuals affected by a decision be given provided with justification and be allowed to contribute to the decision and voice concerns (Greenberg, 1986). Organizational justice was adopted in the IS research and is often used to study fairness perceptions inside organizations and their relationships to technology-driven organizational change (Joshi, 1989; Li et al., 2014; Tarafdar et al., 2015). While all aspects of organizational justice may be affected by algorithmic bias, the distributive aspects receive the most attention (Robert et al., 2020).

# 3. Methodology

We engage in *problematization* so as to establish an informative research agenda for IS and the sociotechnical notion of AF. Problematization is an approach for developing research questions from a body of literature. We explicate the underlying assumptions in existing studies and question them (Alvesson & Sandberg, 2011). This frames research as an ongoing dialogue that relies on challenging the status quo rather than gap-filling. Problematization is a way to facilitate more influential management and organizational literature theories (Alvesson & Sandberg, 2011). It has been promoted in the recent IS literature (Avital et al., 2017; Grover & Lyytinen, 2015; Templier & Paré, 2018) and was adopted in earlier IS studies (Ortiz de Guinea & Webster, 2017). Since the research into AF in IS is still emerging (Dolata & Schwabe, 2021; Ebrahimi & Hassanein, 2019; Haas, 2019; Kordzadeh & Ghasemaghaei, 2021; Marjanovic et al., 2021; Martin, 2019; Rhue, 2019; van den Broek et al., 2019; Wang et al., 2019) and relies on perspectives on AF from reference disciplines (most prominently, computer science), we chose to interrogate assumptions that characterize the AF discourse via a multidisciplinary review. Alvesson and Sandberg (2011) differentiated between various categories of assumptions that differ in depth and scope: in-house, root metaphor, paradigm, ideology, and field assumptions. For more detailed descriptions of each category, we refer the reader to Alvesson and Sandberg (2011). We use this classification to assess the identified assumptions' impacts. This provides a solid base for theorizing AF in sociotechnical perspectives of IS.

    The literature we analyzed was collected from different sources and was then classified. On the one hand, we screened articles from four key AF conferences (KDD, ICML, NeurIPS, FAccT). This led to 166 candidate articles for analysis. On the other hand, we conducted a query-based search across the top 25% of outlets from a multidisciplinary background (e.g., management, psychology, etc., according to Scimago Journal & Country Rank), and conducted a criteria-based selection within them. This led to the selection of 114 more articles. In a subsequent step, we classified all articles by their approaches to AF (technical vs. social), focus (social component, technical component, data and information, adaptation between components, broader context), scope (generic vs. limited to a specific application domain), and methodological paradigm (engineering, exploratory, literature review, critical, behavioral, formal). Further, we listed and analyzed the implicit and explicit assumptions according to the

identified approaches. The Appendix contains complete descriptions of the steps involved in the data collection and analysis.

## 4. Problematizing Algorithmic Fairness

In this section, we problematize AF. We explicate the premises of the papers that were classified as following the technical approach (Section 4.1), and studies that tracked a more social perspective (Section 4.2). We infer assumptions that underlie AF based on a systematic literature review (for details, see the Appendix). Specifically, we identified common assumptions in the literature, classified as having either a technical or a social orientation, and identified articles that can be used to exemplify the assumptions we found (see Tables 1 and 2). When describing the assumptions, we refer to articles from the literature review. The analysis revealed that the articles lacked a shared, coherent agenda. Further, at first sight, the assumptions they make may even contradict one another. Not all the assumptions we list in the following sections co-exist in each paper we considered – the papers differed concerning the assumptions they rely on. While the literature, independent of its core approach, provided valuable inputs, it has not conceptualized AF as a sociotechnical construct, despite the overall goal of preventing technology-based societal discrimination shared across fields, perspectives, or attention foci. We don't intend to invalidate the research in the reviewed studies or to suggest the existence of irreconcilable camps. We argue that AF *should* be seen as a sociotechnical construct (Section 4.3). This perspective can reconcile the approaches to AF.

### 4.1. The Premises of the Technical Perspective

Identification of the assumptions that underlie the perspectives is the first step toward a unified, sociotechnical understanding of AF and an agenda for advancing the research in a more coherent, holistic direction.

The literature on AF has been dominated by the technical perspective. Accordingly, AF has been defined as efforts to "translate regulations mathematically into non-discriminatory constraints, and develop predictive modeling algorithms that take into account those constraints, while at the same time be as accurate as possible" (Žliobaitė, 2017), or as "the aim of assessing and managing various disparities that arise among various demographic groups in connection with the deployment of ML-supported decision systems in various (often allocative) settings." (Fazelpour & Lipton, 2020). While there is nothing wrong with the aim of obeying "non-discrimination constraints" or "managing disparities," this perspective is restricted by the underlying notion of fairness. The technical literature relies on a range of paradigmatic assumptions, i.e., shared beliefs, definitions, and methodological approaches (Alvesson & Sandberg, 2011). We will now review these assumptions (see Table 1) along with appropriate examples from the reviewed studies.

First, the proposed solutions often assume a comprehensive a priori understanding of where and why biases occur (Abràmoff et al., 2020). The studies frame specific biases as problems in search of a solution. In a good engineering way, the studies focus on selecting a notion of fairness that is appropriate to this bias, then develop a mathematical construct to represent it, and use it in the algorithms. They posit that there is a mathematical or formal way to adequately resolve biases without generating new, previously unknown biases (Dutta et al., 2020). The studies follow the *engineering assumption*: "Each problem has a human-made technical solution. A solution is good when it solves the problem."

Second, technical AF posits a conceptual equivalence of various notions of fairness. There are concurrent fairness ideals (equity, equality, need, etc.), each with varying social connotations (Binns, 2018; Narayanan, 2018; Rawls & Kelly, 2003). Still, most technical researchers treat them synonymously and select among them as if they were interchangeable. There is no consistency concerning what should be considered when selecting a fairness ideal, how the selection should be conducted, or what the long-term social consequences are of employing any of the measures. For instance, some studies have simply referred to court cases or have repeated arguments from ethics or political philosophy (Fazelpour & Lipton, 2020). Others have advocated including the selection process for an adequate notion of fairness in a participatory design process (Ahmad et al., 2020) or a survey (Srivastava et al., 2019b). These approaches have moved the decision on an adequate notion of fairness away from the designer or the researcher to the broader public, who is presented with various notions of fairness and is asked to choose among them. Yet, even in these cases, the implications of the choice have barely been considered. Depending on the population sampling method, this method can even exhibit bias. Further, the thought experiments and hypothetical situations used in these approaches were shown to often fail (Gendler, 2014). Overall, each selection approach makes an *equivalence assumption*, assuming that fairness ideals are to some extent equivalent and that choosing between them takes place in a closed environment, as opposed to – open-ended – reality.

Third, having chosen a fairness ideal, the technical literature has posited a mathematical operationalization of it. Owing to their general and abstract formulation, many ideals that rely on equity or equality don't directly fit mathematical disparity measures (Fazelpour & Lipton, 2020), because the ambiguity of ethical and legal rules allow human judges to deliberate about a concrete situation, balancing the distribution of goods or rights. Developing a strict metric involves value assessments without a specific context or situation – often presented as a *translation* (Narayanan, 2018; Žliobaitė, 2017). This induces the illusion of mathematically expressing abstract notions without either losses or unexpected consequences. We refer to this as the *translation assumption*.

Fourth, the need to quantify a notion of fairness pushes researchers and practitioners to primarily focus on distributive justice (Fazelpour & Lipton, 2020). Yet fairness may go beyond the distribution of goods, to address interactional or procedural justice. If a system cannot be used by some people of color because facial recognition does not work properly for them, this impacts on their dignity and belongs to the interactional justice dimension (Celis Bueno, 2020; Hanna et al., 2020; Robert et al., 2020). Many technical studies implicitly reformulate this as a distributive justice concern (e.g., the distribution of properly recognized faces across groups). Further, the studies have focused on between-subject justice across groups or individuals (comparison with others) and not on within-subject justice (comparison with the subject's engagement). This can generate perceptions of unfairness. An individual could be treated differently at two different points in time despite behaving in exactly the same way, because others changed *their* behaviors and the system adopted the altered distribution. These examples highlight the consequences of a *distributiveness assumption*, which posits that all fairness issues can be presented as a statistical distribution.

While a statistical engineering approach to AF has dominated the current debate, there are alternative approaches to assuring fairness through technical interventions.

However, they suffer from further assumptions and often commit to one or more of the flawed assumptions (such as the *engineering assumption* or the *translation assumption)*.

The counterfactual fairness discourse has pursued a view of AF that uses directed acyclic graphs to model social biases that may occur in data (Coston et al., 2020; Kusner et al., 2017). This line of research assumes that social bias is a global causal structure element and needs to be explicitly modeled in algorithmic decision-making (Kusner et al., 2017). Alternatively, some researchers see the origins of bias in data and treat the problem as a database repair process. They work to achieve the desired balance and train the models using a sort of ideal, bias-free dataset rather than real data (Salimi et al., 2020). While some biases ("foreigners commit more crime than locals") or protected groups ("people of color") may be easier to identify and explicate, others are more implicit, limited to a local community, or simply rely on urban myths and fake news ("small people drive large cars," "obese people lack self-discipline," etc.). Some members of society may feel stigmatized if social biases need to be explicated and presented in a model or data. Positing that the biases can be comprehensively explicated, many approaches fall victim to an *explicitness assumption*.

Finally, almost all engineering approaches to unfairness share *independence assumptions*. We identified three assumptions concerning independence: (1) context-independence, (2) time-independence, and (3) component-independence, as follows: (1) Context-dependence refers to whether a one-size-fits-all approach is replaced by developing tailored, problem-specific solutions for AF. Here, most technical studies use multiple 'synthetic' or 'publicly available' datasets that represent various independent decision problems (recidivism prediction, diabetes treatment, or creditworthiness). These studies developed and evaluated solutions with these datasets and claim applicability across situations (Celis et al., 2018; Coston et al., 2020; Valera et al., 2018; Zafar et al., 2017). This contradicts the evidence that proved that fairness assessments are highly context-dependent (Srivastava et al., 2019b; Wong, 2020; Zafar et al., 2017). (2) Time-dependence refers to whether AF solutions consider dynamics of the environment. Here, technical studies on AF often rely on data from the past (U.S. Census from the 1990s) to prove their system's fairness, but prescribe future use (Bera et al., 2019). They see data and the decision context as static, committing to a time-independence assumption. (3) Component-dependence denotes that individual components of an AF system interact with the environment. Previous studies that followed a technical perspective on AF focused on improving and testing a single classifier, ignoring other technical components that will interact with it. The classifier will be like a small gear-wheel in a larger technological system involving classifiers, data pre-processors, and interfaces to other systems. In a broader sense, it will become part of a complex sociotechnical system whose characteristics emerge not only from its parts but also from interactions between the parts and the context (Mitchell, 2009). Because "a system is more than the sum of its parts" (Ackoff, 1973), there is no guarantee that a system composed of fair parts will in fact be fair (Dwork & Ilvento, 2018). Thus, these studies commit to a component-independence assumption. The three abovementioned independence assumptions prevail explicitly or implicitly, despite singular studies that paid attention to context (Chouldechova et al., 2018; Kallus & Zhou, 2019; Rahmattalabi et al., 2019), accepted temporal dynamics of fairness issues (D'Amour et al., 2020; Liu et al., 2018), or addressed technical systems in a holistic way (Dwork & Ilvento, 2018). They represent notable exceptions rather than the norm.

*Table 1. The Typical Assumptions of the Technical Approach to Algorithmic Fairness with Exemplary Quotes from the Reviewed Papers*

| Assumption | Example |
| --- | --- |
| *Engineering assumption:* for any problem, there is a technical solution, and the solution is good if it solves the problem at hand. | "We identify the insufficiency of existing fairness metrics and propose four new metrics that address different forms of unfairness. These fairness metrics can be optimized by adding fairness terms to the learning objective. Experiments on synthetic and real data show that our new metrics can better measure fairness than the baseline, and the fairness objectives effectively help reduce unfairness." (Yao & Huang, 2017) |
| *Equivalence assumption:* operationalizations and notions of fairness are equivalent and can be exchanged based on their performance. | "In legal scholarships, the notion of fairness is evolving and multi-faceted. We set an overarching goal to develop a unified machine learning framework that can handle any definition of fairness, the combinations, and also new definitions that might be stipulated in the future." (Quadrianto & Sharmanska, 2017) |
| | "While the problem of selecting an appropriate fairness metric has gained prominence in recent years, is perhaps best understood as a special case of the task of choosing evaluation metrics in machine learning." (Hiranandani et al., 2020) |
| *Translation assumption:* complex, ambiguous notions of fairness or legal rules can be | "Although the DI doctrine is a law in the United States, violating the DI doctrine is by itself not illegal; it is illegal only if the violation cannot be justified by the decision- |
| translated into mathematical or statistical terms without loss. | maker. In the clustering setting, this translates to the following algorithmic question: what is the loss in quality of the clustering when all protected classes are required to have approximately equal representation in the clusters returned?" (Bera et al., 2019) |
| *Distributiveness assumption*: representing problems of interactional or procedural justice in terms of distributive justice so as to facilitate statistical processing. | "Though helpful in seeing a systematic error, gender, and skin type analysis by themselves do not present the whole story. Is misclassification distributed evenly amongst all females? Are there other factors at play? Likewise, is the misclassification of darker skin uniform across gender?" (Buolamwini & Gebru, 2018) |
| *Explicitness assumption*: existing prejudices, social biases, and protected groups can be known upfront and can be made explicit in the model (affects especially counterfactual modeling of fairness). | "We advocate that, for fairness, society should not be satisfied in pursuing only counterfactually-free guarantees. (…) We experimentally contrasted our approach with previous fairness approaches and show that our explicit causal models capture these social biases and make clear the implicit trade-off between prediction accuracy and fairness in an unfair world. We propose that fairness should be regulated by explicitly modeling the causal structure of the world." (Kusner et al., 2017) |
| *Independence assumptions*:<br><br>• context-independence<br>AF solutions proven to work on a range of datasets can be transferred to applications in different contexts without loss<br>• time-independence<br>AF solutions proven to work on historical data can be transferred to current and future applications without loss<br>• component-independence<br>AF solutions proven to work in isolation from other technical components can be transferred to complex technological artifacts without loss | "Based on an online learning algorithm, we develop an iterative algorithm that provably converges to such a fair and robust solution. Experiments on standard machine learning fairness datasets suggest that, compared to the state-of-the-art fair classifiers, our classifier retains fairness guarantees and test accuracy for a large class of perturbations on the test set." (Mandal et al., 2020)<br><br>"We adopt surrogate functions to smooth the loss function and constraints, and theoretically show that the excess risk of the proposed loss function can be bounded in a form that is the same as that for traditional surrogated loss functions. Experiments using both synthetic and real-world datasets show the effectiveness of our approach." (Y. Hu et al., 2020) |

## 4.2. The Premises of the Social Perspective

The social perspective on algorithmic (un)fairness is becoming increasingly important. Positions with a social perspective on AF are appearing at computer science conferences and in journals (Barabas et al., 2020; Binns, 2018; Mulligan et al., 2019; Robert et al., 2020) as well as in outlets from other disciplines, including the social sciences (Hoffmann, 2019), philosophy (Mohamed et al., 2020; Wong, 2020), and criminology and law (Barocas & Selbst, 2016; Helberger et al., 2020; Završnik, 2019). This makes clear that researchers across disciplinary boundaries are engaging in the social aspects of AF and are trying to understand it as a problem that cannot be completely solved through technology. For instance, they indicate that sources of algorithmic unfairness go beyond the issue of unbalanced data and derive from a lack of political power balance (Barabas et al., 2020; Mohamed et al., 2020) or the lack of political discourse about what fairness is (Wong, 2020; Završnik, 2019). Other studies have described the status quo from the perspectives of social science (Helberger et al., 2020) or organizational science (Robert et al., 2020). The multidisciplinary debate aims to identify and overcome the limitations of the technical approach. Most studies refer to a common-sense image of ML and, in the context of problematizing (Alvesson & Sandberg, 2011), suggest the presence of different assumptions that fall under the concept of root metaphors. The assumptions relate to a general understanding of the subject matter shared beyond a single discipline; in this case, it is the shared notion of ML and algorithms engineering as being solely about data and their processing. We will now review these assumptions, listing them in Table 2 along with examples from the reviewed studies.

The studies that follow a social perspective have addressed differences between the mathematical notions or even methods used to represent fairness in algorithms. While the technical community has developed a wide variety of methods for reducing discrimination in ML (e.g., counterfactual reasoning or debiasing for textual data), they were rarely considered in social discourses as a separate way of operationalizing fairness (Binns, 2018). However, as discussed, these methods may have crucial implications for both the technology's design and for the sociotechnical context: some methods require explications of potential social biases, while others rely on the availability of sensitive data, and yet others manipulate the data to reduce risks of bias. This explicates that social discourse of AF treats technical approaches to AF as a *black box*, without decoding the notions of fairness encoded by the engineers and their social, political, or organizational implications. Although the social perspective acknowledges AF-related progress of the technical perspective, the engagement with the subject matter has remained superficial. It often seemed that technical AF was reduced to a generic *algorithmic* approach. Given that the variety of technical AF and the interplays between specific social and technical measures remain unpacked, we refer to a *black box assumption*.

Some studies have engaged in a *bad actor* debate: they try to identify who is to blame for unfairness in automated decision-making. There are two general lines of argumentation: some argue that the application of ML in high-stakes decision-making is the problem (algorithms or big data is the bad actor) (e.g., Barocas & Selbst, 2016; Završnik, 2019), while others argue against developers, designers, and organizations who provide these solutions (Kuhlman et al., 2020; Mohamed et al., 2020; Wong, 2020). We call the first tendency the *technology agency assumption*, and the second the *human agency only assumption*. Others propose a shared responsibility: "humans and algorithms co-conspire in upholding discrimination." (Hoffmann, 2019).

We agree that understanding liability for discrimination is important to countering it with regulation. It is necessary to understand who or what the source of discrimination is. Nonetheless, some aspects of this debate may benefit from acknowledging recent developments in the technical approach to AF, for instance that decisions concerning fairness measures are often taken through participation of a broader public (M. P. Kim et al., 2020; Srivastava et al., 2019b). While we understand that it is important to identify the origins of biases, the discourse often does not reveal how exactly the bad actor in question impacts on the algorithmic decision-making and which technical components are affected – yet such deliberation could support the technical AF community in approaching the key points (Ågerfalk, 2020; Draude et al., 2019; Hoffmann, 2019; Ziewitz, 2016).

Finally, several authors, especially in legal science, have idealized the status quo in law enforcement and have declined algorithmic support (e.g., Beyleveld & Brownsword, 2019; Huq, 2018; Johnson, 2020; Završnik, 2019), while others have proposed replacing the current system with a 'code-based' sentencing (Chandler, 2019; De Filippi & Hassan, 2018; Kalpokas, 2019; Lessig, 2000). Each side has committed to a *purity assumption*, claiming that only it yields unbiased decisions. However, there is cumulated evidence that unfair decisions have been made in the past, independent of whether or not they relied on human reasoning or algorithms (Gladwell, 2019; O'Neil, 2016; Payne et al., 2017). Biases in law enforcement may emerge from existing training, incentive systems, or archetypes in the organizations (Gladwell, 2019) – and these biases may be present in both persons and in algorithms. Instead of claiming that one or the other is better or fairer, one must acknowledge that "accountability mechanisms and legal standards that govern decision processes have not kept pace with technology." (Kroll et al., 2016). We claim that coordinating algorithm development with overall justice system development may lead to law enforcement with fewer systematic biases.

*Table 2. Typical Assumptions of the Social Approach to Algorithmic Fairness with Exemplary Quotes from the Reviewed Papers*

| Assumption | Examples |
|---|---|
| *Black box assumption*: Seeing the technical perspective on AF as a coherent field without acknowledging differences between the approaches and notions, and without attending to these choices' social, organizational, or ethical consequences; also, suggesting improvements based on the social perspective without attending to the technical implications thereof | "Research in algorithmic fairness has recognized that efforts to generate a fair classifier can still lead to discriminatory or unethical outcomes for marginalized groups, depending on the underlying dynamics of power, because a "true" definition of fairness is often a function of political and social factors. Quijano (2000) again speaks to us, posing questions of who is protected by mainstream notions of fairness, and to understand the exclusion of certain groups as 'continuities and legacies of colonialism embedded in modern structures of power, control, and hegemony'." (Mohamed et al., 2020) |
| *Bad actor assumption*:<ul><li>technology agency unfairness emerges as the result of applying technology to the taking of decisions</li><li>human agency only unfairness always emerges because of humans; technology just perpetuates human biases or a power imbalance</li></ul> | "Of course, not all work in this area reduces discrimination entirely to some set of 'blameworthy' humans behind the machine. Many discussions make clear that algorithmic discrimination can happen in ways that are unintentional or difficult to account for, for example when upstream social biases are reflected in training data in ways that may be difficult to predict. In these cases, biases are said to 'sneak in', 'whether on purpose or by accident', or in ways that only emerge over time." (Hoffmann, 2019) |
| *Purity assumption*: seeing either the justice system or algorithmic decision-making as superior and as the reference point for fairness | "Second, it shows why automated predictive decision-making tools are often at variance with fundamental liberties and also with the established legal doctrines and concepts of criminal procedure law." (Završnik, 2019) |

### 4.3. The Need for a Sociotechnical Perspective

Overall, the current AF discourse suffers from assumptions that render exchanges between various approaches difficult. Especially the relationships between social and technical aspects of AF have largely been overlooked. The social and the technical perspectives both provide valid points and remedies. Although some articles have addressed the relationship between the social and the technical, they have rarely moved beyond identifying problems in each area. Thus, the efforts have not added up to a holistic and comprehensive solution. In our view, this results from a selective perception of what algorithmic (un)fairness is: some researchers see it as a technical phenomenon and seek solutions in technology; others see it as a symptom of discrimination in society and seek a remedy in changing the social structures that enable discrimination. Both approaches make valid points. They also don't directly contradict each other: at first glance, enhancing algorithms or manipulating data does not interfere much with political or social agendas. However, this development may confuse practitioners, decision-makers, and society. Apart from clearly misguided decisions, such as when the lack of sensitive attributes in the data (a political decision) disables the use of most effective AF solutions (a technical decision), other problems may also emerge. Thus, we address why a sociotechnical perspective is needed: it discusses practical consequences of a potential sociotechnical framing of AF and therefore supports the overall statement that we urgently need a sociotechnical perspective.

Without a coherent perspective that acknowledges the interdependencies between the social and the technical aspects of AF, organizations may be reluctant to effectively tackle this problem. If they treat algorithmic (un)fairness as a purely technical problem, they may assume that adding a social element will sufficiently solve unfairness. They may believe that positioning an employee as a control instance who needs to sign off decisions made by the algorithm will sufficiently mitigate discrimination. This aligns with the *human-in-the-loop* claim, according to which introducing human control into algorithmic decision-making will prevent or limit unintended consequences of purely algorithmic decision processes (Brockman, 2019; Marjanovic et al., 2021). Following the sociotechnical perspective, we argue that this reasoning is problematic. It assumes that the human and the algorithm are distinct moral agents capable of making an independent decision, and it implies a picture of a righteous and critical person who is able to question algorithmic output or assess the focal situation. We argue that the algorithm and the person are not as independent as it may seem. Algorithms empower and constrain persons: they may direct human attention to only some aspects to be considered; they may require a specific decision output format; and they may require the person to take decisions in a decontextualized environment. Likewise, persons may inappropriately interpret the algorithm's output or take random, uninformed decisions, following the illusion that their opinion is just one out of several votes. For instance, it is known that employees rely on decisions taken by the algorithms and rationalize or explain them rather than controlling their quality and bias (Rhue, 2019). Thus, rather than calling for *a human in the loop*, we need to gain an understanding of human-algorithm ensembles as collective moral agents and thus respect the complex mutual influences between the ensemble's subparts (Verbeek, 2011, 2014). In our view, the sociotechnical perspective on AF is the first step in this direction.

The evaluation of complex decision processes involving persons and algorithms as parts of an ensemble requires an overall approach to assess whether the ultimate outcomes produced by the sociotechnical system are fair and to identify reasons for potential unfairness. If one focuses on the technical and the human components separately, they ignore unexpected interferences between the components, risking an unfair final decision. Finally, without holistic guidance, companies risk choosing a combination of incompatible technical and organizational fairness measures, especially if such decisions are made by different units (e.g., IT and human resources). Accordingly, we argue that it is crucial to view AF as a sociotechnical construct.

IS has a long tradition of systemic sociotechnical approaches to solving urgent and important problems. For instance, IS encapsulated the social and psychological effects of organizational implementation of new technologies in the sociotechnical concept of *technostress* (Tarafdar et al., 2017). IS is also concerned with supporting collaboration that uses a mix of technical and social components – *collaboration engineering* (Briggs et al., 2003). Finally, IS offers a holistic understanding of *trust* as a phenomenon not limited to persons but also emerging in relation to technology (Söllner et al., 2012, 2018). We see great potential for approaching fairness as a sociotechnical phenomenon. While decisions in all domains will increasingly rely on algorithmic processing of data and will involve ML predictions (Agrawal et al., 2018), they will also involve a person as a (co-)decision-maker, target of the decision, or evaluator. It is essential to understand interactions between technical and social components to embrace the complexity of fairness. While existing research makes valuable contributions to either technical or social aspects, the IS discipline – owing to its sociotechnical anchoring – can better understand how these efforts complement, depend on, and mutually influence one another. Nonetheless, the studies of AF in IS have resorted to the technical perspective and have focused on ways to improve algorithms through for instance more adequate scoring methods (Wang et al., 2019), better assessment of tradeoffs (Haas, 2019), or better understanding of current applications (van den Broek et al., 2019). This has led to some very recent suggestions or calls to extend the notion of AF to embrace the behavioral, procedural, and contextual aspects of algorithmic decision-making (Kordzadeh & Ghasemaghaei, 2021; Marjanovic et al., 2021). Although those extensions and calls are a step in the right direction, they have neither explicitly problematized the assumptions that underlie existing formulations of AF, nor addressed the dynamics of interaction and balance between the social and the technical aspects of AF. In our view, IS needs a reorientation toward the sociotechnical perspective if it is to provide a holistic understanding of AF as a phenomenon.

## 5. A Sociotechnical Perspective on Algorithmic Fairness

Clearly, neither a technical nor a social view alone is sufficient. We will now position AF as a sociotechnical phenomenon.

The sociotechnical perspective has formed the foundation of IS research for decades (Davison & Tarafdar, 2018; Lee et al., 2015; Sarker et al., 2019). It builds on the key insight that work involves interactions between persons and technology. Persons, including individuals and collectives, as well as the relationships among them or attributes thereof – including structures, cultures, economic systems, rituals, best practices, organizations, or social capital – form the *social component* (Lee, 2004; Sarker et al., 2019). The technology, including human-made hardware, software, data sources, and techniques that describe ways of using them to achieve human goals or serve human purposes form the *technical component* of a sociotechnical system (Lee, 2004; Sarker et al., 2019). The sociotechnical view stresses the *mutual interdependency* between the components, so much so that connections between them are reciprocal and iterative, but neither incidental nor nominal (Lee, 2004). The social and technical components engage in *joint optimization* to create a productive sociotechnical system (Sarker et al., 2019). The IS tradition has also acknowledged that components should be treated *equivalently* regarding importance and impact (Beath et al., 2013). A sociotechnical account of AF requires careful consideration of how machines and persons can and should co-engage or collaborate to achieve fairness.

We will now first examine the characteristics of AF that suggest the sociotechnical lens is most appropriate to attend to it, providing arguments for why it is a sociotechnical rather than a social or a technical phenomenon (Section 5.1). We will then show how the sociotechnical view addresses the limitations of existing AF research (Section 5.2). We will refer to the rolling example of a recidivism prediction system and will use this example at various points to ease the understanding of an abstract matter.

**5.1. Why is Algorithmic Fairness a Sociotechnical Phenomenon?**

Multiple characteristics of AF position it as a sociotechnical phenomenon. First, the algorithm creation process is a social practice. Developing algorithms is to some extent a research activity driven by epistemic values, including consistency, accuracy, or generalizability (Laudan, 1968). Similarly, contextual values that replicate the developer's personal or humanist concerns are equally important (Friedman et al., 2013; Kincaid et al., 2007; Nissenbaum, 2001; van de Poel & Kroes, 2014). The developer's background may impact on their perception of what is fair and for whom. Since definitions of fairness relate to the stakeholders' interests, developers could tend to prefer some fairness measures over others (Narayanan, 2018; Wong, 2020). Second, algorithms inevitably impact on the lives of individuals, groups, and societies based on where they are used and what they are used for (Draude et al., 2019; Mohamed et al., 2020; O'Neil, 2016). A widely used algorithm for selecting healthcare system entry was found to discriminate against racial minorities, thereby affecting thousands of people (Obermeyer et al., 2019). Finally, algorithms have and are becoming the object of public debate around AF (O'Neil, 2016). Algorithms have long been an object of sociotechnical practice.

At the same time, algorithms are involved in fairness assessments. Decades ago, the justice system moved from narrative-based consideration of cases to prosecution that relies on ML techniques (Aas, 2006; Harcourt, 2015). For instance, algorithms are used to predict areas in need of policing, so as to automatically identify potentially criminal individuals online, or for analysis of biological or computer data acquired during prosecution (Harcourt, 2015). All these applications bear risks of discriminating: these systems' accuracy may be higher for some types of crimes or for some ethnic groups. Similarly, digital technology was shown to restrict freedom of public administration – the mythical 'computer' rather than every officer or the organization was taking decisions about what a fair welfare subsidy is (Dolata et al., 2020; Landsbergen, 2004). Although the publicity identified humans as decision-makers accountable for fairness, in fact, the work was and continues to be distributed between social and technical components. As presented here, there is not only a need to consider AF as a sociotechnical phenomenon, but there are good reasons to do so.

### 5.2. Developing a Sociotechnical Perspective for Algorithmic Fairness

We will now develop a sociotechnical perspective for AF, and start by discussing basic constructs relevant to the sociotechnical view of AF. We will also provide examples from the case of recidivism risk assessment (Dieterich et al., 2016).

The sociotechnical perspective focuses on interactions between the *technical* and *social components* of an IS. For instance, the recidivism prediction case involves judges, penitentiary workers, inmates, and their attorneys, as well as institutions and law enforcement rules. All these individuals and collectives form the social component of the prison release decision system – they take decisions or are directly affected by them. When individuals take decisions in companies or organizations, they widely rely on decision support systems. Such systems, like any technological system, have various components. In the AF case, the component for ML is particularly relevant (i.e., as it predicts the likelihood of a person committing a crime again after release). Through *reciprocal interactions*, the components achieve coherence (harmony, fit, joint optimization), which results in an effective IS (Lee, 2004). All the individuals involved in penitentiary processes and the tools they use are engaged in continual adaptation. They establish new work practices that allow them to take better decisions, while the algorithms are retrained based on decisions taken or are simply changed to reflect new rules or routines. Based on this, we propose that a sociotechnical view of AF assumes complex relationships between social and technical components, such that the working of the overall system cannot be derived from structure or internal processes of its components. This specifically implies that we cannot predict that an overall IS will become fairer by only improving the technical component's fairness.

An effective IS should lead to better *instrumental* and *humanistic outputs* (Sarker et al., 2019). Decision systems are typically employed to improve the decision accuracy while reducing the decision costs (Power, 2008). For instance, a recidivism risk analysis system should relieve the overcrowded justice system, reduce processing time for jail release applications, provide prison inmates with earlier decisions, and lead to more frequent application processing cycles. At the same time, it must obey ethical norms, including fairness. Thus, a sociotechnical view of AF assumes multiple interrelated or even contradictory outcomes beyond fairness. This implies that fairness cannot be seen as a unique goal. Instead, the overall system should be evaluated against multiple goals, including fairness, where fairness is a necessary condition but is not sufficient to ensure that the system is useful.

An IS is embedded in an *environment* – a larger social, economic, regulatory, or material context, which offers structures for the IS's operation (Briggs et al., 2010; Dourish, 2001). COMPAS, the aforementioned recidivism prediction system, was improved based on societal pressure from NGOs (Washington, 2018). Based on this, we propose a sociotechnical view of AF to assume a dynamic and mutual interaction with the context. This implies that the IS needs mechanisms to interpret and process inputs (e.g., a changing notion of fairness) or feedback (e.g., changes in the environment and reactions caused by its past outputs).

Apart from the classical elements of a sociotechnical perspective of an IS, we followed Chatterjee et al. (2021) by also considering *information* as a core element within a sociotechnical system. Given that data's role for achieving AF becomes inevitable, in our view, it complements the overall sociotechnical perspective we are pursuing here. The recidivism prediction system relies on data about past inmates (personal data, criminal history, education, etc.) and their offenses. It also mines data about the inmate under consideration, as well as statistical and ML models that link the data. We propose a sociotechnical view of AF to assume information as a key to steer the interaction between the social and the technical components. Since data are neither neutral nor independent, they require a critical approach.

In sum, the proposed perspective on AF positions the decision-making as involving humans and algorithms at the very core of the system. The reciprocal interactions between these individuals or collectives and the technology are what enable the system to yield a decision. Information provided to the individuals and to the algorithms is what underlies and structures the decision-making. The system is embedded in the environment, which is affected by the decisions taken by the system and reacts to them. The proposed sociotechnical view provides tools to explore relevant aspects of AF.

### 5.3. How Does a Sociotechnical Perspective Surpass Existing Premises?

The sociotechnical framing of AF not only establishes a tool set to precisely understand and describe the nature and mechanisms of algorithmic discrimination. It also helps to overcome the limitations of previous approaches, as discussed in Sections 4.1 and 4.2 and as presented below. This follows the strategy suggested by Alvesson and Sandberg (2011), who suggest reconsidering the problematized assumptions in light of a new theoretical perspective. We will now revisit premises identified in the literature and compare them to the sociotechnical view.

First, a sociotechnical perspective suggests that solutions to algorithmic unfairness problems may not function properly if they don't factor in social components and dynamics of adaptation among the components. This wholly contradicts the *engineering assumption*: presenting a classifier that produces a less biased output is not yet a solution. We can claim that a solution is successful only if the entire sociotechnical system achieves a state of coherence in which it generates fewer biases.

The proposed sociotechnical lens approaches *equivalence assumptions* by acknowledging that notions of fairness and their operationalizations could be adequate depending on the overall system's state. The technical literature often sets out to identify the best notion among many. This attempt is destined to fail in complex and dynamic environments where the operationalizations' fits may vary. The proposed perspective makes it clear that it is not possible to explicate biases to be addressed upfront (as with the *explicitness assumption* or the *engineering assumption*). Algorithmic unfairness emerges as an undesired system output. Measuring performance against biases that were known a priori (i.e., came as the input) is also not sufficient. An audit against all possible biases would be necessary for the output.

Finally, the sociotechnical perspective overthrows all the *independence assumptions* and the *bad actor assumption*, because all components (including the environment) are involved in creating unfairness or assuring fairness. Through ongoing mutual adaptation, the process is highly dynamic. The process of analyzing and assigning responsibility for insufficient fairness cannot be reduced to a single component.

*Translation* and *distributiveness assumptions* relate to mathematics as the primary tool for representing fairness. While this may be true and necessary for the technical component, humans have dealt poorly with mathematical formalizations, especially in the context of distributions (Kahneman, 2011). Since perceptions of fairness are at the heart of being human (as claimed in anthropology and neurology), humans are highly unlikely to engage in mathematical calculations while making fairness assessments. However, when technical components 'speak' mathematics and humans do not, mutual adaptation can be hindered and can prevent the system from being jointly optimized. While the proposed framework does not directly relax this assumption, it makes clear that this point requires attention.

Sociotechnical systems require understanding all the processes involving data, technology, and humans. These components will differ in every application of ML and AF. Human practices and behaviors are situated (Draude et al., 2019; Orlikowski, 2008), which makes a great difference in the fairness of a recidivism prediction system or other systems, for instance for assessing creditworthiness. Because the environments differ, the components differ; thus, the outcomes also differ. This makes clear that reducing AF to a single approach, following *black box assumptions*, is inadequate, because it omits numerous technical approaches and because suggests that these situations are comparable. Finally, our proposed perspective interrogates the *purity assumption*. The technical and societal contexts provide inputs that prescribe the workings of a system and observe system outcomes. However, in many cases, the reactions of the society or the justice system to the occurrence of algorithmic unfairness was sluggish. Unfair systems are in constant use. Thus, it is misleading to idealize technological solutions or parts of society as being under the influence of a purity assumption.

The assumptions we identified negatively influence the variety and applicability of measures against unfairness. The sociotechnical perspective reveals that social and technical components are equally involved in discriminating and in assuring fairness. Only when one attends to how these components interact – i.e., how decisions emerge from the mutual adaptation and optimization between humans and algorithms, and how these interactions relate to the instrumental and humanist outcomes the system should produce – will we be able to effectively address AF. To overcome the current lack of a sociotechnical perspective on AF, we will now propose research directions for creating a sociotechnical knowledge base on fairness.

## 6. Directions for Sociotechnical Research into Algorithmic Fairness

The sociotechnical perspective on AF overcomes assumptions present in the literature and offers a framework for identifying sources of algorithmic bias. To date, sources of bias were either related to the ML workflow (Feuerriegel et al., 2020; Vokinger et al., 2021; von Zahn et al., 2021) or ascribed to social biases (Mohamed et al., 2020; Wong, 2020). We will now address sources of bias identified in the literature (see Appendix) and classify them according to the components of a sociotechnical system or the interrelationships between them. As presented below, the origins of algorithmic unfairness are distributed across the entire sociotechnical system. We identify directions for IS research to eliminate various algorithmic discrimination types at their source. However, these research directions are rarely limited to one component of the sociotechnical system, but instead call for a holistic approach to AF. For instance, to eradicate biases that emerge through low-quality information, one must address the social and organizational processes related to data generation and management. Similarly, to overcome technological limitations, one may need to ascertain the perspectives of various stakeholders. Accordingly, we will now address the identified sources of algorithmic unfairness, discussing how a holistic, sociotechnical perspective helps to resolve them. Specifically, technology may be responsible for specific algorithmic discrimination types but, as opposed to the technical perspective, we claim that potential solutions relate to social and organizational practices or procedures. We claim that algorithmic unfairness can only be effectively addressed from a sociotechnical perspective and propose lines of research to substantiate this standpoint.

## 6.1. Technology as a Source of Algorithmic Unfairness

Algorithmic unfairness can emerge when the technology produces a biased output. Most papers pinpointed algorithmic bias in the technological subsystem. Algorithms with and without wrong or insufficient operationalization of fairness *are the origin of discrimination in AF discourse*. Technical AF progresses by proposing new algorithms that use better operationalizations of fairness or other constraints and techniques to guarantee that outputs will be fair. The problem is they have rarely if ever moved beyond conceptual modeling and experimentation with decontextualized data. In the recidivism prediction case, the algorithm used in COMPAS exhibited racial bias. COMPAS's developers claimed to have implemented a fairness mechanism (Dieterich et al., 2016), yet evaluation in the scientific community showed that this was insufficient (Corbett-Davies et al., 2017). Specific reasons for algorithmic unfairness within the technological subsystem may include but are not limited to: missing or insufficient fairness constraints, the impossibility of simultaneously achieving multiple notions of fairness, defective pre-processing and post-processing routines, and biased procedures hardcoded into the algorithm. Further, bias can also emerge through the interactions between several tools.

Possible research lines are emerging from the insight that a technological artifact can be inherently unfair. It remains hard to evaluate the fairness of technological components. Where evaluation of single classifiers in technical studies is done by comparison to a baseline, this is not applicable to more complex technological artifacts, nor is it feasible for fields where no data exist or where no baseline was established by previous research. Except for prominent cases such as COMPAS, researchers and practitioners lack real-world evaluations of AF. They lack external validation that fairness constraints work in practice, especially in a complex and fuzzy context. IS has always promoted research with high external validity, seeking systematic and realistic tests beyond confined computer science playgrounds (Mettler et al., 2014; Sonnenberg & Brocke, 2012; Venable et al., 2014). IS has traditionally focused on evaluating one or two declared and positive goals (e.g., improving decision quality, improving adoption, improving usability, etc.). Fairness and other fundamental values have types of constraints and technological artifacts that should generate desired effects without violating them. This possibility challenges existing IS design and evaluation frameworks. Most prominently, IS is predestined to evaluate AF solutions in a real-world context, concerning organizational objectives, free from the technical assumptions of ML discourse.

*What sorts of core theories should be employed to inform the design and evaluation of technological artifacts that then need to obey fairness and potentially other ethical values? Which measures should be used to ensure that the system does no harm? How can we standardize evaluations and make them applicable to both practitioners and researchers? How do we conduct those evaluations in realistic contexts without risking real harm to test subjects?*

The evaluation and auditing of technical solutions requires both appropriate metrics and procedures as well as insights into the algorithms and their decision routines. Based on these insights, the designers could identify risks of bias even before testing the system. End-users could easily review system decisions based on the applied procedures and the considered data. The explainability or interpretability of ML is a subdiscipline of its own (Molnar, 2020), and despite many efforts from several scientific communities and practitioners, its progress and practical uptake remain limited (Caruana et al., 2020; Mitchell, 2019; Sejnowski, 2018). While the discourse on explainable AI has reached IS (Ågerfalk, 2020; B. Kim et al., 2020), it remains nascent. The fairness example makes it clear that we need technological systems that *reflect and explain* themselves and their actions. Research outside IS has focused on two stakeholder groups that particularly benefit from ML explainability: developers, who – based on insights – can improve their model without costly experimentation (primarily in the ML literature) and end-users, who should become able to enter a dialogue with the machine (primarily in the human-computer interaction literature). Here, the existing studies have identified additional stakeholders who depend on a deep understanding of what algorithms do: (1) high-stakes decision-makers who rely on predictions, such as governments who try to adapt plans to worldwide developments such as pandemics or global warming and refer to ML-based predictions, but also need to explain their decisions to the public; (2) auditors of medical application tools, who decide whether or not an application is a medical product, (3) judges and others in the justice system. IS can contribute to this discourse by identifying and specifying domain-dependent and context-dependent requirements for explainable ML.

*Who are the stakeholders that rely on a deep understanding of ML technologies? What explanation type is needed for each of them? How can a system reflect on its predictions concerning fairness or other fundamental values? How do we combine the autonomy of a technological system with the requirement to make everything explicable?*

### 6.2. Information as a Source of Algorithmic Unfairness

The information that embrace data and models is another technical source of algorithmic unfairness. Data not only perpetuate social biases encoded in the data generation process, but may also introduce new biases (through missing entries, an imbalanced representation of features, or inadequate scales to represent the features) (Buolamwini & Gebru, 2018; Saleiro et al., 2020). Many studies have proposed solutions such as re-weighting or filtering and balancing the datasets (Kallus & Zhou, 2018; Yang et al., 2020) or adapting algorithms, including specific pre-processing and post-processing steps (Gebru, 2020; Samadi et al., 2018). They attempt to mitigate unfairness in the information subsystem by adaptation in the information/technology subsystem. In recidivism prediction, unbalanced data introduced more racial bias than biased data (Biswas et al., 2020). However, balanced and complete data are not easily available. ML solutions require vast amounts of data and often rely on datasets that were neither developed for use in ML nor with a specific focus on fairness, but that were collected for other documentation purposes.

While organizations globally possess, collect, and process large volumes of data, only a fraction of the data is accessible to ML researchers and practitioners. This creates obstacles to the development and evaluation of fair models and algorithms. While data governance and management have a long tradition in IS (Counihan et al., 2002; Goodhue et al., 1988; Kettinger & Marchand, 2011; Otto, 2011), the proposed framework and practices pertain mostly to issues of investment and value generation in the organization. They rarely thematize how to lever the value of data via ML and how to ensure that outcomes are fair. AF discourse calls for putting these issues on the agenda so as to extend the data governance and management literature in IS.

*Which data management practices are adequate to generate fair data or make existing data fair? How do we incentivize organizations to follow those practices? How do we ensure fairness in combination with other desired features such as consistency, integrity,*

*or security? How can existing data be made available to researchers engaging in the study of AI for humanist goals?*

## 6.3. Interactions between Social and Technical Components as a Source of Algorithmic Unfairness

Algorithmic unfairness emerges owing to mutual adaptation between social and technical components. Humans often buy into the myth of infallible or objective algorithms and data. In other cases, we deliberately move the responsibility for decisions to 'the computer' and rarely question it (Orr & Davis, 2020). In doing so, they co-create unfairness (Hoffmann, 2019). However, adaptation in the opposite direction can also generate discrimination. Data, models (including causal models), algorithms, and pre-processing and post-processing routines are created and curated by humans and can perpetuate their individual social biases. COMPAS's model was trained using historical data from a law enforcement system dominated by white officers, who may have made decisions using an implicit or explicit racial pattern (Gladwell, 2019; Starr, 2014). The model adapted to this pattern and reproduced it more broadly (Kirkpatrick, 2017). Without systematic human audits, given the human tendency to outsource tough decisions, the system discriminated against prison inmates based on their race.

Interactions between social and technical components yield appealing fields for IS research. The IS community is interested in studying the division of work between humans and AI, using terms such as "machines as teammates" (Seeber et al., 2020) or the "future of work." (Ågerfalk et al., 2020). While this research often presents AI as a way to empower human workers (Nolte et al., 2020), AF requires both empowerment by machines and empowerment *against* machines. While we don't see machines as evil, there are situations in which humans require a sense of self-efficacy and sovereignty to disagree with and push back against a machine. This can be achieved through adequate control mechanisms in machines, through assigning the "last word" to humans, or through partnership and dialogue between the two.

*What is the appropriate division of work between humans and machines in high-stakes decision-making? How do we establish a partnership decision-making process between humans and machines? How do we incentivize workers and organizations to critically*

*question machine predictions? How do we prevent 'groupthink' in human-machine teams?*

Answering these questions can contribute to a more differentiated discourse on machines as teammates; it can also establish a more dynamic adaptation relationship in the ML domain.

Similarly, IS has addressed technology development as a sociotechnical endeavor (Hassan & Mathiassen, 2018) in which the values of involved individuals become embedded in the created artifacts (Chatterjee et al., 2009; Friedman et al., 2013; van de Poel & Kroes, 2014). AF emphasizes that IS development is not value-neutral and requires careful consideration of implicit and explicit humanist values and operational objectives during system development, data generation, model generation, evaluation, and organizational implementation. It is necessary to extend the discourse beyond design to the entire lifecycle of a system and autonomous technologies. The resulting contributions may extend the discourse on value-sensitive design as well as IS development. For AF, this would involve answering questions such as:

*How does one involve all relevant stakeholders in the process of selecting adequate notions of fairness for the given context? How does one ensure that the mathematical notion of fairness aligns with the values of users or affected stakeholders? How does one ensure that fairness and other relevant values are considered throughout the system's lifecycle?*

Further, IS has for decades dealt with dynamics and agency in complex systems, with a special focus on AI (Ågerfalk, 2020; Fang et al., 2018; Wastell, 1996). Autonomous technologies as parts of sociotechnical systems introduce a new agency type beyond humans (e.g., actor-network theory) (Ågerfalk, 2020; Rose et al., 2005). Many other frameworks and theories for analyzing behaviors, social practices, or changes in an organization ascribe agency solely to humans (Karanasios, 2018; Karanasios & Allen, 2018; Rose et al., 2005). The AF discourse supports the need to review and update frameworks and provide robust definitions of a nonhuman agency so as to facilitate the discourse within IS and adjacent disciplines. This can help to turn AF discourse away from bad actor arguments toward a shared solution-oriented effort.

*How do the agencies of human and nonhuman entities impact on the generation of values in sociotechnical systems? How do the agencies interfere with each other? What are the intended and unintended consequences of nonhuman agency? How should fairness be guaranteed, despite agency being distributed across multiple human and nonhuman actors?*

Finally, IS intends to develop stable systems, i.e., ones that reach the optimal fit between the technical and the social components. Borrowing the notion of entropy from thermodynamics, Chatterjee et al. (2021) framed the level of discrepancy or non-alignment between the components in an IS as *entropy*. In the case of AF, entropy grows when the technical component does not yield fair results as expected by the social component. This mismatch leads to an unstable, incoherent overall system. Following the thermodynamics metaphor, one could say that the system *heats up*. This relationship points to a larger discourse relating to the application of AI: *value alignment* (Russell, 2019). Whenever the values that underlie the technical and the social components differ, the performance of the overall sociotechnical system declines. Chatterjee et al. (2021) claimed that *information* acts as a moderator in this situation: coherent and useful information can *cool down* the system, while incoherent or contradictory information may make the entropy rise even further. In short, the quality of information introduced in the system is crucial for reaching a stable and coherent state. The literature suggests that AF requires high-quality data. Rather than adding more training data, it is more important to provide information with specific characteristics. On the one hand, this information may embrace carefully selected fairness constraints that fit the social components' expectations (Binns, 2018; Srivastava et al., 2019a). On the other hand, it may mean selecting or manipulating the training data accordingly by filling blind spots, improving the variety of the dataset, or ensuring that the training data fit the application environment (Kazemi et al., 2018; Miron et al., 2020). Other information types can likely help to achieve a fair state of the overall system, which offers potential for IS research.

*Which information types reduce entropy in sociotechnical systems? How does one identify the entropy level in algorithmic decision-making systems? How does entropy change over time in such systems? How can information transfer the desired notion of fairness, apart from mathematical formulas?*

## 6.4. The Environment as a Source of Algorithmic Unfairness

Finally, interactions with the environment may introduce or perpetuate algorithmic unfairness. This can emerge as a result of governance issues (L. Hu & Chen, 2020; Kuhlman et al., 2020; Noriega-Campero et al., 2020), social order (Lepri et al., 2018; Mohamed et al., 2020; Rosenberger, 2020; Wellner & Rothman, 2020; Yarger et al., 2019), or public discourse around AI (Araujo et al., 2020; Helberger et al., 2020). For instance, Kuhlman et al. (2020) identified the lack of cultural diversity among ML researchers as the reason for algorithmic bias. He recommended exchange and feedback between ML researchers and members of underrepresented or protected groups as a remedy; they address the broader societal context's impacts on the sociotechnical system. Here, influence in the opposite direction is also likely. Given that the notion of fairness is not static and is constantly being formed and disputed (Degoey, 2000), decisions taken by technological artifacts can change what society considers fair over the long term. This exchange needs careful analysis.

It is common knowledge in IS that the environment of an IS affects how technology is utilized and such utilizations' outcomes (Dennis et al., 2001; DeSanctis & Poole, 1994; Orlikowski, 2008). Notions of fairness are ingrained in these structures (Hufnagel & Birnberg, 1989) and are subject to change when structures change. IS has studied how work practices, organizational hierarchies, and economic structures change owing to technological innovation (Allen et al., 2013; Avgerou, 2001; Heracleous & Barrett, 2001). It is now necessary to understand how fundamental values change through the introduction of new technologies. Understanding the forces involved in this evolution are necessary to be able to predict how technologies will impact on society. While computer science researchers and practitioners are often confronted with accusations of developing and rolling out technologies without considering their negative effects (Tarafdar et al., 2013), they often lack tools to predict and analyze undesired consequences. The origins of AF are the best showcase for this. Thus, the IS discourse on the implications of technology use needs to be updated and needs to focus more on *predicting* rather than reacting to undesired developments. Having acknowledged the complexity of changes through technological progress, the IS community needs to step in as a moderator of these changes.

*How do technology and sociotechnical systems impact on the fundamental values in organizations and societies? How can we govern technologies' impacts on society? How do sociotechnical systems interact with one anther to establish shared values? How do notions of fairness change through these interactions?*

Feedback loops can introduce another source of unfairness. After deployment, an IS's output can influence the environment, which impacts on future inputs to the system. It may introduce bias into the system or may have unintended long-term consequences for the environment (Barocas & Selbst, 2016). In our previous example, the system for assessing recidivism risk may draw on socioeconomic variables such as the previous income so as to predict a defendant's risk of committing another crime. Such a system may introduce a feedback loop: when identifying a high recidivism risk, the system may prolong an offender's sentence, and this longer sentence could negatively affect this individual's socioeconomic status. In this case, one can expect a lower income and thus an increased likelihood of a new crime. When the individual then commits a new crime, the system for assessing recidivism risk may then rely on the lower income and may again recommend a prolonged sentence, thereby diminishing the chances of the individual's early release. As seen in this example, the feedback loop is reinforced by the underlying ML system (Barocas & Selbst, 2016). Similar examples of feedback loops have been documented in online advertising, with female candidates being shown fewer ads for high-paid jobs (Datta et al., 2018).

Feedback loops arise from interactions over time and are therefore evident only after deployment. This makes them challenging for detection by IS practitioners. For instance, empirical evidence for the above examples in recidivism risk assessment and online advertising were reported only very recently. A question for IS practitioners is how to detect feedback loops. Here, a practical approach is offered by the above formulation of AF through an IS artifact. Since the sociotechnical perspective often refers to general systems theory (Kast & Rosenzweig, 1972; Matook & Brown, 2017), it allows us to follow established procedures from systems theory to identify feedback loops. An example is shown in Figure 2. In systems theory, the relationships between input and output are formalized in diagrams, with a positive relationship marked by a + and a negative relationship by a -. Relationships that form closed cycles point to potential feedback loops that propagate bias. Through such closed cycles, input deviations are directly propagated into output and fed again into the system. This has direct implications for IS practice, because an existing formalization of system dynamics can now be checked against potential feedback loops.

*What are the long-term implications of AF for economic welfare? How can we understand and formalize the 'economics of AF'? How can we effectively mitigate feedback loops in IS? How can we address feedback loops that arise from human interactions with IS? How can human interactions be designed that warrant constraints from AF?*

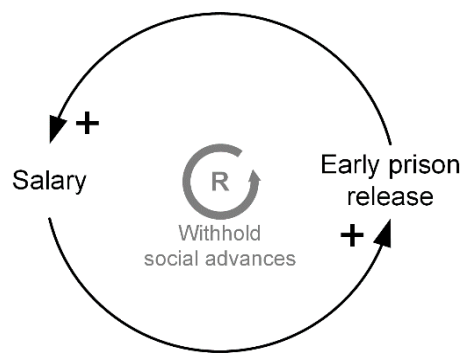

*Figure 2. Example of a Feedback Loop Visualized through a Systems Theory Diagram*

*Notes:* A system for recidivism risk assessment is shown here. The system grants an early prison release by using ML, with a higher salary (or income) predicting a lower recidivism risk, leading to an early release (shown by +; bottom). An early prison release also increases chances for social advancement and thus a higher salary (shown by +; top). This leads to a reinforcing behavior (indicated by R).

In sum, algorithmic unfairness does not emerge solely from algorithms or data used for decision-making. It can arise out of ineffective or 'lazy' adaptations between technological and social components, or interactions within a larger context. Yet the research has focused on technical solutions and the solving of issues that emerge from interactions between components via technical solutions, rather than addressing the entire sociotechnical system or its context. As noted, a sociotechnical approach requires that one acknowledge the interactive and equal roles of social and technical components.

# 7. Implications

Positioning AF as a holistic, sociotechnical phenomenon has implications for IS practice and research. We see the following implications as particularly relevant and impactful. First, AF should be seen as a multidisciplinary endeavor in which researchers transcend the boundaries of individual disciplines, combining strengths from different disciplines to advance the interface of social and technical systems at risk of algorithmic (un)fairness. Second, AF is complex. Singular and punctual interventions work only for a short time until data, algorithms, or human perceptions change. It is crucial that legislative bodies and legal practitioners understand this. Otherwise, they risk multiplying specific rules on which data can be stored and how, or which measure to apply in the evaluation. This may not lead to de facto humanist goals and values such as fairness. Measuring outcomes of sociotechnical systems is more appropriate than intervening in low-level processes within a sociotechnical system. Third, AF is a valuable objective and has real impacts on organizations. If the sociotechnical systems reach a coherent state without discrepancies between the social and the technical, the overall system's productivity will increase. This is only possible if notions of fairness align across the IS. If employees struggle with notions of fairness in the software they use, they could refuse it or employ workarounds. Organizations should implement fairness-oriented solutions carefully, not just for society but also for themselves.

AF themes have implications for IS. Engineering education should both equip students with the necessary technological understanding and should cover humanistic, social, and behavioral dimensions. Practitioners should judge their artifacts critically. For academia, it is important to provide detailed descriptions of all datasets. This helps to assess the risk of biases introduced during data collection that can adversely affect decisions' fairness.

The insights we have presented have limitations that are typical of any literature review. Field portrayal is limited by the databases and keywords used, along with the filtering used in selection processes. Further, the summary of the articles we presented is limited; a complete list of references appears in the Appendix, and we encourage the reader to consider them in detail.

# 8. Conclusion

We claim that AF is just a precursor to a larger debate on value alignment relating to autonomous or semi-autonomous technologies. While an abstract value alignment discourse is emerging outside IS (Russell, 2019), the AF example shows that problems can become complex and fuzzy once confronted with reality. In particular, the reciprocal interactions between the social and the technical components, as well as their embedding in a larger environment, disrupt theoretically valid ideas. This is exemplified by the purely technical approach to fairness in automated decision-making. Here, state-of-the-art technical approaches cannot guarantee a fair outcome at scale. Similar effects may be expected when algorithms begin to directly affect other humanist values. Thus, it is crucial to formulate the problems of value alignment in a sociotechnical way from the outset. This means considering how social and technical components will change depending on the desired humanist outcomes, how the interdependencies between them can be employed to prevent undesired outcomes, how they can complement one another, and which moderators can effectively help them achieve a coherent state of low entropy, and how the broader environment is affected by the outcomes the system produces. To date, the debate on value alignment has focused more on creating rules to assure that algorithms don't overrule humans and obey human values, whatever they may be. This formulation forgets that social values and specifications thereof will change through interactions with technology, and will evolve and undergo adaptation, such that the social and the technical components will experience states of low and high entropy. The AF case makes clear that the system is dynamic and complex, and human values have not yet been articulated in a way to make algorithms simply obey them. AF could not be achieved with a simple constraint, with a range of constraints, or with supervised and unsupervised approaches. The debate on value alignment requires that one acknowledge that values are shaped and negotiated in sociotechnical processes. IS is predestined to contribute to this discussion its practical orientation, technical understanding, and sensitivity to societal progress. IS should pursue further research into sociotechnical AF before algorithmic (un)fairness turns into another 'dark side' of IT (Tarafdar et al., 2013).

# References

Aas, K. F. (2006). 'The body does not lie': Identity, risk and trust in technoculture. *Crime, Media, Culture*, *2*(2), 143–158.

Abràmoff, M. D., Tobey, D., & Char, D. S. (2020). Lessons Learned About Autonomous AI: Finding a Safe, Efficacious, and Ethical Path Through the Development Process. *American Journal of Ophthalmology*, *214*, 134–142. https://doi.org/10.1016/j.ajo.2020.02.022

Ackoff, R. L. (1973). Science in the Systems Age: Beyond IE, OR, and MS. *Operations Research*, *21*(3), 661–671. https://doi.org/10.1287/opre.21.3.661

Adams, J. S. (1963). Towards an understanding of inequity. *The Journal of Abnormal and Social Psychology*, *67*(5), 422–436. https://doi.org/10.1037/h0040968

Ågerfalk, P. J. (2020). Artificial intelligence as digital agency. *European Journal of Information Systems*, *29*(1), 1–8. https://doi.org/10.1080/0960085X.2020.1721947

Ågerfalk, P. J., Conboy, K., Crowston, K., Jarvenpaa, S., Lundström, J. E., Mikalef, P., & Ram, S. (2020). Artificial Intelligence – Beyond the Hype. *ICIS 2020 Proceedings*. https://aisel.aisnet.org/icis2020/pdws/pdws/1

Agrawal, A., Gans, J., & Goldfarb, A. (2018). *Prediction machines: The simple economics of artificial intelligence*. Harvard Business Review Press.

Ahmad, M. A., Patel, A., Eckert, C., Kumar, V., & Teredesai, A. (2020). Fairness in Machine Learning for Healthcare. *Proceedings of the 26th ACM SIGKDD International Conference on Knowledge Discovery & Data Mining*, 3529–3530. https://doi.org/10.1145/3394486.3406461

Aizenberg, E., & van den Hoven, J. (2020). Designing for human rights in AI. *Big Data & Society*, *7*(2), 205395172094956. https://doi.org/10.1177/2053951720949566

Allen, D. K., Brown, A., Karanasios, S., & Norman, A. (2013). How Should Technology-Mediated Organizational Change Be Explained? A Comparison of the Contributions of Critical Realism and Activity Theory. *MIS Quarterly*, *37*(3), 835–854. JSTOR.

Alvesson, M., & Sandberg, J. (2011). Generating research questions through problematization. *The Academy of Management Review*, *36*(2), 247–271. JSTOR.

Angwin, J., Larson, J., Mattu, S., & Kirchner, L. (2016, May 23). Machine Bias: There's software used across the country to predict future criminals. And it's biased against blacks. *ProPublica*. https://www.propublica.org/article/machine-bias-risk-assessments-in-criminal-sentencing

Araujo, T., Helberger, N., Kruikemeier, S., & De Vreese, C. H. (2020). In AI we trust? Perceptions about automated decision-making by artificial intelligence. *AI & SOCIETY*, 1–13.

Avgerou, C. (2001). The significance of context in information systems and organizational change. *Information Systems Journal*, *11*(1), 43–63. https://doi.org/10.1046/j.1365-2575.2001.00095.x

Avital, M., Mathiassen, L., & Schultze, U. (2017). Alternative genres in information systems research. *European Journal of Information Systems*, *26*(3), 240–247. https://doi.org/10.1057/s41303-017-0051-4

Barabas, C., Doyle, C., Rubinovitz, J., & Dinakar, K. (2020). Studying up: Reorienting the study of algorithmic fairness around issues of power. *Proceedings of the 2020 Conference on Fairness, Accountability, and Transparency*, 167–176. https://doi.org/10.1145/3351095.3372859

Bargh, J. A. (2017). *Before you know it: The unconscious reasons we do what we do*. Touchstone.

Barocas, S., & Selbst, A. D. (2016). Big Data's Disparate Impact. *California Law Review*, *104*, 671.

Beath, C., Berente, N., Gallivan, M. J., & Lyytinen, K. (2013). Expanding the frontiers of information systems research: Introduction to the special issue. *Journal of the Association for Information Systems*, *14*(4), 4.

Bera, S., Chakrabarty, D., Flores, N., & Negahbani, M. (2019). Fair Algorithms for Clustering. In H. Wallach, H. Larochelle, A. Beygelzimer, F. d\textquotesingle Alché-Buc, E. Fox, & R. Garnett (Eds.), *Advances in Neural Information Processing Systems 32* (pp. 4954–4965). Curran Associates, Inc. http://papers.nips.cc/paper/8741-fair-algorithms-for-clustering.pdf

Beyleveld, D., & Brownsword, R. (2019). Punitive and preventive justice in an era of profiling, smart prediction and practical preclusion: Three key questions. *International Journal of Law in Context*, *15*(2), 198–218. https://doi.org/10.1017/S1744552319000120

Bies, R. J. (2001). Interactional (in) justice: The sacred and the profane. *Advances in Organizational Justice*, *89118*.

Binns, R. (2018). Fairness in Machine Learning: Lessons from Political Philosophy. *Conference on Fairness, Accountability and Transparency*, 149–159. http://proceedings.mlr.press/v81/binns18a.html

Biswas, A., Kolczynska, M., Rantanen, S., & Rozenshtein, P. (2020). The Role of In-Group Bias and Balanced Data: A Comparison of Human and Machine Recidivism Risk Predictions. *Proceedings* *of the 3rd ACM SIGCAS Conference on Computing and Sustainable Societies*, 97–104. https://doi.org/10.1145/3378393.3402507

Briggs, R. O., De Vreede, G.-J., & Nunamaker Jr, J. F. (2003). Collaboration Engineering with ThinkLets to Pursue Sustained Success with Group Support Systems. *Journal of Management Information Systems*, *19*(4), 31–64. https://doi.org/10.1080/07421222.2003.11045743

Briggs, R. O., Nunamaker, J. F., & Sprague, R. H. (2010). Special Section: Social Aspects of Sociotechnical Systems. *Journal of Management Information Systems*, *27*(1), 13–16. https://doi.org/10.2753/MIS0742-1222270101

Brockman, J. (Ed.). (2019). *Possible minds: Twenty-five ways of looking at AI*. Penguin Press.

Brosnan, S. F. (2013). Justice- and fairness-related behaviors in nonhuman primates. *Proceedings of the National Academy of Sciences of the United States of America (PNAS)*, *110*(Supplement 2), 10416–10423. https://doi.org/10.1073/pnas.1301194110

Brosnan, S. F., & de Waal, F. B. (2014). Evolution of responses to (un) fairness. *Science*, *346*(6207), 1251776.

Buolamwini, J., & Gebru, T. (2018). Gender Shades: Intersectional Accuracy Disparities in Commercial Gender Classification. *Conference on Fairness, Accountability and Transparency*, 77–91. http://proceedings.mlr.press/v81/buolamwini18a.html

Caruana, R., Lundberg, S., Ribeiro, M. T., Nori, H., & Jenkins, S. (2020). Intelligible and Explainable Machine Learning: Best Practices and Practical Challenges. *Proceedings of the 26th ACM SIGKDD International Conference on Knowledge Discovery & Data Mining*, 3511–3512. https://doi.org/10.1145/3394486.3406707

Celis Bueno, C. (2020). The Face Revisited: Using Deleuze and Guattari to Explore the Politics of Algorithmic Face Recognition. *Theory, Culture & Society*, *37*(1), 73–91. https://doi.org/10.1177/0263276419867752

Celis, E., Keswani, V., Straszak, D., Deshpande, A., Kathuria, T., & Vishnoi, N. (2018). Fair and diverse DPP-Based data summarization. In J. Dy & A. Krause (Eds.), *Proceedings of the 35th international conference on machine learning* (Vol. 80, pp. 716–725). PMLR. http://proceedings.mlr.press/v80/celis18a.html

Chandler, D. (2019). *Digital governance in the Anthropocene: The rise of the correlational machine*.

Chatterjee, S., Sarker, S., & Fuller, M. (2009). A Deontological Approach to Designing Ethical Collaboration. *Journal of the Association for Information Systems*, *10*(3), 138–169. https://doi.org/10.17705/1jais.00190

Chatterjee, S., Sarker, S., Lee, M. J., Xiao, X., & Elbanna, A. (2021). A possible conceptualization of the information systems (IS) artifact: A general systems theory perspective1. *Information Systems Journal*, *31*(4), 550–578. https://doi.org/10.1111/isj.12320

Chouldechova, A. (2017). Fair prediction with disparate impact: A study of bias in recidivism prediction instruments. *Big Data*, *5*(2), 153–163. https://doi.org/10.1089/big.2016.0047

Chouldechova, A., Benavides-Prado, D., Fialko, O., & Vaithianathan, R. (2018). A case study of algorithm-assisted decision making in child maltreatment hotline screening decisions. *Conference on Fairness, Accountability and Transparency*, 134–148. http://proceedings.mlr.press/v81/chouldechova18a.html

Chouldechova, A., & Roth, A. (2018). The Frontiers of Fairness in Machine Learning. *ArXiv:1810.08810 [Cs, Stat]*. http://arxiv.org/abs/1810.08810

Clark, J. N. (2008). The three Rs: Retributive justice, restorative justice, and reconciliation. *Contemporary Justice Review*, *11*(4), 331–350. https://doi.org/10.1080/10282580802482603

Cohn, A., Fehr, E., & Maréchal, M. A. (2014). Business culture and dishonesty in the banking industry. *Nature*, *516*(7529), 86–89. https://doi.org/10.1038/nature13977

Corbett-Davies, S., Pierson, E., Feller, A., Goel, S., & Huq, A. (2017). Algorithmic decision making and the cost of fairness. *Proceedings of the 23rd Acm Sigkdd International Conference on Knowledge Discovery and Data Mining*, 797–806.

Coston, A., Mishler, A., Kennedy, E. H., & Chouldechova, A. (2020). Counterfactual risk assessments, evaluation, and fairness. *Proceedings of the 2020 Conference on Fairness, Accountability, and Transparency*, 582–593. https://doi.org/10.1145/3351095.3372851

Counihan, A., Finnegan, P., & Sammon, D. (2002). Towards a framework for evaluating investments in data warehousing. *Information Systems Journal*, *12*(4), 321–338. https://doi.org/10.1046/j.1365-2575.2002.00134.x

D'Amour, A., Srinivasan, H., Atwood, J., Baljekar, P., Sculley, D., & Halpern, Y. (2020). Fairness is not static: Deeper understanding of long term fairness via simulation studies. *Proceedings of the 2020 Conference on Fairness, Accountability, and Transparency*, 525–534. https://doi.org/10.1145/3351095.3372878

Datta, A., Datta, A., Makagon, J., Mulligan, D. K., & Tschantz, M. C. (2018). Discrimination in Online Advertising: A Multidisciplinary Inquiry. *Conference on Fairness, Accountability and Transparency*, 20–34. http://proceedings.mlr.press/v81/datta18a.html

Davison, R. M., & Tarafdar, M. (2018). Shifting baselines in information systems research threaten our future relevance. *Information Systems Journal*, 28(4), 587–591. https://doi.org/10.1111/isj.12197

De Filippi, P., & Hassan, S. (2018). Blockchain Technology as a Regulatory Technology: From Code is Law to Law is Code. *ArXiv:1801.02507 [Cs]*. http://arxiv.org/abs/1801.02507

Decety, J., & Yoder, K. J. (2016). Empathy and motivation for justice: Cognitive empathy and concern, but not emotional empathy, predict sensitivity to injustice for others. *Social Neuroscience*, 11(1), 1–14. https://doi.org/10.1080/17470919.2015.1029593

Degoey, P. (2000). Contagious Justice: Exploring The Social Construction of Justice in Organizations. *Research in Organizational Behavior*, 22, 51–102. https://doi.org/10.1016/S0191-3085(00)22003-0

Dennis, A. R., Wixom, B. H., & Vandenberg, R. J. (2001). Understanding Fit and Appropriation Effects in Group Support Systems via Meta-Analysis. *MIS Quarterly*, 25(2), 167–193. https://doi.org/10.2307/3250928

DeSanctis, G., & Poole, M. S. (1994). Capturing the complexity in advanced technology use: Adaptive structuration theory. *Organization Science*, 5(2), 121–147.

Dieterich, W., Mendoza, C., & Brennan, T. (2016). COMPAS risk scales: Demonstrating accuracy equity and predictive parity. *Northpointe Inc.*

Dolata, M., Schenk, B., Fuhrer, J., Marti, A., & Schwabe, G. (2020). When the system does not fit: Coping strategies of employment consultants. *Computer Supported Cooperative Work*, 29(6), 657–696. https://doi.org/10.1007/s10606-020-09377-x

Dolata, M., & Schwabe, G. (2021). How Fair Is IS Research? In S. Aier, P. Rohner, & J. Schelp (Eds.), *Engineering the Transformation of the Enterprise: A Design Science Research Perspective* (pp. 37–49). Springer International Publishing. https://doi.org/10.1007/978-3-030-84655-8_3

Dourish, P. (2001). *Where the action is: The foundations of embodied interaction*. MIT Press.

Draude, C., Klumbyte, G., Lücking, P., & Treusch, P. (2019). Situated algorithms: A sociotechnical systemic approach to bias. *Online Information Review*, 44(2), 325–342. https://doi.org/10.1108/OIR-10-2018-0332

Dutta, S., Wei, D., Yueksel, H., Chen, P.-Y., Liu, S., & Varshney, K. (2020). Is There a Trade-Off Between Fairness and Accuracy? A Perspective Using Mismatched Hypothesis Testing. *International Conference on Machine Learning*, 2803–2813. http://proceedings.mlr.press/v119/dutta20a.html

Dwork, C., Hardt, M., Pitassi, T., Reingold, O., & Zemel, R. (2012). Fairness through awareness. *Proceedings of the 3rd Innovations in Theoretical Computer Science Conference on - ITCS '12*, 214–226. https://doi.org/10.1145/2090236.2090255

Dwork, C., & Ilvento, C. (2018). Fairness Under Composition. *ArXiv:1806.06122 [Cs, Stat]*, 20 pages. https://doi.org/10.4230/LIPIcs.ITCS.2019.33

Ebrahimi, S., & Hassanein, K. (2019, November 6). Empowering Users to Detect Data Analytics Discriminatory Recommendations. *Proc. Intl. Conf. Information Systems*. Intl. Conf. Infromation Systems, Munich, Germany. https://aisel.aisnet.org/icis2019/cyber_security_privacy_ethics_IS/cyber_security_privacy/39

Fang, Y., Lim, K., Qian, Y., & Feng, B. (2018). System dynamics modeling for information systems research: Theory development and practical application. *MIS Quarterly: Management Information Systems*, 42, 1303–1329. https://doi.org/10.25300/MISQ/2018/12749

Fazelpour, S., & Lipton, Z. C. (2020). Algorithmic Fairness from a Non-ideal Perspective. *Proceedings of the AAAI/ACM Conference on AI, Ethics, and Society*, 57–63. https://doi.org/10.1145/3375627.3375828

Feuerriegel, S., Dolata, M., & Schwabe, G. (2020). Fair AI. *Business & Information Systems Engineering*, 62(4), 379–384. https://doi.org/10.1007/s12599-020-00650-3

Friedler, S. A., Scheidegger, C., Venkatasubramanian, S., Choudhary, S., Hamilton, E. P., & Roth, D. (2019). A comparative study of fairness-enhancing interventions in machine learning. *Proceedings of the Conference on Fairness, Accountability, and Transparency*, 329–338. https://doi.org/10.1145/3287560.3287589

Friedman, B., Kahn, P. H., Borning, A., & Huldtgren, A. (2013). Value Sensitive Design and Information Systems. In N. Doorn, D. Schuurbiers, I. van de Poel, & M. E. Gorman (Eds.), *Early engagement and new technologies: Opening up the laboratory* (pp. 55–95). Springer Netherlands. https://doi.org/10.1007/978-94-007-7844-3_4

Gebru, T. (2020). Lessons from Archives: Strategies for Collecting Sociocultural Data in Machine Learning. *Proceedings of the 26th ACM SIGKDD International Conference on Knowledge Discovery & Data Mining*, 3609. https://doi.org/10.1145/3394486.3409559

Gendler, T. S. (2014). *Thought Experiment: On the Powers and Limits of Imaginary Cases*. Routledge.

Gladwell, M. (2019). *Talking to Strangers: What We Should Know about the People We Don't Know*. Penguin UK.

Goodhue, D., Quillard, J., & Rockart, J. (1988). Managing the Data Resource: A Contingency Perspective. *Management Information Systems Quarterly*, *12*(3). https://aisel.aisnet.org/misq/vol12/iss3/2

Greenberg, J. (1986). Determinants of perceived fairness of performance evaluations. *Journal of Applied Psychology*, *71*(2), 340–342. https://doi.org/10.1037/0021-9010.71.2.340

Grover, V., & Lyytinen, K. (2015). New state of play in information systems research: The push to the edges. *MIS Quarterly*, *39*(2), 271–296. https://doi.org/10.25300/MISQ/2015/39.2.01

Haas, C. (2019, November 6). The Price of Fairness—A Framework to Explore Trade-Offs in Algorithmic Fairness. *Proc. Intl. Conf. Information Systems*. Intl. Conf. Information Systems, Munich, Germany. https://aisel.aisnet.org/icis2019/data_science/data_science/19

Hanna, A., Denton, E., Smart, A., & Smith-Loud, J. (2020). Towards a critical race methodology in algorithmic fairness. *Proceedings of the 2020 Conference on Fairness, Accountability, and Transparency*, 501–512. https://doi.org/10.1145/3351095.3372826

Harcourt, B. E. (2015). Risk as a Proxy for RaceThe Dangers of Risk Assessment. *Federal Sentencing Reporter*, *27*(4), 237–243. https://doi.org/10.1525/fsr.2015.27.4.237

Hardt, M., & Price, E. (2016). Equality of opportunity in supervised learning. *Advances in Neural Information Processing Systems (NIPS)*.

Hassan, N. R., & Mathiassen, L. (2018). Distilling a body of knowledge for information systems development. *Information Systems Journal*, *28*(1), 175–226. https://doi.org/10.1111/isj.12126

Helberger, N., Araujo, T., & de Vreese, C. H. (2020). Who is the fairest of them all? Public attitudes and expectations regarding automated decision-making. *Computer Law & Security Review*, *39*, 105456. https://doi.org/10.1016/j.clsr.2020.105456

Heracleous, L., & Barrett, M. (2001). Organizational Change as Discourse: Communicative Actions and Deep Structures in the Context of Information Technology Implementation. *Academy of Management Journal*, *44*(4), 755–778. https://doi.org/10.2307/3069414

Hiranandani, G., Narasimhan, H., & Koyejo, O. O. (2020). Fair Performance Metric Elicitation. In *Advances in Neural Information Processing Systems* (Vol. 33). https://proceedings.neurips.cc/paper/2020/hash/7ec2442aa04c157590b2fa1a7d093a33-Abstract.html

Hoffmann, A. L. (2019). Where fairness fails: Data, algorithms, and the limits of antidiscrimination discourse. *Information, Communication & Society*, *22*(7), 900–915. https://doi.org/10.1080/1369118X.2019.1573912

Hu, L., & Chen, Y. (2020). Fair classification and social welfare. *Proceedings of the 2020 Conference on Fairness, Accountability, and Transparency*, 535–545. https://doi.org/10.1145/3351095.3372857

Hu, Y., Wu, Y., Zhang, L., & Wu, X. (2020). Fair Multiple Decision Making Through Soft Interventions. In *Advances in Neural Information Processing Systems* (Vol. 33). https://proceedings.neurips.cc/paper/2020/hash/d0921d442ee91b896ad95059d13df618-Abstract.html

Hufnagel, E. M., & Birnberg, J. G. (1989). Perceived Chargeback System Fairness in Decentralized Organizations: An Examination of the Issues. *MIS Quarterly*, *13*(4), 415–430. https://doi.org/10.2307/248726

Huq, A. Z. (2018). Racial Equity in Algorithmic Criminal Justice. *Duke Law Journal*, *68*, 1043.

Johnson, G. M. (2020). Algorithmic bias: On the implicit biases of social technology. *Synthese*. https://doi.org/10.1007/s11229-020-02696-y

Joshi, K. (1989). The Measurement of Fairness or Equity Perceptions of Management Information Systems Users. *MIS Quarterly*, *13*(3), 343–358. https://doi.org/10.2307/249010

Kahneman, D. (2011). *Thinking, Fast and Slow*. Macmillan.

Kallus, N., & Zhou, A. (2019). The Fairness of Risk Scores Beyond Classification: Bipartite Ranking and the XAUC Metric. In H. Wallach, H. Larochelle, A. Beygelzimer, F. d\textquotesingle Alché-Buc, E. Fox, & R. Garnett (Eds.), *Advances in Neural Information Processing Systems 32* (pp. 3438–3448). Curran Associates, Inc. http://papers.nips.cc/paper/8604-the-fairness-of-risk-scores-beyond-classification-bipartite-ranking-and-the-xauc-metric.pdf

Kallus, N., & Zhou, A. (2018). Residual unfairness in fair machine learning from prejudiced data. In J. Dy & A. Krause (Eds.), *Proceedings of the 35th international conference on machine learning* (Vol. 80, pp. 2439–2448). PMLR. http://proceedings.mlr.press/v80/kallus18a.html

Kalpokas, I. (2019). The Code That Is Law. In I. Kalpokas (Ed.), *Algorithmic Governance: Politics and Law in the Post-Human Era* (pp. 27–47). Springer International Publishing. https://doi.org/10.1007/978-3-030-31922-9_3

Karanasios, S. (2018). Toward a unified view of technology and activity: The contribution of activity theory to information systems research. *Information Technology & People*, *31*(1), 134–155. https://doi.org/10.1108/ITP-04-2016-0074

Karanasios, S., & Allen, D. (2018). Activity theory in Information Systems Research. *Information Systems Journal*, *28*(3), 439–441. https://doi.org/10.1111/isj.12184

Kazemi, E., Zadimoghaddam, M., & Karbasi, A. (2018). Scalable deletion-robust submodular maximization: Data summarization with privacy and fairness constraints. In J. Dy & A. Krause (Eds.), *Proceedings of the 35th international conference on machine learning* (Vol. 80, pp. 2544–2553). PMLR. http://proceedings.mlr.press/v80/kazemi18a.html

Kettinger, W. J., & Marchand, D. A. (2011). Information management practices (IMP) from the senior manager's perspective: An investigation of the IMP construct and its measurement. *Information Systems Journal*, *21*(5), 385–406. https://doi.org/10.1111/j.1365-2575.2011.00376.x

Kim, B., Park, J., & Suh, J. (2020). Transparency and accountability in AI decision support: Explaining and visualizing convolutional neural networks for text information. *Decision Support Systems*, *134*. Scopus. https://doi.org/10.1016/j.dss.2020.113302

Kim, M. P., Korolova, A., Rothblum, G. N., & Yona, G. (2020). Preference-informed fairness. *Proceedings of the 2020 Conference on Fairness, Accountability, and Transparency*, 546. https://doi.org/10.1145/3351095.3373155

Kincaid, H., Dupre, J., & Wylie, A. (2007). *Value-Free Science: Ideals and Illusions?* Oxford University Press.

Kirkpatrick, K. (2017). It's not the algorithm, it's the data. *Communications of the ACM*, *60*(2), 21–23. https://doi.org/10.1145/3022181

Kleinberg, J., Mullainathan, S., & Raghavan, M. (2017). Inherent trade-offs in the fair determination of risk scores. *Conference on Innovations in Theoretical Computer Science (ITCS)*.

Kordzadeh, N., & Ghasemaghaei, M. (2021). Algorithmic bias: Review, synthesis, and future research directions. *European Journal of Information Systems*, 1–22. https://doi.org/10.1080/0960085X.2021.1927212

Kroll, J. A., Barocas, S., Felten, E. W., Reidenberg, J. R., Robinson, D. G., & Yu, H. (2016). Accountable Algorithms. *University of Pennsylvania Law Review*, *165*, 633.

Kuhlman, C., Jackson, L., & Chunara, R. (2020). No Computation without Representation: Avoiding Data and Algorithm Biases through Diversity. *Proceedings of the 26th ACM SIGKDD International Conference on Knowledge Discovery & Data Mining*, 3593. https://doi.org/10.1145/3394486.3411074

Kusner, M. J., Loftus, J., Russell, C., & Silva, R. (2017). Counterfactual Fairness. In I. Guyon, U. V. Luxburg, S. Bengio, H. Wallach, R. Fergus, S. Vishwanathan, & R. Garnett (Eds.), *Advances in Neural Information Processing Systems 30* (pp. 4066–4076). Curran Associates, Inc. http://papers.nips.cc/paper/6995-counterfactual-fairness.pdf

Landsbergen, D. (2004). Screen level bureaucracy: Databases as public records. *Government Information Quarterly*, *21*(1), 24–50. https://doi.org/10.1016/j.giq.2003.12.009

Laudan, L. (1968). Theories of Scientific Method from Plato to Mach: A Bibliographical Review. *History of Science*, *7*(1), 1–63. https://doi.org/10.1177/007327536800700101

Lee, A. S. (2004). Thinking about social theory and philosophy for information systems. In J. Mingers & L. Willcocks (Eds.), *Social theory and philosophy for information systems* (pp. 1–26). John Wiley & Sons.

Lee, A. S., Thomas, M., & Baskerville, R. L. (2015). Going back to basics in design science: From the information technology artifact to the information systems artifact. *Information Systems Journal*, *25*(1), 5–21. https://doi.org/10.1111/isj.12054

Lepri, B., Oliver, N., Letouzé, E., Pentland, A., & Vinck, P. (2018). Fair, Transparent, and Accountable Algorithmic Decision-making Processes. *Philosophy & Technology*, *31*(4), 611–627. https://doi.org/10.1007/s13347-017-0279-x

Lessig, L. (2000). Code is law. *Harvard Magazine*, *1*, 2000.

Li, H., Sarathy, R., Zhang, J., & Luo, X. (2014). Exploring the effects of organizational justice, personal ethics and sanction on internet use policy compliance. *Information Systems Journal*, *24*(6), 479–502. https://doi.org/10.1111/isj.12037

Lind, E. A., Kray, L., & Thompson, L. (1998). The Social Construction of Injustice: Fairness Judgments in Response to Own and Others' Unfair Treatment by Authorities. *Organizational Behavior and Human Decision Processes*, *75*(1), 1–22. https://doi.org/10.1006/obhd.1998.2785

Liu, L. T., Dean, S., Rolf, E., Simchowitz, M., & Hardt, M. (2018). Delayed impact of fair machine learning. In J. Dy & A. Krause (Eds.), *Proceedings of the 35th international conference on machine learning* (Vol. 80, pp. 3150–3158). PMLR. http://proceedings.mlr.press/v80/liu18c.html

Mandal, D., Deng, S., Jana, S., Wing, J., & Hsu, D. J. (2020). Ensuring Fairness Beyond the Training Data. In *Advances in Neural Information Processing Systems* (Vol. 33). https://proceedings.neurips.cc/paper/2020/hash/d6539d3b57159babf6a72e106beb45bd-Abstract.html

Marjanovic, O., Cecez-Kecmanovic, D., & Vidgen, R. (2021). Theorising Algorithmic Justice. *European Journal of Information Systems*, 1–19. https://doi.org/10.1080/0960085X.2021.1934130

Martin, K. (2019). Designing Ethical Algorithms. *MIS Quarterly Executive*, 129–142. https://doi.org/10.17705/2msqe.00012

Mettler, T., Eurich, M., & Winter, R. (2014). On the Use of Experiments in Design Science Research: A Proposition of an Evaluation Framework. *Communications of the AIS*, *34*(1), 223.

Miron, M., Tolan, S., Gómez, E., & Castillo, C. (2020). Evaluating causes of algorithmic bias in juvenile criminal recidivism. *Artificial Intelligence and Law*. https://doi.org/10.1007/s10506-020-09268-y

Mitchell, M. (2009). *Complexity: A guided tour*. Oxford University Press.

Mitchell, M. (2019). *Artificial intelligence: A guide for thinking humans*.

Mohamed, S., Png, M.-T., & Isaac, W. (2020). Decolonial AI: Decolonial Theory as Sociotechnical Foresight in Artificial Intelligence. *Philosophy & Technology*, *33*(4), 659–684. https://doi.org/10.1007/s13347-020-00405-8

Molnar, C. (2020). *Interpretable Machine Learning*. Leanpub. https://leanpub.com/interpretable-machine-learning

Mulligan, D. K., Kroll, J. A., Kohli, N., & Wong, R. Y. (2019). This Thing Called Fairness: Disciplinary Confusion Realizing a Value in Technology. *Proceedings of the ACM on Human-Computer Interaction*, *3*(CSCW), 119:1-119:36. https://doi.org/10.1145/3359221

Narayanan, A. (2018). Translation tutorial: 21 fairness definitions and their politics. *Proc. Conf. Fairness Accountability Transp., New York, USA*, 1170.

Nissenbaum, H. (2001). How computer systems embody values. *Computer*, *34*(3), 120, 118–119. https://doi.org/10.1109/2.910905

Nolte, F., Guhr, N., & Richter, A. (2020). The Journey towards Digital Work Empowerment—Conceptualizing IS-Induced Change on the Shop Floor. *ICIS 2020 Proceedings*. https://aisel.aisnet.org/icis2020/is_workplace_fow/is_workplace_fow/17

Noriega-Campero, A., Garcia-Bulle, B., Cantu, L. F., Bakker, M. A., Tejerina, L., & Pentland, A. (2020). Algorithmic targeting of social policies: Fairness, accuracy, and distributed governance. *Proceedings of the 2020 Conference on Fairness, Accountability, and Transparency*, 241–251. https://doi.org/10.1145/3351095.3375784

Obermeyer, Z., Powers, B., Vogeli, C., & Mullainathan, S. (2019). Dissecting racial bias in an algorithm used to manage the health of populations. *Science*, *366*(6464), 447–453. https://doi.org/10.1126/science.aax2342

O'Neil, C. (2016). *Weapons of math destruction: How big data increases inequality and threatens democracy* (First edition). Crown.

Orlikowski, W. J. (2008). Using Technology and Constituting Structures: A Practice Lens for Studying Technology in Organizations. In *Resources, Co-Evolution and Artifacts* (pp. 255–305). Springer, London. https://doi.org/10.1007/978-1-84628-901-9_10

Orr, W., & Davis, J. L. (2020). Attributions of ethical responsibility by Artificial Intelligence practitioners. *Information, Communication & Society*, *23*(5), 719–735. https://doi.org/10.1080/1369118X.2020.1713842

Ortiz de Guinea, A., & Webster, J. (2017). Combining variance and process in information systems research: Hybrid approaches. *Information and Organization*, *27*(3), 144–162. https://doi.org/10.1016/j.infoandorg.2017.06.002

Otto, B. (2011). Data Governance. *Business & Information Systems Engineering*, *3*(4), 241–244.

Payne, B. K., Vuletich, H. A., & Lundberg, K. B. (2017). The Bias of Crowds: How Implicit Bias Bridges Personal and Systemic Prejudice. *Psychological Inquiry*, *28*(4), 233–248. https://doi.org/10.1080/1047840X.2017.1335568

Pessach, D., & Shmueli, E. (2020). Algorithmic Fairness. *ArXiv:2001.09784 [Cs, Stat]*. http://arxiv.org/abs/2001.09784

Power, D. J. (2008). Decision Support Systems: A Historical Overview. In F. Burstein & C. W. Holsapple (Eds.), *Handbook on Decision Support Systems 1: Basic Themes* (pp. 121–140). Springer. https://doi.org/10.1007/978-3-540-48713-5_7

Quadrianto, N., & Sharmanska, V. (2017). Recycling Privileged Learning and Distribution Matching for Fairness. In I. Guyon, U. V. Luxburg, S. Bengio, H. Wallach, R. Fergus, S. Vishwanathan, & R. Garnett (Eds.), *Advances in Neural Information Processing Systems 30* (pp. 677–688). Curran Associates, Inc. http://papers.nips.cc/paper/6670-recycling-privileged-learning-and-distribution-matching-for-fairness.pdf

Rahmattalabi, A., Vayanos, P., Fulginiti, A., Rice, E., Wilder, B., Yadav, A., & Tambe, M. (2019). Exploring Algorithmic Fairness in Robust Graph Covering Problems. In H. Wallach, H. Larochelle, A. Beygelzimer, F. d\textquotesingle Alché-Buc, E. Fox, & R. Garnett (Eds.), *Advances in Neural Information Processing Systems 32* (pp. 15776–15787). Curran Associates,

Inc. http://papers.nips.cc/paper/9707-exploring-algorithmic-fairness-in-robust-graph-covering-problems.pdf

Rawls, J., & Kelly, E. (Eds.). (2003). *Justice as fairness: A restatement* (3rd ed.). Harvard University Press.

Rhue, L. (2019). Beauty's in the AI of the Beholder: How AI Anchors Subjective and Objective Predictions. *ICIS 2019 Proceedings*. https://aisel.aisnet.org/icis2019/future_of_work/future_work/15

Robert, L. P., Pierce, C., Marquis, L., Kim, S., & Alahmad, R. (2020). Designing fair AI for managing employees in organizations: A review, critique, and design agenda. *Human–Computer Interaction*, *35*(5–6), 545–575. https://doi.org/10.1080/07370024.2020.1735391

Rose, J., Jones, M., & Truex, D. (2005). Socio-theoretic accounts of IS: The problem of agency. *Scandinavian Journal of Information Systems*, *17*(1), 8.

Rosenberger, R. (2020). "But, That's Not Phenomenology!": A Phenomenology of Discriminatory Technologies. *Techné: Research in Philosophy and Technology*, *24*(1/2), 83–113.

Russell, S. (2019). *Human compatible: Artificial intelligence and the problem of control*. Allen Lane, an imprint of Penguin Books.

Saleiro, P., Rodolfa, K. T., & Ghani, R. (2020). Dealing with Bias and Fairness in Data Science Systems: A Practical Hands-on Tutorial. *Proceedings of the 26th ACM SIGKDD International Conference on Knowledge Discovery & Data Mining*, 3513–3514. https://doi.org/10.1145/3394486.3406708

Salimi, B., Howe, B., & Suciu, D. (2020). Database Repair Meets Algorithmic Fairness. *ACM SIGMOD Record*, *49*(1), 34–41.

Samadi, S., Tantipongpipat, U., Morgenstern, J. H., Singh, M., & Vempala, S. (2018). The Price of Fair PCA: One Extra dimension. In S. Bengio, H. Wallach, H. Larochelle, K. Grauman, N. Cesa-Bianchi, & R. Garnett (Eds.), *Advances in Neural Information Processing Systems 31* (pp. 10976–10987). Curran Associates, Inc. http://papers.nips.cc/paper/8294-the-price-of-fair-pca-one-extra-dimension.pdf

Sarker, S., Chatterjee, S., Xiao, X., & Elbanna, A. (2019). The Sociotechnical Axis of Cohesion for the IS Discipline: Its Historical Legacy and its Continued Relevance. *MIS Quarterly*, *43*(3), 695–719. https://doi.org/10.25300/MISQ/2019/13747

Seeber, I., Bittner, E., Briggs, R. O., de Vreede, T., de Vreede, G.-J., Elkins, A., Maier, R., Merz, A. B., Oeste-Reiß, S., Randrup, N., Schwabe, G., & Söllner, M. (2020). Machines as teammates: A research agenda on AI in team collaboration. *Information & Management*, *57*(2), 103174. https://doi.org/10.1016/j.im.2019.103174

Sejnowski, T. J. (2018). *The deep learning revolution*. The MIT Press.

Senoner, J., Netland, T., & Feuerriegel, S. (2021). Using explainable artificial intelligence to improve process quality: Evidence from semiconductor manufacturing. *Management Science*. https://www.research-collection.ethz.ch/handle/20.500.11850/492028

Söllner, M., Benbasat, I., Gefen, D., Leimeister, J. M., & Pavlou, P. A. (2018, June). *Trust—Research Curation*. MIS Quarterly. https://www.misqresearchcurations.org/blog/2017/5/10/trust-1

Söllner, M., Hoffmann, A., Hoffmann, H., & Wacker, Arno & Leimeister, Jan Marco (Eds.). (2012). *Understanding the formation of trust in IT artifacts*.

Sonnenberg, C., & Brocke, J. vom. (2012). Evaluation Patterns for Design Science Research Artefacts. In M. Helfert & B. Donnellan (Eds.), *Practical Aspects of Design Science* (pp. 71–83). Springer Berlin Heidelberg. http://link.springer.com/chapter/10.1007/978-3-642-33681-2_7

Srivastava, M., Heidari, H., & Krause, A. (2019a). Mathematical Notions vs. Human Perception of Fairness: A Descriptive Approach to Fairness for Machine Learning. *ArXiv:1902.04783 [Cs]*. http://arxiv.org/abs/1902.04783

Srivastava, M., Heidari, H., & Krause, A. (2019b). Mathematical Notions vs. Human Perception of Fairness: A Descriptive Approach to Fairness for Machine Learning. *Proceedings of the 25th ACM SIGKDD International Conference on Knowledge Discovery & Data Mining*, 2459–2468. https://doi.org/10.1145/3292500.3330664

Stahl, B. C. (2021). *Artificial Intelligence for a Better Future: An Ecosystem Perspective on the Ethics of AI and Emerging Digital Technologies* (1st ed. 2021 Edition). Springer.

Starr, S. B. (2014). Evidence-based sentencing and the scientific rationalization of discrimination. *Stan. L. Rev.*, *66*, 803.

Tarafdar, M., Cooper, C. L., & Stich, J.-F. (2017). The technostress trifecta - techno eustress, techno distress and design: Theoretical directions and an agenda for research. *Information Systems Journal*, *29*(1), 6–42. https://doi.org/10.1111/isj.12169

Tarafdar, M., Gupta, A., & Turel, O. (2013). The dark side of information technology use. *Information Systems Journal*, *23*(3), 269–275. https://doi.org/10.1111/isj.12015

Tarafdar, M., Gupta, A., & Turel, O. (2015). Editorial. *Information Systems Journal*, *25*(3), 161–170. https://doi.org/10.1111/isj.12070

Templier, M., & Paré, G. (2018). Transparency in literature reviews: An assessment of reporting practices across review types and genres in top IS journals. *European Journal of Information Systems*, *27*(5), 503–550. https://doi.org/10.1080/0960085X.2017.1398880

Valera, I., Singla, A., & Gomez Rodriguez, M. (2018). Enhancing the Accuracy and Fairness of Human Decision Making. In S. Bengio, H. Wallach, H. Larochelle, K. Grauman, N. Cesa-Bianchi, & R. Garnett (Eds.), *Advances in Neural Information Processing Systems 31* (pp. 1769–1778). Curran Associates, Inc. http://papers.nips.cc/paper/7448-enhancing-the-accuracy-and-fairness-of-human-decision-making.pdf

van de Poel, I., & Kroes, P. (2014). Can Technology Embody Values? In P. Kroes & P.-P. Verbeek (Eds.), *The Moral Status of Technical Artefacts* (pp. 103–124). Springer Netherlands. https://doi.org/10.1007/978-94-007-7914-3_7

van den Broek, E., Sergeeva, A., & Huysman, M. (2019, November 6). Hiring Algorithms: An Ethnography of Fairness in Practice. *Proc. Intl. Conf. Information Systems*. Intl. Conf. Information Systems, Munich, Germany. https://aisel.aisnet.org/icis2019/future_of_work/future_work/6

Vavra, P., van Baar, J., & Sanfey, A. (2017). The Neural Basis of Fairness. In M. Li & D. P. Tracer (Eds.), *Interdisciplinary Perspectives on Fairness, Equity, and Justice*. Springer International Publishing. https://www.springer.com/gp/book/9783319589923

Venable, J., Pries-Heje, J., & Baskerville, R. (2014). FEDS: A Framework for Evaluation in Design Science Research. *European Journal of Information Systems*. https://doi.org/10.1057/ejis.2014.36

Verbeek, P.-P. (2011). *Moralizing technology: Understanding and designing the morality of things*. The University of Chicago Press.

Verbeek, P.-P. (2014). Some Misunderstandings About the Moral Significance of Technology. In P. Kroes & P.-P. Verbeek (Eds.), *The Moral Status of Technical Artefacts* (pp. 75–88). Springer Netherlands. https://doi.org/10.1007/978-94-007-7914-3_5

Verma, S., & Rubin, J. (2018). Fairness definitions explained. *Proceedings of the International Workshop on Software Fairness - FairWare '18*, 1–7. https://doi.org/10.1145/3194770.3194776

Vokinger, K. N., Feuerriegel, S., & Kesselheim, A. S. (2021). Mitigating bias in machine learning for medicine. *Communications Medicine*, *1*(1), 1–3. https://doi.org/10.1038/s43856-021-00028-w

vom Brocke, J., Simons, A., Niehaves, B., Reimer, K., Plattfaut, R., & Cleven, A. (2009, January 1). Reconstructing the giant: On the importance of rigour in documenting the literature search process. *Proc. European Conf. on Information Systems*. http://aisel.aisnet.org/ecis2009/161

von Zahn, M., Feuerriegel, S., & Kuehl, N. (2021). The cost of fairness in AI: Evidence from e-commerce. *Business & Information Systems Engineering*. https://www.research-collection.ethz.ch/handle/20.500.11850/488982

Wang, H., Li, C., Gu, B., & Min, W. (2019). Does AI-based credit scoring improve financial inclusion? Evidence from online payday lending. *Proc. Intl. Conf. on Information Systems*, 9.

Washington, A. L. (2018). How to Argue with an Algorithm: Lessons from the COMPAS-ProPublica Debate. *Colorado Technology Law Journal*, *17*, 131.

Wastell, D. (1996). Human-machine dynamics in complex information systems: The 'microworld' paradigm as a heuristic tool for developing theory and exploring design issues. *Information Systems Journal*, *6*(4), 245–260. https://doi.org/10.1111/j.1365-2575.1996.tb00017.x

Wellner, G., & Rothman, T. (2020). Feminist AI: Can We Expect Our AI Systems to Become Feminist? *Philosophy & Technology*, *33*(2), 191–205. https://doi.org/10.1007/s13347-019-00352-z

White & Case. (2017). *Algorithms and bias: What lenders need to know*. https://www.whitecase.com/sites/whitecase/files/files/download/publications/algorithm-risk-thought-leadership.pdf

Wong, P.-H. (2020). Democratizing Algorithmic Fairness. *Philosophy & Technology*, *33*(2), 225–244. https://doi.org/10.1007/s13347-019-00355-w

Yang, K., Qinami, K., Fei-Fei, L., Deng, J., & Russakovsky, O. (2020). Towards fairer datasets: Filtering and balancing the distribution of the people subtree in the ImageNet hierarchy. *Proceedings of the 2020 Conference on Fairness, Accountability, and Transparency*, 547–558. https://doi.org/10.1145/3351095.3375709

Yao, S., & Huang, B. (2017). Beyond Parity: Fairness Objectives for Collaborative Filtering. In I. Guyon, U. V. Luxburg, S. Bengio, H. Wallach, R. Fergus, S. Vishwanathan, & R. Garnett (Eds.), *Advances in Neural Information Processing Systems 30* (pp. 2921–2930). Curran Associates, Inc. http://papers.nips.cc/paper/6885-beyond-parity-fairness-objectives-for-collaborative-filtering.pdf

Yarger, L., Cobb Payton, F., & Neupane, B. (2019). Algorithmic equity in the hiring of underrepresented IT job candidates. *Online Information Review*, *44*(2), 383–395. https://doi.org/10.1108/OIR-10-2018-0334

Zafar, M. B., Valera, I., Rodriguez, M., Gummadi, K., & Weller, A. (2017). From Parity to Preference-based Notions of Fairness in Classification. In I. Guyon, U. V. Luxburg, S. Bengio, H. Wallach, R. Fergus, S. Vishwanathan, & R. Garnett (Eds.), *Advances in Neural Information Processing*

*Systems 30* (pp. 229–239). Curran Associates, Inc. http://papers.nips.cc/paper/6627-from-parity-to-preference-based-notions-of-fairness-in-classification.pdf

Završnik, A. (2019). Algorithmic justice: Algorithms and big data in criminal justice settings. *European Journal of Criminology*, 1477370819876762. https://doi.org/10.1177/1477370819876762

Ziewitz, M. (2016). Governing Algorithms: Myth, Mess, and Methods. *Science, Technology, & Human Values*, *41*(1), 3–16. https://doi.org/10.1177/0162243915608948

Žliobaitė, I. (2017). Measuring discrimination in algorithmic decision making. *Data Mining and Knowledge Discovery*, *31*(4), 1060–1089. https://doi.org/10.1007/s10618-017-0506-1

# Appendix

The Appendix details the procedure used in our systematic literature review. The overall objective was to characterize the discourse on AF and to identify potential limitations and assumptions typical for the discourse. We conducted the following steps: literature search and selection, classification, and analysis. At the end of the Appendix, we provide the list of all considered articles.

**Literature Search and Selection**

We selected literature from two sources: (A) Key machine learning (ML) conferences, which explicitly address the topic of algorithmic fairness (AF) to capture the most recent developments in AF, and (B) Query-based search in a multidisciplinary scientific database to capture the discourse on AF beyond computer science. We refer to literature from (A) and (B) as the *conference set* and the *multidisciplinary set*, respectively. They were merged before analysis. We only considered peer-reviewed articles and restricted our time span to January 2017 through December 2020.

To create the *conference set*, we proceeded as follows. (1) We selected outlets with key importance to the AF community in ML by consulting non-IS colleagues and the most recent peer-reviewed overviews of AF (Chouldechova & Roth, 2018; Pessach & Shmueli, 2020). We settled on the following four conferences: ACM Conference on Fairness, Accountability, and Transparency (FAccT; formerly FAT*); ACM SIGKDD International Conference on Knowledge Discovery & Data Mining (KDD); International Conference on Machine Learning (ICML); and Conference on Neural Information Processing Systems (NeurIPS; formerly NIPS). (2) We reviewed all proceedings of these conferences between January 2017 and December 2020 including overall 9,392 articles (FAccT: 127 items, KDD: 1,286, ICML: 2,919, and NeurIPS: 5,060). For this, we screened the titles of all articles published in these conferences for the occurrences of words suggesting relevance to AF discourse, such as "fair," "justice," "bias," or "discriminate" (and their various morphological forms) leading to a pre-selection of 187 articles. (3) Based on the review of the abstracts, we selected articles that contribute to AF discourse; we also dismissed articles submitted as tutorials. The manual screening of titles and abstracts yielded 166 relevant articles. These articles were included in the subsequent analysis.

To create the *multidisciplinary set*, we employed the following procedure inspired by the systematic approaches to literature studies in IS (vom Brocke et al., 2009). (1) We composed a broad search query to accept all potentially relevant articles. The query we used looks as follows (presented using the Scopus syntax): (("*fair*" PRE/1 ("ML" OR "machine learning" OR "AI" OR "artificial intelligence")) OR (("algorithmic*" OR "AI" OR "ML" OR "machine learning" OR "artificial intelligence") PRE/1 ("fair*" OR "justi*" OR "bias*" OR "unfair*"))). This query accepts various phrases including "fair AI," "fairness-constrained ML," "algorithmic fairness," or "AI bias." (2) We applied the search query to titles, abstracts, and keywords in Elsevier Scopus. The search was limited to articles published between January 2017 and December 2020 in one of the 25% best outlets in each discipline (i.e., 1st quartile according to Scimago Journal & Country Rank listing 7600 journals or conference proceedings with over 2,17 million articles). The search yielded 149 potentially relevant articles. (3) We reviewed the articles based on their title, abstract, and introduction and, based on predefined criteria (see below), selected 69 for further processing. These were included in the *multidisciplinary set*. (4.) Using those 69 articles, through backward and forward search, we composed a list of 287 other potentially relevant articles. (5) We reviewed the articles from this list according to the same criteria as above and selected further 45 articles to be included in the *multidisciplinary set*. Altogether, the *multidisciplinary set* is composed of 114 articles from various disciplines (e.g., humanities and arts, social sciences, computer science, or nursing).

The following criteria were used for selection of the relevant articles: (1) Is it an article with an individual contribution? (eliminating editorials, commentaries, calls for papers, or tutorials). (2) Does it use the words used in the query with the intended meaning? (eliminating articles using "fair" to mean satisfactory or bright, mentioning "affairs," or using the abbreviation "ML" to refer to, e.g., maximum likelihood or milliliters). (3) Does it refer to fairness, justice, or discrimination? (eliminating articles referring to algorithmic or social bias in terms of systematic deviation without implying unfair treatment of anyone). (4) Does it refer to machine learning, artificial intelligence, or algorithmic decision making as a source, a remedy, or an aspect of discrimination?

(eliminating articles referring to unfairness in general terms without establishing a link between technology and discrimination). (5) Does it make contribution towards AF or discusses it as a core aspect? (eliminating articles which refer to AF only as an area for future research or as motivation). Overall, the selection criteria were formulated and applied with caution to guarantee that the selected literature represents the broad discourse about algorithmic fairness.

Overall, 280 articles from various disciplines and including diverse viewpoints form the basis of this critical review: 166 in the conference set and 114 in the multidisciplinary set. The articles were subsequently analyzed as presented in the next subsection.

**Literature Classification**

Each article was classified according to four dimensions: (1) fairness perspective, (2) IS component, (3) methodological paradigm, and (4) scope. The classification was primarily conducted by a senior researcher, who was supported by two other senior researchers / faculty members. In particular, the researchers discussed and established the classification schema together, identified core examples for the classes, and decided on edge and vague cases. They discussed the results in several workshops and interpreted them together.

(1) The *fairness perspective* describes the approach towards or framing of fairness that dominates in the given article. Here we differentiate between social and technical perspectives. Some articles treat fairness as a *social* phenomenon and acknowledge the human origin of fairness. Others see it as a *technical* phenomenon, that is, as something that originates in the data or can be expressed in statistical or mathematical terms. We did not encounter a paper that commits to the sociotechnical perspective as presented in Section 5 and characterized by mutual interdependency, joint optimization, and equivalency between the technical and social components (Sarker et al., 2019). Several articles nominally refer to a *sociotechnical* perspective but they discuss AF in relation to the whole society, position it on a political level, and refer to frameworks provided by gender studies (Draude et al., 2019), colonialism (Mohamed et al., 2020), feminism (Wellner & Rothman, 2020), democracy (Wong, 2020), or human rights (Aizenberg & van den Hoven, 2020). Frequently, these papers use the term "sociotechnical" to highlight the contrast of their perspective with the technical discourse, but they focus on the social aspect of AF by framing fairness as an aspect of human society. Upon careful consideration, we thus decided to count such articles towards the social perspective. Overall, the technical perspective dominates with 210 items (148 in the *conference set*, 62 in the *multidisciplinary set*), while 70 items characterized as following a social perspective (18 in the *conference set*, 52 in the *multidisciplinary set*).

(2) The addressed *IS component* describes the contribution of an article by referring to the component of the IS artifact model (Chatterjee et al., 2021) primarily affected by the contribution. For instance, if an article proposes and tests a new algorithmic metric to constrain fairness of an algorithm, it is classified as *technology subsystem*. If it proposes a new procedure to generate a balanced data set or identifies and describes bias in existing data sets widely used in ML, it is classified as *information subsystem*. If it analyzes understanding of technical fairness metrics by ML developers, it is classified as addressing the link between the *social subsystem* and the *technology subsystem* (*social subsystem* ⇔ *technology subsystem*). If it discusses societal consequences or regulatory aspects of algorithmic discrimination, it falls into the category *broader context*. Other classes were defined accordingly. An overview is in Figure A.

The framework we use for classification of the literature relies on the IS artifact concept offered by Chatterjee et al. (2021) for two reasons. First, it foresees the connection of inputs and outputs through processing within the system and a feedback loop with the environment (Chatterjee et al., 2021). Given the importance of the broad societal impact of the decision algorithms and claims that through interaction with the social environment those algorithms might reinforce disparity (O'Neil, 2016), differentiating between humans who are directly affected by the system and the general society might be useful for better characterizing the existing contributions. Second, the role of data for AF becomes inevitable, such that Chatterjee et al.'s (2021) attempt to include information as a component seemed beneficial for our purposes. We claim the agency of data becomes obvious in AF, where data frequently and without a clear provenance or an explicit creator "decides" upon an individual or a group's treatment (cf. open corpora such as ImageNet).

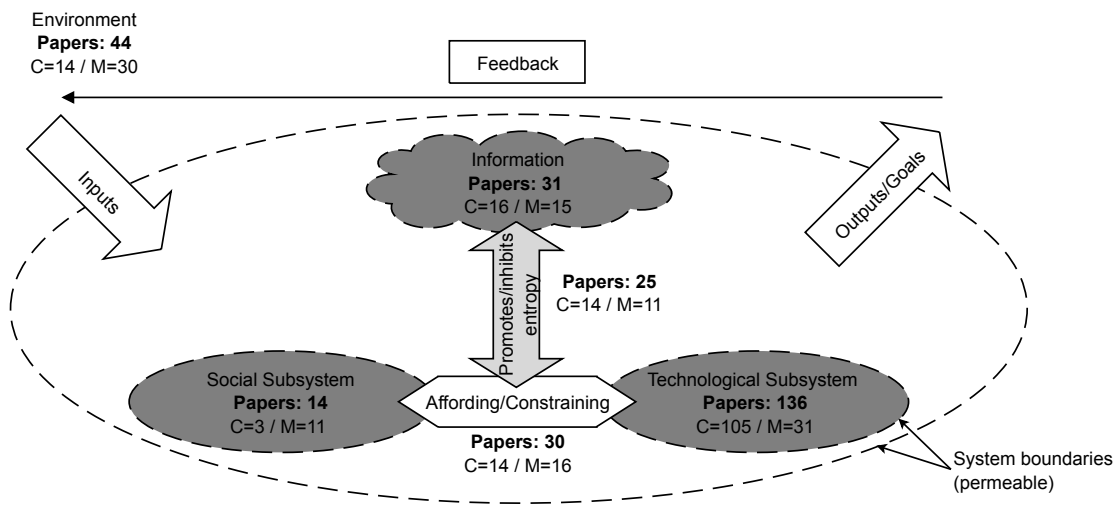

*Figure A. Overview on the analyzed literature mapped onto the sociotechnical framework for theorizing AF. Reported are the overall number of applicable papers from our systematic literature search, as well as those in the conference set (C) and in the multidisciplinary set (M).*

(3) The *methodological paradigm* describes the overall scientific approach of an article. The most frequent approach is subsumed under *engineering*, which involves formulation of a problem, conceptual or formal development of a solution, and evaluation of this solution against the identified problem and comparison with other possible solutions. Overall, 151 studies follow this approach (*conference set*: 121, *multidisciplinary set*: 30). 37 studies focus on *exploring bias* in a specific application domain, a data set, or a case (*conference set*: 16, *multidisciplinary set*: 21). These studies contribute understanding concerning the sources of bias or characterize it quantitatively and qualitatively. In addition, 30 articles rely on literature review as their main evidence (*conference set*: 8, *multidisciplinary set*: 22). Most of them compare various notions of fairness presented in AF, politics, philosophy, or law. Twenty-five studies have *critical* or *argumentative* character (*conference set*: 2, *multidisciplinary set*: 23). They use previous results from various disciplines to propose a new view on AF, and in most cases, these studies are critical of using algorithms for decision making. Twenty-two studies use *behavioral* methods (varying mix of qualitative methods like interviews, surveys, or user experiments) to capture the notion of fairness among study participants or to compare those notions with mathematical operationalization of fairness (*conference set*: 5, *multidisciplinary set*: 17). Finally, 15 studies can be classified as formal or mathematical *proofs* which use logics or mathematical axioms to confirm or disprove a hypothesis (*conference set*: 14, *multidisciplinary set*: 1). Overall, the studies use a wide range of available methods, but more than 50 percent focus on engineering new approaches to address algorithmic bias.

(4) Finally, the *scope* describes whether the research was done in a particular application domain, i.e., is *domain specific*, or whether the paper claims *generic* insights going beyond a particular domain. Overall, we identified 77 domain-specific articles (*conference set*: 30, *multidisciplinary set*: 47). The domains include, among others, health, criminal justice, and loan allocation. The remaining 203 articles all make generic claims (*conference set*: 136, *multidisciplinary set*: 67). For instance, they propose a new technique or metric to prevent algorithmic bias, test it on multiple datasets, and offer it as a context-independent solution to algorithmic discrimination.

**Literature Analysis**

Having classified the literature according to the above categories, we then reviewed the papers to identify assumptions they rely on. We focused on the clusters emerging from the classification: starting with the largest ones (e.g., engineering papers with focus on technology and information subsystems and technical perspective on fairness) moving to medium ones (e.g., critical and argumentative papers with focus on broader context and/or social subsystem and social perspective on fairness) and on down to individual cases (e.g., an engineering paper designing a platform for use by the developers of AF solutions, i.e., social subsystem, to compare across various notions of fairness, i.e., technical perspective). We identified typical assumptions for the clusters and concluded that the assumptions map to the perspective on fairness that dominates in the papers. This mapping is reflected in the structure of the paper, which differentiates between the assumptions of the technical perspective and the assumptions of the social perspective. We grouped similar or overlapping assumptions to offer a comprehensive presentation to the reader.

**Analyzed Articles**

Conference Set

| Article (Conference Set) | Fairness Perspective | IS Component | Methodological Paradigm | Scope |
|---|---|---|---|---|
| Agarwal, A., Beygelzimer, A., Dudik, M., Langford, J., Wallach, H. (2018). A reductions approach to fair classification. In: Proceedings of the 35th international conference on machine learning, 8060-69 | technical | technology | engineering | generic |
| Agarwal, A., Dudik, M., Wu, Zhiwei S. (2019). Fair regression: Quantitative definitions and reduction-based algorithms. In: Proceedings of the 36th international conference on machine learning, 97120-129 | technical | technology | engineering | generic |
| Ahmadian, S., Epasto, A., Knittel, M., Kumar, R., Mahdian, M., Moseley, B., Pham, P., Vassilvitskii, S., Wang, Y. (2020). Fair Hierarchical Clustering. In: Advances in Neural Information Processing Systems, 33 | technical | technology | engineering | generic |
| Ahmadian, S., Epasto, A., Kumar, R., Mahdian, M. (2019). Clustering without Over-Representation. In: Proceedings of the 25th ACM SIGKDD International Conference on Knowledge Discovery & Data Mining, 267–275 | technical | technology | engineering | generic |
| Aivodji, U., Arai, H., Fortineau, O., Gambs, S., Hara, S., Tapp, A. (2019). Fairwashing: the risk of rationalization. In: Proceedings of the 36th international conference on machine learning, 97161-170 | technical | social <> technology | engineering | generic |
| Albarghouthi, A., Vinitsky, S. (2019). Fairness-Aware Programming. In: Proceedings of the Conference on Fairness, Accountability, and Transparency, 211–219 | technical | social | engineering | generic |
| Anders, C., Pasliev, P., Dombrowski, Ann-K., Müller, K-R., Kessel, P. (2020). Fairwashing explanations with off-manifold detergent. In: International Conference on Machine Learning, 314-323 | technical | technology | proof | generic |
| Babaioff, M., Nisan, N., Talgam-Cohen, I. (2019). Fair Allocation through Competitive Equilibrium from Generic Incomes. In: | technical | technology | engineering | domain-specific |

| Article (Conference Set) | Fairness Perspective | IS Component | Methodological Paradigm | Scope |
|---|---|---|---|---|
| Proceedings of the Conference on Fairness, Accountability, and Transparency, 180 | | | | |
| Backurs, A., Indyk, P., Onak, K., Schieber, B., Vakilian, A., Wagner, T. (2019). Scalable fair clustering. In: Proceedings of the 36th international conference on machine learning, 97405-413 | technical | technology | engineering | generic |
| Barabas, C., Doyle, C., Rubinovitz, J., Dinakar, K. (2020). Studying up: reorienting the study of algorithmic fairness around issues of power. In: Proceedings of the 2020 Conference on Fairness, Accountability, and Transparency, 167–176 | social | social | literature | domain-specific |
| Bechavod, Y., Jung, C., Wu, Steven Z. (2020). Metric-Free Individual Fairness in Online Learning. In: Advances in Neural Information Processing Systems, 33 | technical | technology | engineering | generic |
| Bello, K., Honorio, J. (2020). Fairness constraints can help exact inference in structured prediction. In: Advances in Neural Information Processing Systems, 33 | technical | technology | proof | generic |
| Benthall, S., Haynes, B.D. (2019). Racial categories in machine learning. In: Proceedings of the Conference on Fairness, Accountability, and Transparency, 289–298 | technical | technology | engineering | domain-specific |
| Bera, S., Chakrabarty, D., Flores, N., Negahbani, M. (2019). Fair Algorithms for Clustering. In: Advances in Neural Information Processing Systems 32, 4954–4965 | technical | technology | engineering | generic |
| Beutel, A., Chen, J., Doshi, T., Qian, H., Wei, L., Wu, Y., Heldt, L., Zhao, Z., Hong, L., Chi, E.H., Goodrow, C. (2019). Fairness in Recommendation Ranking through Pairwise Comparisons. In: Proceedings of the 25th ACM SIGKDD International Conference on Knowledge Discovery & Data Mining, 2212–2220 | technical | technology | engineering | domain-specific |
| Binns, R. (2018). Fairness in Machine Learning: Lessons from Political Philosophy. In: Conference on Fairness, Accountability and Transparency, 149-159 | social | social <> technology | literature | generic |
| Binns, R. (2020). On the apparent conflict between individual and group | social | broader | literature | generic |

| Article (Conference Set) | Fairness Perspective | IS Component | Methodological Paradigm | Scope |
|---|---|---|---|---|
| fairness. In: Proceedings of the 2020 Conference on Fairness, Accountability, and Transparency, 514–524 | | context | | |
| Bose, A., Hamilton, W. (2019). Compositional fairness constraints for graph embeddings. In: Proceedings of the 36th international conference on machine learning, 97715-724 | technical | technology | engineering | generic |
| Brubach, B., Chakrabarti, D., Dickerson, J., Khuller, S., Srinivasan, A., Tsepenekas, L. (2020). A Pairwise Fair and Community-preserving Approach to k-Center Clustering. In: International Conference on Machine Learning, 1178-1189 | technical | technology | engineering | generic |
| Buolamwini, J., Gebru, T. (2018). Gender Shades: Intersectional Accuracy Disparities in Commercial Gender Classification. In: Conference on Fairness, Accountability and Transparency, 77-91 | technical | information <> technology | exploring bias | generic |
| Burke, R., Sonboli, N., Ordonez-Gauger, A. (2018). Balanced Neighborhoods for Multi-sided Fairness in Recommendation. In: Conference on Fairness, Accountability and Transparency, 202-214 | technical | technology | engineering | domain-specific |
| Canetti, R., Cohen, A., Dikkala, N., Ramnarayan, G., Scheffler, S., Smith, A. (2019). From Soft Classifiers to Hard Decisions: How fair can we be?. In: Proceedings of the Conference on Fairness, Accountability, and Transparency, 309–318 | technical | technology | engineering | generic |
| Caruana, R. (2019). Friends Don't Let Friends Deploy Black-Box Models: The Importance of Intelligibility in Machine Learning. In: Proceedings of the 25th ACM SIGKDD International Conference on Knowledge Discovery & Data Mining, 3174 | technical | social <> technology | engineering | generic |
| Cayci, S., Gupta, S., Eryilmaz, A. (2020). Group-Fair Online Allocation in Continuous Time. In: Advances in Neural Information Processing Systems, 33 | technical | technology | engineering | generic |
| Celis, E., Keswani, V., Straszak, D., Deshpande, A., Kathuria, T., Vishnoi, N. (2018). Fair and diverse DPP-Based data summarization. In: Proceedings of the 35th international conference on machine learning, 80716-725 | technical | technology | engineering | generic |

Authors' manuscript accepted for publication in Information Systems Journal in 2021| Article (Conference Set) | Fairness Perspective | IS Component | Methodological Paradigm | Scope |
|---|---|---|---|---|
| Celis, L. E., Mehrotra, A., Vishnoi, Nisheeth K. (2020). Interventions for ranking in the presence of implicit bias. In: Proceedings of the 2020 Conference on Fairness, Accountability, and Transparency, 369–380 | technical | technology | engineering | generic |
| Chaibub Neto, E. (2020). A Causal Look at Statistical Definitions of Discrimination. In: Proceedings of the 26th ACM SIGKDD International Conference on Knowledge Discovery & Data Mining, 873–881 | technical | technology | engineering | generic |
| Chen, J., Kallus, N., Mao, X., Svacha, G., Udell, M. (2019). Fairness Under Unawareness: Assessing Disparity When Protected Class Is Unobserved. In: Proceedings of the Conference on Fairness, Accountability, and Transparency, 339–348 | technical | information | engineering | generic |
| Chen, X., Fain, B., Lyu, L., Munagala, K. (2019). Proportionally fair clustering. In: Proceedings of the 36th international conference on machine learning, 971032-1041 | technical | technology | engineering | generic |
| Chierichetti, F., Kumar, R., Lattanzi, S., Vassilvitskii, S. (2017). Fair Clustering Through Fairlets. In: Advances in Neural Information Processing Systems 30, 5029–5037 | technical | technology | engineering | generic |
| Chiplunkar, A., Kale, S., Ramamoorthy, Sivaramakrishnan N. (2020). How to Solve Fair k-Center in Massive Data Models. In: International Conference on Machine Learning, 1877-1886 | technical | technology | engineering | generic |
| Cho, J., Hwang, G., Suh, C. (2020). A Fair Classifier Using Kernel Density Estimation. In: Advances in Neural Information Processing Systems, 33 | technical | technology | engineering | generic |
| Choi, K., Grover, A., Singh, T., Shu, R., Ermon, S. (2020). Fair Generative Modeling via Weak Supervision. In: International Conference on Machine Learning, 1887-1898 | technical | technology | engineering | generic |
| Chouldechova, A., Benavides-Prado, D., Fialko, O., Vaithianathan, R. (2018). A case study of algorithm-assisted decision making in child maltreatment hotline screening decisions. In: Conference on Fairness, Accountability and Transparency, 134-148 | social | social <> technology | exploring bias | domain-specific |
| Chzhen, E., Denis, C., Hebiri, M., Oneto, L., Pontil, M. (2019). | technical | information | engineering | generic |

| Article (Conference Set) | Fairness Perspective | IS Component | Methodological Paradigm | Scope |
|---|---|---|---|---|
| Leveraging Labeled and Unlabeled Data for Consistent Fair Binary Classification. In: Advances in Neural Information Processing Systems 32, 12760–12770 | | | | |
| Chzhen, E., Denis, C., Hebiri, M., Oneto, L., Pontil, M. (2020). Fair regression via plug-in estimator and recalibration with statistical guarantees. In: Advances in Neural Information Processing Systems, 33 | technical | information <> technology | engineering | generic |
| Chzhen, E., Denis, C., Hebiri, M., Oneto, L., Pontil, M. (2020). Fair regression with Wasserstein barycenters. In: Advances in Neural Information Processing Systems, 33 | technical | technology | engineering | generic |
| Coston, A., Mishler, A., Kennedy, E.H., Chouldechova, A. (2020). Counterfactual risk assessments, evaluation, and fairness. In: Proceedings of the 2020 Conference on Fairness, Accountability, and Transparency, 582–593 | technical | technology | engineering | domain-specific |
| Cotter, A., Gupta, M., Jiang, H., Srebro, N., Sridharan, K., Wang, S., Woodworth, B., You, S. (2019). Training well-generalizing classifiers for fairness metrics and other data-dependent constraints. In: Proceedings of the 36th international conference on machine learning, 971397-1405 | technical | technology | engineering | generic |
| Creager, E., Madras, D., Jacobsen, Joern-H., Weis, M., Swersky, K., Pitassi, T., Zemel, R. (2019). Flexibly fair representation learning by disentanglement. In: Proceedings of the 36th international conference on machine learning, 971436-1445 | technical | information <> technology | engineering | generic |
| Creager, E., Madras, D., Pitassi, T., Zemel, R. (2020). Causal Modeling for Fairness In Dynamical Systems. In: International Conference on Machine Learning, 2185-2195 | technical | technology | engineering | generic |
| D'Amour, A., Srinivasan, H., Atwood, J., Baljekar, P., Sculley, D., Halpern, Y. (2020). Fairness is not static: deeper understanding of long term fairness via simulation studies. In: Proceedings of the 2020 Conference on Fairness, Accountability, and Transparency, 525–534 | technical | information <> technology | engineering | generic |
| Datta, A., Datta, A., Makagon, J., Mulligan, Deirdre K., Tschantz, | technical | broader | exploring bias | domain- |

| Article (Conference Set) | Fairness Perspective | IS Component | Methodological Paradigm | Scope |
|---|---|---|---|---|
| Michael C. (2018). Discrimination in Online Advertising: A Multidisciplinary Inquiry. In: Conference on Fairness, Accountability and Transparency, 20-34 | | context | | specific |
| De-Arteaga, M., Romanov, A., Wallach, H., Chayes, J., Borgs, C., Chouldechova, A., Geyik, S., Kenthapadi, K., Kalai, Adam T. (2019). Bias in Bios: A Case Study of Semantic Representation Bias in a High-Stakes Setting. In: Proceedings of the Conference on Fairness, Accountability, and Transparency, 120–128 | technical | information | exploring bias | domain-specific |
| DiCiccio, C., Vasudevan, S., Basu, K., Kenthapadi, K., Agarwal, D. (2020). Evaluating Fairness Using Permutation Tests. In: Proceedings of the 26th ACM SIGKDD International Conference on Knowledge Discovery & Data Mining, 1467–1477 | technical | social <> technology | engineering | generic |
| Donahue, K., Kleinberg, J. (2020). Fairness and utilization in allocating resources with uncertain demand. In: Proceedings of the 2020 Conference on Fairness, Accountability, and Transparency, 658–668 | technical | technology | engineering | generic |
| Donini, M., Oneto, L., Ben-David, S., Shawe-Taylor, John S., Pontil, M. (2018). Empirical Risk Minimization Under Fairness Constraints. In: Advances in Neural Information Processing Systems 31, 2791–2801 | technical | technology | engineering | generic |
| Dutta, S., Wei, D., Yueksel, H., Chen, Pin-Y., Liu, S., Varshney, K. (2020). Is There a Trade-Off Between Fairness and Accuracy? A Perspective Using Mismatched Hypothesis Testing. In: International Conference on Machine Learning, 2803-2813 | technical | technology | proof | generic |
| Ekstrand, M.D., Tian, M., Azpiazu, Ion M., Ekstrand, J.D.. Anuyah, O., McNeill, D., Pera, Maria S. (2018). All The Cool Kids, How Do They Fit In?: Popularity and Demographic Biases in Recommender Evaluation and Effectiveness. In: Conference on Fairness, Accountability and Transparency, 172-186 | technical | information <> technology | exploring bias | domain-specific |
| Ekstrand, Michael D., Joshaghani, R., Mehrpouyan, H. (2018). Privacy for All: Ensuring Fair and Equitable Privacy Protections. In: Conference on Fairness, Accountability and Transparency, 35-47 | technical | information <> technology | literature | generic |

| Article (Conference Set) | Fairness Perspective | IS Component | Methodological Paradigm | Scope |
|---|---|---|---|---|
| El Halabi, M., Mitrović, S., Norouzi-Fard, A., Tardos, J., Tarnawski, J M. (2020). Fairness in Streaming Submodular Maximization: Algorithms and Hardness. In: Advances in Neural Information Processing Systems, 33 | technical | technology | engineering | generic |
| Elzayn, H., Jabbari, S., Jung, C., Kearns, M., Neel, S., Roth, A., Schutzman, Z. (2019). Fair Algorithms for Learning in Allocation Problems. In: Proceedings of the Conference on Fairness, Accountability, and Transparency, 170–179 | technical | technology | engineering | domain-specific |
| Esmaeili, S., Brubach, B., Tsepenekas, L., Dickerson, J. (2020). Probabilistic Fair Clustering. In: Advances in Neural Information Processing Systems, 33 | technical | technology | engineering | generic |
| Friedler, S.A., Scheidegger, C., Venkatasubramanian, S., Choudhary, S., Hamilton, E.P., Roth, D. (2019). A comparative study of fairness-enhancing interventions in machine learning. In: Proceedings of the Conference on Fairness, Accountability, and Transparency, 329–338 | technical | technology | engineering | generic |
| Gebru, T. (2020). Lessons from Archives: Strategies for Collecting Sociocultural Data in Machine Learning. In: Proceedings of the 26th ACM SIGKDD International Conference on Knowledge Discovery & Data Mining, 3609 | social | social | critical | generic |
| Geyik, Sahin C., Ambler, S., Kenthapadi, K. (2019). Fairness-Aware Ranking in Search & Recommendation Systems with Application to LinkedIn Talent Search. In: Proceedings of the 25th ACM SIGKDD International Conference on Knowledge Discovery & Data Mining, 2221–2231 | technical | technology | engineering | domain-specific |
| Gillen, S., Jung, C., Kearns, M., Roth, A. (2018). Online Learning with an Unknown Fairness Metric. In: Advances in Neural Information Processing Systems 31, 2600–2609 | technical | technology | engineering | generic |
| Glymour, B., Herington, J. (2019). Measuring the Biases that Matter: The Ethical and Casual Foundations for Measures of Fairness in Algorithms. In: Proceedings of the Conference on Fairness, Accountability, and Transparency, 269–278 | social | broader context | literature | generic |

| Article (Conference Set) | Fairness Perspective | IS Component | Methodological Paradigm | Scope |
|---|---|---|---|---|
| Goelz, P., Kahng, A., Procaccia, Ariel D (2019). Paradoxes in Fair Machine Learning. In: Advances in Neural Information Processing Systems 32, 8342–8352 | technical | technology | engineering | generic |
| Gordaliza, P., Barrio, Eustasio D., Fabrice, G., Loubes, Jean-M. (2019). Obtaining fairness using optimal transport theory. In: Proceedings of the 36th international conference on machine learning, 972357-2365 | technical | technology | engineering | generic |
| Hanna, A., Denton, E., Smart, A., Smith-Loud, J. (2020). Towards a critical race methodology in algorithmic fairness. In: Proceedings of the 2020 Conference on Fairness, Accountability, and Transparency, 501–512 | social | broader context | literature | generic |
| Har-Peled, S., Mahabadi, S. (2019). Near Neighbor: Who is the Fairest of Them All?. In: Advances in Neural Information Processing Systems 32, 13176–13187 | technical | technology | engineering | generic |
| Harb, E., Lam, Ho S. (2020). KFC: A Scalable Approximation Algorithm for $k$−center Fair Clustering. In: Advances in Neural Information Processing Systems, 33 | technical | technology | engineering | generic |
| Harrison, G., Hanson, J., Jacinto, C., Ramirez, J., Ur, B. (2020). An empirical study on the perceived fairness of realistic, imperfect machine learning models. In: Proceedings of the 2020 Conference on Fairness, Accountability, and Transparency, 392–402 | social | social <> technology | behavioral | generic |
| Hashimoto, T., Srivastava, M., Namkoong, H., Liang, P. (2018). Fairness without demographics in repeated loss minimization. In: Proceedings of the 35th international conference on machine learning, 801929-1938 | technical | technology | engineering | generic |
| Heidari, H., Ferrari, C., Gummadi, K., Krause, A. (2018). Fairness Behind a Veil of Ignorance: A Welfare Analysis for Automated Decision Making. In: Advances in Neural Information Processing Systems 31, 1265–1276 | technical | technology | engineering | generic |
| Heidari, H., Loi, M., Gummadi, K.P., Krause, A. (2019). A Moral Framework for Understanding Fair ML through Economic Models of | social | social <> technology | engineering | generic |

| Article (Conference Set) | Fairness Perspective | IS Component | Methodological Paradigm | Scope |
|---|---|---|---|---|
| Equality of Opportunity. In: Proceedings of the Conference on Fairness, Accountability, and Transparency, 181–190 | | | | |
| Heidari, H., Nanda, V., Gummadi, K. (2019). On the long-term impact of algorithmic decision policies: Effort unfairness and feature segregation through social learning. In: Proceedings of the 36th international conference on machine learning, 972692-2701 | social | broader context | exploring bias | generic |
| Hiranandani, G., Narasimhan, H., Koyejo, Oluwasanmi O. (2020). Fair Performance Metric Elicitation. In: Advances in Neural Information Processing Systems, 33 | technical | technology | engineering | generic |
| Hu, L., Chen, Y. (2020). Fair classification and social welfare. In: Proceedings of the 2020 Conference on Fairness, Accountability, and Transparency, 535–545 | technical | social <> technology | engineering | domain-specific |
| Hu, L., Kohler-Hausmann, I. (2020). What's sex got to do with machine learning?. In: Proceedings of the 2020 Conference on Fairness, Accountability, and Transparency, 513 | technical | technology | engineering | generic |
| Hu, Y., Wu, Y., Zhang, L., Wu, X. (2020). Fair Multiple Decision Making Through Soft Interventions. In: Advances in Neural Information Processing Systems, 33 | technical | technology | engineering | generic |
| Huang, L., Jiang, S., Vishnoi, N. (2019). Coresets for Clustering with Fairness Constraints. In: Advances in Neural Information Processing Systems 32, 7589–7600 | technical | technology | engineering | generic |
| Huang, L., Vishnoi, N. (2019). Stable and fair classification. In: Proceedings of the 36th international conference on machine learning, 972879-2890 | technical | technology | engineering | generic |
| Hutchinson, B., Mitchell, M. (2019). 50 Years of Test (Un)fairness: Lessons for Machine Learning. In: Proceedings of the Conference on Fairness, Accountability, and Transparency, 49–58 | social | broader context | literature | generic |
| Ilvento, C., Jagadeesan, M., Chawla, S. (2020). Multi-category fairness in sponsored search auctions. In: Proceedings of the 2020 Conference on Fairness, Accountability, and Transparency, 348–358 | technical | technology | engineering | domain-specific |

| Article (Conference Set) | Fairness Perspective | IS Component | Methodological Paradigm | Scope |
|---|---|---|---|---|
| Jabbari, S., Joseph, M., Kearns, M., Morgenstern, J., Roth, A. (2017). Fairness in reinforcement learning. In: Proceedings of the 34th international conference on machine learning, 701617-1626 | technical | technology | engineering | generic |
| Jagielski, M., Kearns, M., Mao, J., Oprea, A., Roth, A., Malvajerdi, S.S., Ullman, J. (2019). Differentially private fair learning. In: Proceedings of the 36th international conference on machine learning, 973000-3008 | technical | technology | engineering | generic |
| Ji, D., Smyth, P., Steyvers, M. (2020). Can I Trust My Fairness Metric? Assessing Fairness with Unlabeled Data and Bayesian Inference. In: Advances in Neural Information Processing Systems, 33 | technical | information | engineering | generic |
| Jiang, J., Lu, Z. (2019). Learning Fairness in Multi-Agent Systems. In: Advances in Neural Information Processing Systems 32, 13854–13865 | technical | technology | engineering | generic |
| Jones, M., Nguyen, H., Nguyen, T. (2020). Fair k-Centers via Maximum Matching. In: International Conference on Machine Learning, 4940-4949 | technical | technology | engineering | generic |
| Kallus, N., Zhou, A. (2018). Residual unfairness in fair machine learning from prejudiced data. In: Proceedings of the 35th international conference on machine learning, 802439-2448 | technical | broader context | exploring bias | domain-specific |
| Kallus, N., Zhou, A. (2019). The Fairness of Risk Scores Beyond Classification: Bipartite Ranking and the XAUC Metric. In: Advances in Neural Information Processing Systems 32, 3438–3448 | technical | technology | engineering | domain-specific |
| Kang, J., He, J., Maciejewski, R., Tong, H. (2020). InFoRM: Individual Fairness on Graph Mining. In: Proceedings of the 26th ACM SIGKDD International Conference on Knowledge Discovery & Data Mining, 379–389 | technical | technology | engineering | generic |
| Kazemi, E., Zadimoghaddam, M., Karbasi, A. (2018). Scalable deletion-robust submodular maximization: Data summarization with privacy and fairness constraints. In: Proceedings of the 35th international conference on machine learning, 802544-2553 | technical | information | engineering | generic |
| Kearns, M., Neel, S., Roth, A., Wu, Zhiwei S. (2018). Preventing fairness gerrymandering: Auditing and learning for subgroup fairness. | technical | technology | engineering | generic |

| Article (Conference Set) | Fairness Perspective | IS Component | Methodological Paradigm | Scope |
|---|---|---|---|---|
| In: Proceedings of the 35th international conference on machine learning, 802564-2572 | | | | |
| Kearns, M., Neel, S., Roth, A., Wu, Zhiwei S. (2019). An Empirical Study of Rich Subgroup Fairness for Machine Learning. In: Proceedings of the Conference on Fairness, Accountability, and Transparency, 100–109 | technical | technology | engineering | generic |
| Kearns, M., Roth, A., Wu, Zhiwei S. (2017). Meritocratic fairness for cross-population selection. In: Proceedings of the 34th international conference on machine learning, 701828-1836 | technical | technology | engineering | generic |
| Kilbertus, N., Gascon, A., Kusner, M., Veale, M., Gummadi, K., Weller, A. (2018). Blind justice: Fairness with encrypted sensitive attributes. In: Proceedings of the 35th international conference on machine learning, 802630-2639 | technical | information <> technology | engineering | generic |
| Kim, Joon S., Chen, J., Talwalkar, A. (2020). FACT: A Diagnostic for Group Fairness Trade-offs. In: International Conference on Machine Learning, 5264-5274 | technical | technology | engineering | generic |
| Kim, M., Reingold, O., Rothblum, G. (2018). Fairness Through Computationally-Bounded Awareness. In: Advances in Neural Information Processing Systems 31, 4842–4852 | technical | technology | engineering | generic |
| Kim, M.P.. Korolova, A., Rothblum, Guy N., Yona, G. (2020). Preference-informed fairness. In: Proceedings of the 2020 Conference on Fairness, Accountability, and Transparency, 546 | technical | social <> technology | engineering | domain-specific |
| Kleindessner, M., Awasthi, P., Morgenstern, J. (2019). Fair k-Center clustering for data summarization. In: Proceedings of the 36th international conference on machine learning, 973448-3457 | technical | technology | engineering | generic |
| Kleindessner, M., Samadi, S., Awasthi, P., Morgenstern, J. (2019). Guarantees for spectral clustering with fairness constraints. In: Proceedings of the 36th international conference on machine learning, 973458-3467 | technical | technology | engineering | generic |
| Kobren, A., Saha, B., McCallum, A. (2019). Paper Matching with Local | technical | technology | engineering | domain- |

Authors' manuscript accepted for publication in Information Systems Journal in 2021

| Article (Conference Set) | Fairness Perspective | IS Component | Methodological Paradigm | Scope |
|---|---|---|---|---|
| Fairness Constraints. In: Proceedings of the 25th ACM SIGKDD International Conference on Knowledge Discovery & Data Mining, 1247–1257 | | | | specific |
| Komiyama, J., Takeda, A., Honda, J., Shimao, H. (2018). Nonconvex optimization for regression with fairness constraints. In: Proceedings of the 35th international conference on machine learning, 802737-2746 | technical | technology | engineering | generic |
| Kuhlman, C., Jackson, L., Chunara, R. (2020). No Computation without Representation: Avoiding Data and Algorithm Biases through Diversity. In: Proceedings of the 26th ACM SIGKDD International Conference on Knowledge Discovery & Data Mining, 3593 | social | broader context | exploring bias | generic |
| Kulynych, B., Overdorf, R., Troncoso, C., Gürses, S. (2020). POTs: protective optimization technologies. In: Proceedings of the 2020 Conference on Fairness, Accountability, and Transparency, 177–188 | technical | technology | engineering | generic |
| Kusner, M J., Loftus, J., Russell, C., Silva, R. (2017). Counterfactual Fairness. In: Advances in Neural Information Processing Systems 30, 4066–4076 | technical | technology | engineering | generic |
| Lahoti, P., Beutel, A., Chen, J., Lee, K., Prost, F., Thain, N., Wang, X., Chi, E. (2020). Fairness without Demographics through Adversarially Reweighted Learning. In: Advances in Neural Information Processing Systems, 33 | technical | information | engineering | generic |
| Lamy, A., Zhong, Z., Menon, A.K., Verma, N. (2019). Noise-tolerant fair classification. In: Advances in Neural Information Processing Systems 32, 294–306 | technical | information <> technology | engineering | generic |
| Lei, H., Zhao, Y., Cai, L. (2020). Multi-objective Optimization for Guaranteed Delivery in Video Service Platform. In: Proceedings of the 26th ACM SIGKDD International Conference on Knowledge Discovery & Data Mining, 3017–3025 | technical | technology | engineering | domain-specific |
| Lesmana, N.S., Zhang, X., Bei, X. (2019). Balancing Efficiency and Fairness in On-Demand Ridesourcing. In: Advances in Neural Information Processing Systems 32, 5309–5319 | technical | technology | engineering | domain-specific |

| Article (Conference Set) | Fairness Perspective | IS Component | Methodological Paradigm | Scope |
|---|---|---|---|---|
| Li, Y., Sun, H., Wang, Wendy H. (2020). Towards Fair Truth Discovery from Biased Crowdsourced Answers. In: Proceedings of the 26th ACM SIGKDD International Conference on Knowledge Discovery & Data Mining, 599–607 | technical | information | engineering | generic |
| Liu, L T., Dean, S., Rolf, E., Simchowitz, M., Hardt, M. (2018). Delayed impact of fair machine learning. In: Proceedings of the 35th international conference on machine learning, 803150-3158 | technical | broader context | exploring bias | generic |
| Liu, L.T., Simchowitz, M., Hardt, M. (2019). The implicit fairness criterion of unconstrained learning. In: Proceedings of the 36th international conference on machine learning, 974051-4060 | technical | technology | proof | generic |
| Locatello, F., Abbati, G., Rainforth, T., Bauer, S., Schölkopf, B., Bachem, O. (2019). On the Fairness of Disentangled Representations. In: Advances in Neural Information Processing Systems 32, 14611–14624 | technical | information | exploring bias | generic |
| Lohaus, M., Perrot, M., Luxburg, Ulrike V. (2020). Too Relaxed to Be Fair. In: International Conference on Machine Learning, 6360-6369 | technical | technology | engineering | generic |
| Lundgard, A. (2020). Measuring justice in machine learning. In: Proceedings of the 2020 Conference on Fairness, Accountability, and Transparency, 680 | social | broader context | critical | generic |
| Madras, D., Creager, E., Pitassi, T., Zemel, R. (2018). Learning adversarially fair and transferable representations. In: Proceedings of the 35th international conference on machine learning, 803384-3393 | technical | technology | engineering | generic |
| Madras, D., Creager, E., Pitassi, T., Zemel, R. (2019). Fairness through Causal Awareness: Learning Causal Latent-Variable Models for Biased Data. In: Proceedings of the Conference on Fairness, Accountability, and Transparency, 349–358 | technical | technology | engineering | generic |
| Madras, D., Pitassi, T., Zemel, R. (2018). Predict Responsibly: Improving Fairness and Accuracy by Learning to Defer. In: Advances in Neural Information Processing Systems 31, 6147–6157 | technical | social <> technology | engineering | generic |
| Mahabadi, S., Vakilian, A. (2020). Individual Fairness for k-Clustering. | technical | technology | engineering | generic |

| Article (Conference Set) | Fairness Perspective | IS Component | Methodological Paradigm | Scope |
|---|---|---|---|---|
| In: International Conference on Machine Learning, 6586-6596 | | | | |
| Malgieri, G. (2020). The concept of fairness in the GDPR: a linguistic and contextual interpretation. In: Proceedings of the 2020 Conference on Fairness, Accountability, and Transparency, 154–166 | social | broader context | exploring bias | generic |
| Mandal, D., Deng, S., Jana, S., Wing, J., Hsu, Daniel J. (2020). Ensuring Fairness Beyond the Training Data. In: Advances in Neural Information Processing Systems, 33 | technical | information <> technology | engineering | generic |
| Mary, J., Calauzènes, C., Karoui, Noureddine E. (2019). Fairness-aware learning for continuous attributes and treatments. In: Proceedings of the 36th international conference on machine learning, 974382-4391 | technical | information <> technology | engineering | generic |
| Menon, Aditya K., Williamson, Robert C. (2018). The cost of fairness in binary classification. In: Conference on Fairness, Accountability and Transparency, 107-118 | technical | technology | proof | generic |
| Metevier, B., Giguere, S., Brockman, S., Kobren, A., Brun, Y., Brunskill, E., Thomas, Philip S. (2019). Offline Contextual Bandits with High Probability Fairness Guarantees. In: Advances in Neural Information Processing Systems 32, 14922–14933 | technical | technology | engineering | generic |
| Mouzannar, H., Ohannessian, Mesrob I., Srebro, N. (2019). From Fair Decision Making To Social Equality. In: Proceedings of the Conference on Fairness, Accountability, and Transparency, 359–368 | technical | technology | proof | generic |
| Mozannar, H., Ohannessian, M., Srebro, N. (2020). Fair Learning with Private Demographic Data. In: International Conference on Machine Learning, 7066-7075 | technical | information <> technology | engineering | generic |
| Mukherjee, D., Yurochkin, M., Banerjee, M., Sun, Y. (2020). Two Simple Ways to Learn Individual Fairness Metrics from Data. In: International Conference on Machine Learning, 7097-7107 | technical | information <> technology | engineering | generic |
| Nabi, R., Malinsky, D., Shpitser, I. (2019). Learning optimal fair policies. In: Proceedings of the 36th international conference on machine learning, 974674-4682 | technical | technology | engineering | generic |

| Article (Conference Set) | Fairness Perspective | IS Component | Methodological Paradigm | Scope |
|---|---|---|---|---|
| Namaki, Mohammad H., Floratou, A., Psallidas, F., Krishnan, S., Agrawal, A., Wu, Y., Zhu, Y., Weimer, M. (2020). Vamsa: Automated Provenance Tracking in Data Science Scripts. In: Proceedings of the 26th ACM SIGKDD International Conference on Knowledge Discovery & Data Mining, 1542–1551 | technical | information | engineering | generic |
| Nasr, M., Tschantz, Michael C. (2020). Bidding strategies with gender nondiscrimination constraints for online ad auctions. In: Proceedings of the 2020 Conference on Fairness, Accountability, and Transparency, 337–347 | technical | technology | proof | domain-specific |
| Noriega-Campero, A., Garcia-Bulle, B., Cantu, Luis F., Bakker, M.A., Tejerina, L., Pentland, A. (2020). Algorithmic targeting of social policies: fairness, accuracy, and distributed governance. In: Proceedings of the 2020 Conference on Fairness, Accountability, and Transparency, 241–251 | technical | social <> technology | behavioral | domain-specific |
| Obermeyer, Z., Mullainathan, S. (2019). Dissecting Racial Bias in an Algorithm that Guides Health Decisions for 70 Million People. In: Proceedings of the Conference on Fairness, Accountability, and Transparency, 89 | technical | technology | exploring bias | domain-specific |
| Oneto, L., Donini, M., Luise, G., Ciliberto, C., Maurer, A., Pontil, M. (2020). Exploiting MMD and Sinkhorn Divergences for Fair and Transferable Representation Learning. In: Advances in Neural Information Processing Systems, 33 | technical | technology | proof | generic |
| Passi, S., Barocas, S. (2019). Problem Formulation and Fairness. In: Proceedings of the Conference on Fairness, Accountability, and Transparency, 39–48 | social | broader context | behavioral | generic |
| Pleiss, G., Raghavan, M., Wu, F., Kleinberg, J., Weinberger, Kilian Q (2017). On Fairness and Calibration. In: Advances in Neural Information Processing Systems 30, 5680–5689 | technical | technology | proof | generic |
| Pujol, D., McKenna, R., Kuppam, S., Hay, M., Machanavajjhala, A., Miklau, G. (2020). Fair decision making using privacy-protected data. In: Proceedings of the 2020 Conference on Fairness, Accountability, | technical | information | exploring bias | domain-specific |

| Article (Conference Set) | Fairness Perspective | IS Component | Methodological Paradigm | Scope |
|---|---|---|---|---|
| and Transparency, 189–199 | | | | |
| Quadrianto, N., Sharmanska, V. (2017). Recycling Privileged Learning and Distribution Matching for Fairness. In: Advances in Neural Information Processing Systems 30, 677–688 | technical | technology | engineering | generic |
| Rahmattalabi, A., Vayanos, P., Fulginiti, A., Rice, E., Wilder, B., Yadav, A., Tambe, M. (2019). Exploring Algorithmic Fairness in Robust Graph Covering Problems. In: Advances in Neural Information Processing Systems 32, 15776–15787 | technical | technology | engineering | domain-specific |
| Roh, Y., Lee, K., Whang, S., Suh, C. (2020). FR-Train: A Mutual Information-Based Approach to Fair and Robust Training. In: International Conference on Machine Learning, 8147-8157 | technical | information | engineering | generic |
| Ruoss, A., Balunovic, M., Fischer, M., Vechev, M. (2020). Learning Certified Individually Fair Representations. In: Advances in Neural Information Processing Systems, 33 | technical | technology | engineering | generic |
| Russell, C., Kusner, M J., Loftus, J., Silva, R. (2017). When Worlds Collide: Integrating Different Counterfactual Assumptions in Fairness. In: Advances in Neural Information Processing Systems 30, 6414–6423 | technical | technology | engineering | generic |
| Sabato, S., Yom-Tov, E. (2020). Bounding the fairness and accuracy of classifiers from population statistics. In: International Conference on Machine Learning, 8316-8325 | technical | technology | engineering | generic |
| Saha, D., Schumann, C., Mcelfresh, D., Dickerson, J., Mazurek, M., Tschantz, M. (2020). Measuring Non-Expert Comprehension of Machine Learning Fairness Metrics. In: International Conference on Machine Learning, 8377-8387 | social | social <> technology | behavioral | generic |
| Samadi, S., Tantipongpipat, U., Morgenstern, J H., Singh, M., Vempala, S. (2018). The Price of Fair PCA: One Extra dimension. In: Advances in Neural Information Processing Systems 31, 10976–10987 | technical | information | engineering | generic |
| Selbst, A.D., Boyd, D., Friedler, S.A., Venkatasubramanian, S., Vertesi, J. (2019). Fairness and Abstraction in Sociotechnical Systems. In: Proceedings of the Conference on Fairness, Accountability, and | social | broader context | literature | generic |

| Article (Conference Set) | Fairness Perspective | IS Component | Methodological Paradigm | Scope |
|---|---|---|---|---|
| Transparency, 59–68 | | | | |
| Shang, J., Sun, M., Lam, Nina S.N. (2020). List-wise Fairness Criterion for Point Processes. In: Proceedings of the 26th ACM SIGKDD International Conference on Knowledge Discovery & Data Mining, 1948–1958 | technical | technology | engineering | generic |
| Siddique, U., Weng, P., Zimmer, M. (2020). Learning Fair Policies in Multi-Objective (Deep) Reinforcement Learning with Average and Discounted Rewards. In: International Conference on Machine Learning, 8905-8915 | technical | technology | engineering | generic |
| Singh, A., Joachims, T. (2018). Fairness of Exposure in Rankings. In: Proceedings of the 24th ACM SIGKDD International Conference on Knowledge Discovery & Data Mining, 2219–2228 | technical | technology | engineering | generic |
| Singh, A., Joachims, T. (2019). Policy Learning for Fairness in Ranking. In: Advances in Neural Information Processing Systems 32, 5426–5436 | technical | technology | engineering | generic |
| Slack, D., Friedler, Sorelle A., Givental, E. (2020). Fairness warnings and fair-MAML: learning fairly with minimal data. In: Proceedings of the 2020 Conference on Fairness, Accountability, and Transparency, 200–209 | technical | information <> technology | engineering | generic |
| Speicher, T., Ali, M., Venkatadri, G., Ribeiro, Filipe N., Arvanitakis, G., Benevenuto, F., Gummadi, K P., Loiseau, P., Mislove, A. (2018). Potential for Discrimination in Online Targeted Advertising. In: Conference on Fairness, Accountability and Transparency, 43586 | technical | social <> technology | exploring bias | domain-specific |
| Speicher, T., Heidari, H., Grgic-Hlaca, N., Gummadi, K P., Singla, A., Weller, A., Zafar, Muhammad B. (2018). A Unified Approach to Quantifying Algorithmic Unfairness: Measuring Individual &Group Unfairness via Inequality Indices. In: Proceedings of the 24th ACM SIGKDD International Conference on Knowledge Discovery & Data Mining, 2239–2248 | technical | technology | proof | generic |
| Srivastava, M., Heidari, H., Krause, A. (2019). Mathematical Notions vs. Human Perception of Fairness: A Descriptive Approach to Fairness for | social | broader context | behavioral | generic |

| Article (Conference Set) | Fairness Perspective | IS Component | Methodological Paradigm | Scope |
|---|---|---|---|---|
| Machine Learning. In: Proceedings of the 25th ACM SIGKDD International Conference on Knowledge Discovery & Data Mining, 2459–2468 | | | | |
| Sühr, T., Biega, A.J., Zehlike, M., Gummadi, Krishna P., Chakraborty, A. (2019). Two-Sided Fairness for Repeated Matchings in Two-Sided Markets: A Case Study of a Ride-Hailing Platform. In: Proceedings of the 25th ACM SIGKDD International Conference on Knowledge Discovery & Data Mining, 3082–3092 | technical | social <> technology | engineering | domain-specific |
| Sweeney, C., Najafian, M. (2020). Reducing sentiment polarity for demographic attributes in word embeddings using adversarial learning. In: Proceedings of the 2020 Conference on Fairness, Accountability, and Transparency, 359–368 | technical | information | exploring bias | domain-specific |
| Tantipongpipat, U., Samadi, S., Singh, M., Morgenstern, J.H., Vempala, S. (2019). Multi-Criteria Dimensionality Reduction with Applications to Fairness. In: Advances in Neural Information Processing Systems 32, 15161–15171 | technical | information <> technology | engineering | generic |
| Ustun, B., Liu, Y., Parkes, D. (2019). Fairness without harm: Decoupled classifiers with preference guarantees. In: Proceedings of the 36th international conference on machine learning, 976373-6382 | technical | technology | engineering | domain-specific |
| Valera, I., Singla, A., Gomez Rodriguez, M. (2018). Enhancing the Accuracy and Fairness of Human Decision Making. In: Advances in Neural Information Processing Systems 31, 1769–1778 | technical | technology | engineering | generic |
| Wang, S., Guo, W., Narasimhan, H., Cotter, A., Gupta, M., Jordan, M. (2020). Robust Optimization for Fairness with Noisy Protected Groups. In: Advances in Neural Information Processing Systems, 33 | technical | technology | engineering | generic |
| Wick, M., Panda, S., Tristan, J-B. (2019). Unlocking Fairness: a Trade-off Revisited. In: Advances in Neural Information Processing Systems 32, 8783–8792 | technical | information | engineering | generic |
| Williamson, R., Menon, A. (2019). Fairness risk measures. In: Proceedings of the 36th international conference on machine learning, | technical | technology | proof | generic |

| Article (Conference Set) | Fairness Perspective | IS Component | Methodological Paradigm | Scope |
|---|---|---|---|---|
| 976786-6797 | | | | |
| Wu, Y., Zhang, L., Wu, X. (2018). On Discrimination Discovery and Removal in Ranked Data using Causal Graph. In: Proceedings of the 24th ACM SIGKDD International Conference on Knowledge Discovery & Data Mining, 2536–2544 | technical | technology | engineering | generic |
| Wu, Y., Zhang, L., Wu, X., Tong, H. (2019). PC-Fairness: A Unified Framework for Measuring Causality-based Fairness. In: Advances in Neural Information Processing Systems 32, 3404–3414 | technical | technology | engineering | generic |
| Xu, R., Cui, P., Kuang, K., Li, B., Zhou, L., Shen, Z., Cui, W. (2020). Algorithmic Decision Making with Conditional Fairness. In: Proceedings of the 26th ACM SIGKDD International Conference on Knowledge Discovery & Data Mining, 2125–2135 | technical | technology | engineering | generic |
| Yang, F., Cisse, M., Koyejo, Oluwasanmi O. (2020). Fairness with Overlapping Groups; a Probabilistic Perspective. In: Advances in Neural Information Processing Systems, 33 | technical | technology | engineering | generic |
| Yang, K., Qinami, K., Fei-Fei, L., Deng, J., Russakovsky, O. (2020). Towards fairer datasets: filtering and balancing the distribution of the people subtree in the ImageNet hierarchy. In: Proceedings of the 2020 Conference on Fairness, Accountability, and Transparency, 547–558 | technical | information | exploring bias | generic |
| Yao, S., Huang, B. (2017). Beyond Parity: Fairness Objectives for Collaborative Filtering. In: Advances in Neural Information Processing Systems 30, 2921–2930 | technical | technology | engineering | domain-specific |
| Yona, G., Rothblum, G. (2018). Probably approximately metric-fair learning. In: Proceedings of the 35th international conference on machine learning, 805680-5688 | technical | technology | proof | generic |
| Zafar, M B., Valera, I., Rodriguez, M., Gummadi, K., Weller, A. (2017). From Parity to Preference-based Notions of Fairness in Classification. In: Advances in Neural Information Processing Systems 30, 229–239 | technical | technology | engineering | generic |
| Zhang, X., Khaliligarekani, M., Tekin, C., liu, mingyan (2019). Group Retention when Using Machine Learning in Sequential Decision | technical | information | proof | generic |

| Article (Conference Set) | Fairness Perspective | IS Component | Methodological Paradigm | Scope |
|---|---|---|---|---|
| Making: the Interplay between User Dynamics and Fairness. In: Advances in Neural Information Processing Systems 32, 15269–15278 | | | | |
| Zhang, X., Tu, R., Liu, Y., Liu, M., Kjellstrom, H., Zhang, K., Zhang, C. (2020). How do fair decisions fare in long-term qualification?. In: Advances in Neural Information Processing Systems, 33 | technical | technology | engineering | generic |
| Zhao, H., Gordon, G. (2019). Inherent Tradeoffs in Learning Fair Representations. In: Advances in Neural Information Processing Systems 32, 15675–15685 | technical | technology | proof | generic |

Multidisciplinary Set

| Article (Multidisciplinary Set) | Fairness Perspective | IS Component | Methodological Paradigm | Scope |
|---|---|---|---|---|
| Abràmoff M.D., Tobey D., Char D.S. (2020). Lessons Learned About Autonomous AI: Finding a Safe, Efficacious, and Ethical Path Through the Development Process. American Journal of Ophthalmology, 214(), 134–142 | technical | social | critical | domain-specific |
| Ahn Y., Lin Y.-R. (2020). Fairsight: Visual analytics for fairness in decision making. IEEE Transactions on Visualization and Computer Graphics, 26(1), 1086–1095 | technical | social <> technology | engineering | generic |
| Aizenberg E., van den Hoven J. (2020). Designing for human rights in AI. Big Data and Society, 7(2) | social | social | critical | generic |
| Ali M., Sapiezynski P., Bogen M., Korolova A., Mislove A., Rieke A. (2019). Discrimination through optimization: How Facebook's ad delivery can lead to biased outcomes. Proceedings of the ACM on Human-Computer Interaction, 3(CSCW) | technical | technology | exploring bias | domain-specific |
| Alkhatib A., Bernstein M. (2019). Street–level algorithms: A theory at the gaps between policy and decisions. Conference on Human Factors in Computing Systems - Proceedings | social | social <> technology | critical | domain-specific |
| Altman M., Wood A., Vayena E. (2018). A Harm-Reduction Framework for Algorithmic Fairness. IEEE Security and Privacy, 16(3), 34–45 | technical | technology | engineering | generic |
| Araujo T., Helberger N., Kruikemeier S., de Vreese C.H. (2020). In AI we trust? Perceptions about automated decision-making by artificial intelligence. AI and Society, 35(3), 611–623 | social | broader context | behavioral | generic |
| Barredo Arrieta A., Díaz-Rodríguez N., Del Ser J., Bennetot A., Tabik S., Barbado A., Garcia S., Gil-Lopez S., Molina D., Benjamins R., Chatila R., Herrera F. (2020). Explainable Explainable Artificial Intelligence (XAI): Concepts, taxonomies, opportunities and challenges toward responsible AI. Information Fusion, 58(), 82–115 | technical | social <> technology | literature | generic |
| Berk R., Elzarka A.A. (2020). Almost politically acceptable criminal | social | social <> | exploring bias | domain-specific |

| Article (Multidisciplinary Set) | Fairness Perspective | IS Component | Methodological Paradigm | Scope |
|---|---|---|---|---|
| justice risk assessment. Criminology and Public Policy, 19(4), 1231–1257 | | technology | | |
| Beyleveld D., Brownsword R. (2019). Punitive and preventive justice in an era of profiling, smart prediction and practical preclusion: Three key questions. International Journal of Law in Context, 15(2), 198–218 | social | broader context | critical | domain-specific |
| Binns R. (2018). What Can Political Philosophy Teach Us about Algorithmic Fairness?. IEEE Security and Privacy, 16(3), 73–80 | social | technology | literature | generic |
| Brandão M., Jirotka M., Webb H., Luff P. (2020). Fair navigation planning: A resource for characterizing and designing fairness in mobile robots. Artificial Intelligence, 282() | technical | technology | engineering | domain-specific |
| Brown A., Chouldechova A., Putnam-Hornstein E., Tobin A., Vaithianathan R. (2019). Toward algorithmic accountability in public services a qualitative study of affected community perspectives on algorithmic decision-making in child welfare services. Conference on Human Factors in Computing Systems - Proceedings | social | social | behavioral | domain-specific |
| Burrell J., Kahn Z., Jonas A., Griffin D. (2019). When users control the algorithms: Values expressed in practices on the twitter platform. Proceedings of the ACM on Human-Computer Interaction, 3(CSCW) | social | social | behavioral | domain-specific |
| Casacuberta D., Guersenzvaig A. (2019). Using Dreyfus' legacy to understand justice in algorithm-based processes. AI and Society, 34(2), 313–319 | social | broader context | literature | generic |
| Celis L.E., Keswani V. (2020). Implicit Diversity in Image Summarization. Proceedings of the ACM on Human-Computer Interaction, 4(CSCW2) | technical | information | engineering | generic |
| Char D.S., Abràmoff M.D., Feudtner C. (2020). Identifying Ethical Considerations for Machine Learning Healthcare Applications. American Journal of Bioethics, 20(11), 7–17 | technical | information <> technology | critical | domain-specific |
| Chouldechova A. (2017). Fair Prediction with Disparate Impact: A Study of Bias in Recidivism Prediction Instruments. Big Data, 5(2), 153–163 | technical | technology | proof | domain-specific |

| Article (Multidisciplinary Set) | Fairness Perspective | IS Component | Methodological Paradigm | Scope |
|---|---|---|---|---|
| Dash A., Shandilya A., Biswas A., Ghosh K., Ghosh S., Chakraborty A. (2019). Summarizing User-generated Textual Content: Motivation and methods for fairness in algorithmic summaries. Proceedings of the ACM on Human-Computer Interaction, 3(CSCW) | technical | information <> technology | exploring bias | generic |
| Díaz M., Diakopoulos N. (2019). Whose walkability?: Challenges in algorithmically measuring subjective experience. Proceedings of the ACM on Human-Computer Interaction, 3(CSCW) | social | social | behavioral | domain-specific |
| Díaz M., Johnson I., Lazar A., Piper A.M., Gergle D. (2018). Addressing age-related bias in sentiment analysis. Conference on Human Factors in Computing Systems - Proceedings, 2018-April() | technical | technology | exploring bias | generic |
| Draude C., Klumbyte G., Lücking P., Treusch P. (2019). Situated algorithms: a sociotechnical systemic approach to bias. Online Information Review, 44(2), 325–342 | social | broader context | critical | generic |
| Du M., Liu N., Yang F., Hu X. (2020). Learning credible DNNs via incorporating prior knowledge and model local explanation. Knowledge and Information Systems | technical | information | engineering | generic |
| Du M., Yang F., Zou N., Hu X. (2020). Fairness in Deep Learning: A Computational Perspective. IEEE Intelligent Systems | technical | information <> technology | literature | generic |
| Du S., Xie C. (2020). Paradoxes of artificial intelligence in consumer markets: Ethical challenges and opportunities. Journal of Business Research | social | broader context | literature | domain-specific |
| Favaretto M., De Clercq E., Elger B.S. (2019). Big Data and discrimination: perils, promises and solutions. A systematic review. Journal of Big Data, 6(1) | social | broader context | literature | generic |
| Feuerriegel S., Dolata M., Schwabe G. (2020). Fair AI: Challenges and Opportunities. Business and Information Systems Engineering, 62(4), 379–384 | technical | information <> technology | critical | generic |
| Fitzsimons J., Al Ali A., Osborne M., Roberts S. (2019). A general framework for fair regression. Entropy, 21(8) | technical | technology | engineering | generic |

| Article (Multidisciplinary Set) | Fairness Perspective | IS Component | Methodological Paradigm | Scope |
|---|---|---|---|---|
| Gao Y., Cui Y. (2020). Deep transfer learning for reducing health care disparities arising from biomedical data inequality. Nature Communications, 11(1) | technical | information | exploring bias | domain-specific |
| García-Soriano D., Bonchi F. (2020). Fair-by-design matching. Data Mining and Knowledge Discovery, 34(5), 1291–1335 | technical | technology | engineering | generic |
| Grari V., Ruf B., Lamprier S., Detyniecki M. (2020). Achieving Fairness with Decision Trees: An Adversarial Approach. Data Science and Engineering, 5(2), 99–110 | technical | technology | engineering | generic |
| Gu X., Angelov P.P., Soares E.A. (2020). A self-adaptive synthetic over-sampling technique for imbalanced classification. International Journal of Intelligent Systems, 35(6), 923–943 | technical | information | engineering | generic |
| Hacker P. (2018). Teaching fairness to artificial intelligence: Existing and novel strategies againstalgorithmic discrimination under EU law. Common Market Law Review, 55(4), 1143–1185 | technical | technology | literature | generic |
| Hagendorff T. (2019). From privacy to anti-discrimination in times of machine learning. Ethics and Information Technology, 21(4), 331–343 | technical | broader context | critical | generic |
| Hagendorff T. (2020). The Ethics of AI Ethics: An Evaluation of Guidelines. Minds and Machines, 30(1), 99–120 | social | broader context | literature | generic |
| Helberger N., Araujo T., de Vreese C.H. (2020). Who is the fairest of them all? Public attitudes and expectations regarding automated decision-making. Computer Law and Security Review, 39() | social | broader context | behavioral | generic |
| Helberger N., Huh J., Milne G., Strycharz J., Sundaram H. (2020). Macro and Exogenous Factors in Computational Advertising: Key Issues and New Research Directions. Journal of Advertising, 49(4), 377–393 | technical | broader context | literature | domain-specific |
| Hellman D. (2020). Measuring algorithmic fairness. Virginia Law Review, 106(4), 811–866 | technical | technology | critical | generic |
| Hoffmann A.L. (2019). Where fairness fails: data, algorithms, and the limits of antidiscrimination discourse. Information Communication and | social | broader context | literature | generic |

| Article (Multidisciplinary Set) | Fairness Perspective | IS Component | Methodological Paradigm | Scope |
|---|---|---|---|---|
| Society, 22(7), 900–915 | | | | |
| Holstein K., Vaughan J.W., Daumé H., III, Dudík M., Wallach H. (2019). Improving fairness in machine learning systems: What do industry practitioners need? Conference on Human Factors in Computing Systems - Proceedings | technical | social | behavioral | generic |
| Huq A.Z. (2019). Racial equity in algorithmic criminal justice. Duke Law Journal, 68(6), 1004–1134 | technical | technology | engineering | domain-specific |
| Jacobson N.C., Bentley K.H., Walton A., Wang S.B., Fortgang R.G., Millner A.J., Coombs G., III, Rodman A.M., Coppersmith D.D.L. (2020). Ethical dilemmas posed by mobile health and machine learning in psychiatry research [Les dilemmes éthiques posés par la santé mobile et l'apprentissage automatique dans la recherche en psychiatrie] [Los dilemas éticos planteados por la salud móvil y el aprendizaje automático dentro de la investigación en psiquiatría]. Bulletin of the World Health Organization, 98(4), 270–276 | social | broader context | literature | domain-specific |
| JafariNaimi N. (2018). Our Bodies in the Trolley's Path, or Why Self-driving Cars Must *Not* Be Programmed to Kill. Science Technology and Human Values, 43(2), 302–323 | social | broader context | critical | domain-specific |
| Johndrow J.E., Lum K. (2019). An algorithm for removing sensitive information: Application to race-independent recidivism prediction. Annals of Applied Statistics, 13(1), 189–220 | technical | information | engineering | domain-specific |
| Johnson G.M. (2020). Algorithmic bias: on the implicit biases of social technology. Synthese | social | social <> technology | critical | generic |
| Kamiran F., Mansha S., Karim A., Zhang X. (2018). Exploiting reject option in classification for social discrimination control. Information Sciences, 425(), 18–33 | technical | technology | engineering | generic |
| Kamishima T., Akaho S., Asoh H., Sakuma J. (2018). Model-based and actual independence for fairness-aware classification. Data Mining and Knowledge Discovery, 32(1), 258–286 | technical | technology | engineering | generic |
| Keyes O., Durbin M., Hutson J. (2019). A mulching proposal: Analysing | technical | technology | engineering | domain-specific |

| Article (Multidisciplinary Set) | Fairness Perspective | IS Component | Methodological Paradigm | Scope |
|---|---|---|---|---|
| and improving an algorithmic system for turning the elderly into high-nutrient slurry. Conference on Human Factors in Computing Systems - Proceedings | | | | |
| Khalil A., Ahmed S.G., Khattak A.M., Al-Qirim N. (2020). Investigating Bias in Facial Analysis Systems: A Systematic Review. IEEE Access, 8(), 130751–130761 | technical | information <> technology | literature | generic |
| Köchling A., Riazy S., Wehner M.C., Simbeck K. (2020). Highly Accurate, But Still Discriminatory: A Fairness Evaluation of Algorithmic Video Analysis in the Recruitment Context. Business and Information Systems Engineering | technical | information | exploring bias | domain-specific |
| Koenecke A., Nam A., Lake E., Nudell J., Quartey M., Mengesha Z., Toups C., Rickford J.R., Jurafsky D., Goel S. (2020). Racial disparities in automated speech recognition. Proceedings of the National Academy of Sciences of the United States of America, 117(14), 7684–7689 | technical | technology | exploring bias | domain-specific |
| Kou Y., Gui X. (2020). Mediating Community-AI Interaction through Situated Explanation: The Case of AI-Led Moderation. Proceedings of the ACM on Human-Computer Interaction, 4(CSCW2) | social | social <> technology | exploring bias | domain-specific |
| Kuziemski M., Misuraca G. (2020). AI governance in the public sector: Three tales from the frontiers of automated decision-making in democratic settings. Telecommunications Policy, 44(6) | social | broader context | exploring bias | domain-specific |
| Kyriazanos D.M., Thanos K.G., Thomopoulos S.C.A. (2019). Automated Decision Making in Airport Checkpoints: Bias Detection Toward Smarter Security and Fairness. IEEE Security and Privacy, 17(2), 8–16 | technical | technology | engineering | domain-specific |
| Lahoti P., Gummadi K.P., Weikum G. (2019). Operationalizing individual fairness with pairwise fair representations. Proceedings of the VLDB Endowment, 13(4), 506–518 | technical | technology | engineering | domain-specific |
| Lambrecht A., Tucker C. (2019). Algorithmic bias? An empirical study of apparent gender-based discrimination in the display of stem career | technical | technology | behavioral | domain-specific |

| Article (Multidisciplinary Set) | Fairness Perspective | IS Component | Methodological Paradigm | Scope |
|---|---|---|---|---|
| ads. Management Science, 65(7), 2966–2981 | | | | |
| Lee M.K. (2018). Understanding perception of algorithmic decisions: Fairness, trust, and emotion in response to algorithmic management. Big Data and Society, 5(1) | social | social | behavioral | generic |
| Lee M.K., Jain A., Cha H.J.I.N., Ojha S., Kusbit D. (2019). Procedural justice in algorithmic fairness: Leveraging transparency and outcome control for fair algorithmic mediation. Proceedings of the ACM on Human-Computer Interaction, 3(CSCW) | social | social <> technology | engineering | generic |
| Lee M.K., Kusbit D., Kahng A., Kim J.T., Yuan X., Chan A., See D., Noothigattu R., Lee S., Psomas A., Procaccia A.D. (2019). Webuildai: Participatory framework for algorithmic governance. Proceedings of the ACM on Human-Computer Interaction, 3(CSCW) | technical | social <> technology | engineering | generic |
| Lee M.S.A., Floridi L. (2020). Algorithmic Fairness in Mortgage Lending: from Absolute Conditions to Relational Trade-offs. Minds and Machines | technical | technology | engineering | domain-specific |
| Lepri B., Oliver N., Letouzé E., Pentland A., Vinck P. (2018). Fair, Transparent, and Accountable Algorithmic Decision-making Processes: The Premise, the Proposed Solutions, and the Open Challenges. Philosophy and Technology, 31(4), 611–627 | social | broader context | literature | generic |
| Lim H.S.M., Taeihagh A. (2019). Algorithmic decision-making in AVs: Understanding ethical and technical concerns for smart cities. Sustainability (Switzerland), 11(20) | social | broader context | critical | domain-specific |
| Lin Y.-T., Hung T.-W., Huang L.T.-L. (2020). Engineering Equity: How AI Can Help Reduce the Harm of Implicit Bias. Philosophy and Technology | technical | social <> technology | critical | domain-specific |
| Lysaght T., Lim H.Y., Xafis V., Ngiam K.Y. (2019). AI-Assisted Decision-making in Healthcare: The Application of an Ethics Framework for Big Data in Health and Research. Asian Bioethics Review, 11(3), 299–314 | social | broader context | critical | domain-specific |
| Mann M., Matzner T. (2019). Challenging algorithmic profiling: The | technical | technology | critical | generic |

| Article (Multidisciplinary Set) | Fairness Perspective | IS Component | Methodological Paradigm | Scope |
|---|---|---|---|---|
| limits of data protection and anti-discrimination in responding to emergent discrimination. Big Data and Society, 6(2) | | | | |
| McDonald N., Pan S. (2020). Intersectional AI: A Study of How Information Science Students Think about Ethics and Their Impact. Proceedings of the ACM on Human-Computer Interaction, 4(CSCW2) | social | broader context | behavioral | generic |
| Miron M., Tolan S., Gómez E., Castillo C. (2020). Evaluating causes of algorithmic bias in juvenile criminal recidivism. Artificial Intelligence and Law | technical | information | exploring bias | domain-specific |
| Mohamed S., Png M.-T., Isaac W. (2020). Decolonial AI: Decolonial Theory as Sociotechnical Foresight in Artificial Intelligence. Philosophy and Technology, 33(4), 659–684 | social | broader context | critical | generic |
| Morley J., Floridi L., Kinsey L., Elhalal A. (2020). From What to How: An Initial Review of Publicly Available AI Ethics Tools, Methods and Research to Translate Principles into Practices. Science and Engineering Ethics, 26(4), 2141–2168 | social | social | literature | generic |
| Morley J., Machado C.C.V., Burr C., Cowls J., Joshi I., Taddeo M., Floridi L. (2020). The ethics of AI in health care: A mapping review. Social Science and Medicine, 260() | social | broader context | literature | domain-specific |
| Mulligan D.K., Kroll J.A., Kohli N., Wong R.Y. (2019). This thing called fairness: Disciplinary confusion realizing a value in technology. Proceedings of the ACM on Human-Computer Interaction, 3(CSCW) | social | broader context | exploring bias | domain-specific |
| Ntoutsi E., Fafalios P., Gadiraju U., Iosifidis V., Nejdl W., Vidal M.-E., Ruggieri S., Turini F., Papadopoulos S., Krasanakis E., Kompatsiaris I., Kinder-Kurlanda K., Wagner C., Karimi F., Fernandez M., Alani H., Berendt B., Kruegel T., Heinze C., Broelemann K., Kasneci G., Tiropanis T., Staab S. (2020). Bias in data-driven artificial intelligence systems—An introductory survey. Wiley Interdisciplinary Reviews: Data Mining and Knowledge Discovery, 10(3) | technical | information <> technology | literature | generic |
| Obermeyer Z., Powers B., Vogeli C., Mullainathan S. (2019). Dissecting racial bias in an algorithm used to manage the health of | technical | technology | exploring bias | domain-specific |

| Article (Multidisciplinary Set) | Fairness Perspective | IS Component | Methodological Paradigm | Scope |
|---|---|---|---|---|
| populations. Science, 366(6464), 447–453 | | | | |
| Olhede S.C., Wolfe P.J. (2018). The growing ubiquity of algorithms in society: Implications, impacts and innovations. Philosophical Transactions of the Royal Society A: Mathematical, Physical and Engineering Sciences, 376(2128) | social | information <> technology | literature | generic |
| Oneto L., Donini M., Pontil M., Shawe-Taylor J. (2020). Randomized learning and generalization of fair and private classifiers: From PAC-Bayes to stability and differential privacy. Neurocomputing, 416(), 231–243 | technical | technology | engineering | generic |
| Orr W., Davis J.L. (2020). Attributions of ethical responsibility by Artificial Intelligence practitioners. Information Communication and Society, 23(5), 719–735 | technical | social | behavioral | generic |
| Otterbacher J., Bates J., Clough P. (2017). Competent men and warm women: Gender stereotypes and backlash in image search results. Conference on Human Factors in Computing Systems - Proceedings, 2017-May(), 6620–6631 | social | information | exploring bias | generic |
| Peña Gangadharan S., Niklas J. (2019). Decentering technology in discourse on discrimination *. Information Communication and Society, 22(7), 882–899 | social | broader context | behavioral | generic |
| Rajkomar A., Hardt M., Howell M.D., Corrado G., Chin M.H. (2018). Ensuring fairness in machine learning to advance health equity. Annals of Internal Medicine, 169(12), 866–872 | technical | technology | literature | domain-specific |
| Robert L.P., Pierce C., Marquis L., Kim S., Alahmad R. (2020). Designing fair AI for managing employees in organizations: a review, critique, and design agenda. Human-Computer Interaction, 35(43987), 545–575 | social | information <> technology | literature | generic |
| Rosenberger R. (2020). "but, that's not phenomenology!": A phenomenology of discriminatory technologies. Techne: Research in Philosophy and Technology, 24(43862), 83–113 | social | broader context | critical | generic |
| Rozado D. (2020). Wide range screening of algorithmic bias in word | technical | information | exploring bias | generic |

| Article (Multidisciplinary Set) | Fairness Perspective | IS Component | Methodological Paradigm | Scope |
|---|---|---|---|---|
| embedding models using large sentiment lexicons reveals underreported bias types. PLoS ONE, 15(4) | | | | |
| Salimi B., Rodriguez L., Howe B., Suciu D. (2019). Interventional fairness: Causal database repair for algorithmic fairness. Proceedings of the ACM SIGMOD International Conference on Management of Data, 793–810 | technical | information | engineering | generic |
| Saxena N.A., Huang K., DeFilippis E., Radanovic G., Parkes D.C., Liu Y. (2020). How do fairness definitions fare? Testing public attitudes towards three algorithmic definitions of fairness in loan allocations. Artificial Intelligence, 283() | technical | broader context | behavioral | domain-specific |
| Scheuerman M.K., Wade K., Lustig C., Brubaker J.R. (2020). How We've Taught Algorithms to See Identity: Constructing Race and Gender in Image Databases for Facial Analysis. Proceedings of the ACM on Human-Computer Interaction, 4(CSCW1) | social | information | exploring bias | generic |
| Schönberger D. (2019). Artificial intelligence in healthcare: A critical analysis of the legal and ethical implications. International Journal of Law and Information Technology, 27(2), 171–203 | social | broader context | literature | domain-specific |
| Skeem J., Lowenkamp C. (2020). Using algorithms to address trade-offs inherent in predicting recidivism. Behavioral Sciences and the Law, 38(3), 259–278 | technical | information | exploring bias | domain-specific |
| Sun W., Nasraoui O., Shafto P. (2020). Evolution and impact of bias in human and machine learning algorithm interaction. PLoS ONE, 15(44051) | technical | social <> technology | exploring bias | generic |
| Taati B., Zhao S., Ashraf A.B., Asgarian A., Browne M.E., Prkachin K.M., Mihailidis A., Hadjistavropoulos T. (2019). Algorithmic bias in clinical populations - Evaluating and improving facial analysis technology in older adults with dementia. IEEE Access, 7(), 25527–25534 | technical | information <> technology | exploring bias | domain-specific |
| Tae K.H., Roh Y., Oh Y.H., Kim H., Whang S.E. (2019). Data cleaning for accurate, fair, and robust models: A big data - AI integration | technical | information | engineering | generic |

| Article (Multidisciplinary Set) | Fairness Perspective | IS Component | Methodological Paradigm | Scope |
|---|---|---|---|---|
| approach. Proceedings of the ACM SIGMOD International Conference on Management of Data | | | | |
| Tannenbaum C., Ellis R.P., Eyssel F., Zou J., Schiebinger L. (2019). Sex and gender analysis improves science and engineering. Nature, 575(7781), 137–146 | technical | information <> technology | critical | generic |
| Thiebes S., Lins S., Sunyaev A. (2020). Trustworthy artificial intelligence. Electronic Markets | social | information <> technology | critical | generic |
| Tomašev N., Cornebise J., Hutter F., Mohamed S., Picciariello A., Connelly B., Belgrave D.C.M., Ezer D., Haert F.C., Mugisha F., Abila G., Arai H., Almiraat H., Proskurnia J., Snyder K., Otake-Matsuura M., Othman M., Glasmachers T., Wever W., Teh Y.W., Khan M.E., Winne R.D., Schaul T., Clopath C. (2020). AI for social good: unlocking the opportunity for positive impact. Nature Communications, 11(1) | social | broader context | exploring bias | generic |
| Turner Lee N. (2018). Detecting racial bias in algorithms and machine learning. Journal of Information, Communication and Ethics in Society, 16(3), 252–260 | social | broader context | exploring bias | generic |
| Ugwudike P. (2020). Digital prediction technologies in the justice system: The implications of a 'race-neutral' agenda. Theoretical Criminology, 24(3), 482–501 | social | social <> technology | critical | domain-specific |
| Unceta I., Nin J., Pujol O. (2020). Risk mitigation in algorithmic accountability: The role of machine learning copies. PLoS ONE, 15(44146) | technical | technology | engineering | domain-specific |
| Vaccaro K., Sandvig C., Karahalios K. (2020). "At the End of the Day Facebook Does What It Wants": How Users Experience Contesting Algorithmic Content Moderation. Proceedings of the ACM on Human-Computer Interaction, 4(CSCW2) | social | social <> technology | behavioral | domain-specific |
| Van Berkel N., Goncalves J., Hettiachchi D., Wijenayake S., Kelly R.M., Kostakos V. (2019). Crowdsourcing perceptions of fair predictors for machine learning: A recidivism case study. Proceedings of the ACM on Human-Computer Interaction, 3(CSCW) | social | social <> technology | behavioral | domain-specific |

| Article (Multidisciplinary Set) | Fairness Perspective | IS Component | Methodological Paradigm | Scope |
|---|---|---|---|---|
| van Berkel N., Tag B., Goncalves J., Hosio S. (2020). Human-centred artificial intelligence: a contextual morality perspective. Behaviour and Information Technology | social | broader context | behavioral | generic |
| Van Nuenen T., Ferrer X., Such J.M., Cote M. (2020). Transparency for Whom? Assessing Discriminatory Artificial Intelligence. Computer, 53(11), 36–44 | social | broader context | exploring bias | generic |
| Veale M., Binns R. (2017). Fairer machine learning in the real world: Mitigating discrimination without collecting sensitive data. Big Data and Society, 4(2) | social | information | engineering | generic |
| Veale M., Van Kleek M., Binns R. (2018). Fairness and accountability design needs for algorithmic support in high-stakes public sector decision-making. Conference on Human Factors in Computing Systems - Proceedings, 2018-April() | social | social | behavioral | domain-specific |
| Wang M., Deng W. (2020). Mitigating bias in face recognition using skewness-aware reinforcement learning. Proceedings of the IEEE Computer Society Conference on Computer Vision and Pattern Recognition, 9319–9328 | technical | technology | engineering | generic |
| Wellner G., Rothman T. (2020). Feminist AI: Can We Expect Our AI Systems to Become Feminist?. Philosophy and Technology, 33(2), 191–205 | technical | social <> technology | critical | generic |
| Wexler J., Pushkarna M., Bolukbasi T., Wattenberg M., Viegas F., Wilson J. (2020). The what-if tool: Interactive probing of machine learning models. IEEE Transactions on Visualization and Computer Graphics, 26(1), 56–65 | technical | social <> technology | engineering | generic |
| Wong P.-H. (2020). Democratizing Algorithmic Fairness. Philosophy and Technology, 33(2), 225–244 | social | broader context | critical | generic |
| Woodruff A., Fox S.E., Rousso-Schindler S., Warshaw J. (2018). A qualitative exploration of perceptions of algorithmic fairness. Conference on Human Factors in Computing Systems - Proceedings, 2018-April() | social | social | behavioral | generic |

| Article (Multidisciplinary Set) | Fairness Perspective | IS Component | Methodological Paradigm | Scope |
|---|---|---|---|---|
| Yarger L., Cobb Payton F., Neupane B. (2019). Algorithmic equity in the hiring of underrepresented IT job candidates. Online Information Review, 44(2), 383–395 | social | information | critical | domain-specific |
| Yoon T., Lee J., Lee W. (2020). Joint Transfer of Model Knowledge and Fairness over Domains Using Wasserstein Distance. IEEE Access, 8(), 123783–123798 | technical | technology | engineering | generic |
| Završnik A. (2019). Algorithmic justice: Algorithms and big data in criminal justice settings. European Journal of Criminology | social | social <> technology | literature | domain-specific |
| Zehlike M., Hacker P., Wiedemann E. (2020). Matching code and law: achieving algorithmic fairness with optimal transport. Data Mining and Knowledge Discovery, 34(1), 163–200 | technical | technology | engineering | generic |
| Zhang S., Medo M., Lü L., Mariani M.S. (2019). The long-term impact of ranking algorithms in growing networks. Information Sciences, 488(), 257–271 | technical | technology | engineering | generic |
| Zink A., Rose S. (2020). Fair regression for health care spending. Biometrics, 76(3), 973–982 | technical | technology | engineering | domain-specific |
| Žliobaitė I. (2017). Measuring discrimination in algorithmic decision making. Data Mining and Knowledge Discovery, 31(4), 1060–1089 | technical | technology | literature | generic |